\documentclass[onecolumn]{aastex63}
\usepackage[encapsulated]{CJK}
\usepackage{mathrsfs}
\usepackage{subfigure}
\usepackage{epstopdf}
\usepackage{amsmath}
\usepackage{longtable}
\usepackage{booktabs}
\usepackage{hyperref} 
\usepackage{overpic}
\usepackage{lineno}
\usepackage{ulem}
\usepackage{color}
\usepackage{natbib}
\shorttitle{Rotation in the Pleiades, $\alpha$~Per, and Hyades clusters}
\shortauthors{Hao et al.}

\defcitealias{hao2022b}{Paper\,I}

\definecolor{malachite}{rgb}{0.34, 0.7, 0.22}

\begin{document}

   \title{Probing the nature of rotation in the Pleiades, Alpha Persei, and Hyades clusters}
   
   \correspondingauthor{Ye Xu, Ligang Hou}
   \email{xuye@pmo.ac.cn, lghou@nao.cas.cn}
   
   \author{C. J. Hao}
   \affiliation{Purple Mountain Observatory, Chinese Academy of Sciences, Nanjing 210023, PR China}
   \affiliation{School of Astronomy and Space Science, University of Science and Technology of China, Hefei 230026, PR China}
   \author{Y. Xu}
   \affiliation{Purple Mountain Observatory, Chinese Academy of Sciences, Nanjing 210023, PR China}
   \affiliation{School of Astronomy and Space Science, University of Science and Technology of China, Hefei 230026, PR China}
   \author{L. G. Hou}
   \affiliation{National Astronomical Observatories, Chinese Academy of Sciences, 20A Datun Road, Chaoyang District, Beijing 100101, PR China}
   \author{S. B. Bian}
   \affiliation{Purple Mountain Observatory, Chinese Academy of Sciences, Nanjing 210023, PR China}
   \author{Z. H. Lin}
   \affiliation{Purple Mountain Observatory, Chinese Academy of Sciences, Nanjing 210023, PR China}
   \author{Y. J. Li}
   \affiliation{Purple Mountain Observatory, Chinese Academy of Sciences, Nanjing 210023, PR China}
   \author{Y. W. Dong}
   \affiliation{Purple Mountain Observatory, Chinese Academy of Sciences, Nanjing 210023, PR China}
   \affiliation{School of Astronomy and Space Science, University of Science and Technology of China, Hefei 230026, PR China}
   \author{D. J. Liu}
   \affiliation{Purple Mountain Observatory, Chinese Academy of Sciences, Nanjing 210023, PR China}
   \affiliation{School of Astronomy and Space Science, University of Science and Technology of China, Hefei 230026, PR China}

%-------------------------------------------------------------------
 
\begin{abstract}
Unraveling the internal kinematics of open clusters is crucial  
for understanding their formation and evolution. 
However, there is a dearth of research on this topic, primarily due to
the lack of high-quality kinematic data.
Using the exquisite-precision astrometric parameters and radial velocities 
provided by \emph{Gaia} data release 3, we investigate the internal 
rotation in three of the most nearby and best-studied open clusters, namely 
the Pleiades, Alpha Persei, and Hyades clusters.
Statistical analyses of the residual motions of the member stars clearly indicate 
the presence of three-dimensional rotation in the three clusters.
The mean rotation velocities of the Pleiades, Alpha Persei, and Hyades 
clusters within their tidal radii are estimated to be
0.24 $\pm$ 0.04, 0.43 $\pm$ 0.08, and 0.09 $\pm$ 0.03~km~s$^{-1}$, 
respectively. 
Similar to the Praesepe cluster that we have studied before,
the rotation of the member stars within the tidal radii of 
these three open clusters can be well interpreted by Newton's theorem.
No expansion or contraction is detected in the three clusters either.
Furthermore, we find that the mean rotation velocity of open clusters may be 
positively correlated with the cluster mass, and the rotation is likely to diminish 
as open clusters age.
\end{abstract}

%\maketitle
%
%-------------------------------------------------------------------

\section{Introduction}
\label{intro}

Most stars in the Milky Way are thought to form in star
clusters~\citep{lada2003,bressert2010,megeath2016}. 
Open clusters (OCs), gravitationally bound systems of 
tens to thousands of stars, are the rare survivors of the embedded clusters  
gestated in molecular clouds~\citep{elmegreen1985,bastian2006}, 
with ages spanning a broad range from a few million 
years (Myr) to tens of billions of years.
At present, the formation, evolution, and eventual 
dispersal of OCs are still poorly 
understood~\citep[e.g.,][]{proszkow2009,farias2018,hao2023}, 
although thousands of Galactic OCs have been 
discovered~\citep[e.g.,][]{hao2022a,castro2022,hunt2023}. 
In order to better understand the formation and evolution of OCs, 
one of the key aspects is to obtain a detailed characterization of their internal 
kinematics~\citep[see][and references therein]{dellacroce2023}.
Hydrodynamical simulations of the cluster formation and very early 
evolution suggest that rotation may be common in the progenitors
of OCs~\citep[e.g.,][]{mapelli2017}.
However, unlike the Galactic globular clusters, whose rotation has 
been studied extensively~\citep[e.g.,][]{vanLeeuwen2000,lanzoni2018,leanza2022}, 
our knowledge about the internal rotation of OCs is sorely lacking.

To date, the signals of rotation have been reported in only about a dozen 
of OCs.
By identifying the correlation between the tangential velocities and the 
parallaxes of member stars,
\citet{vereshchagin2013a} and \citet{vereshchagin2013b}  
explored the potential rotation in the Hyades and Pleiades clusters, and 
speculated that their rotation was comparable.
Similarly, \citet{kamann2019} and \citet{healy2021} studied the tangential 
or radial velocities of cluster members as a function of distances to the 
cluster center, and they 
reported the presence of rotational signals in the Preasepe and NGC 6791  
and the absence of rotational signals in the Pleiades and NGC 6819. 
Based on the second data release of \emph{Gaia}~\citep[DR2,][]{gaia2016,gaia2018}, 
\citet{loktin2020}~analyzed the rotation in the Praesepe 
through the residual proper motion or radial velocity methods.
Recently, by using the early data release 3 of \emph{Gaia}~\citep[EDR3,][]{gaia2021}
to perform vector-field inference and create spatio-kinematic maps of OCs,
\citet{guilherme2023} reported the detection of rotation patterns in eight clusters, 
with nine additional objects displaying possible rotation signs.
However, all these studies adopted the methods of projection motion 
(proper motion or radial velocity) to study the rotation of OCs. 
The three-dimensional (3D) rotation of OCs, especially their rotation nature,
remained elusive until the 3D rotation of Praesepe was revealed
by~\citet[][hereafter, Paper I]{hao2022b}.

Observationally, studying the 3D rotation of OCs is challenging.
One of the main difficulties is obtaining a detailed characterization 
of the internal stellar kinematics, which requires high-quality measurements of 
the proper motions and radial velocities for the member stars.
For the OCs with dozens to hundreds of members assembled in crowded fields,
the precise spatial positions, and especially, the parallaxes of the member stars, 
are additional difficulties when rotation is to be inferred.
\emph{Gaia} DR3~\citep{gaia2023} opens the 
possibility of new studies of stellar cluster dynamics.
Considering the data quality of \emph{Gaia} DR3, we focus on the nearby 
OCs that are rich in member stars, 
as \emph{Gaia} DR3 is able to provide high-precision astrometric parameters 
and especially the radial velocities for a sufficient number of their member stars.
In addition, reliable membership determination of an OC is also critical for 
revealing its internal rotation.
After cross-matching a series of large-scale surveys to complement 
the \emph{Gaia} catalog,
\citet{lodieu2019,lodieu2019b}~provided revised membership lists of four 
nearby star-rich OCs based on comprehensive analyses, including the Pleiades, 
Praesepe, Alpha Persei ($\alpha$~Per), and the Hyades.
The ages of these four OCs span tens to hundreds of million years. 
In~\citetalias{hao2022b}, we have developed the data-analyzing method and
studied the Praesepe.
In this work, we aim to assess the 3D rotation of the Pleiades, $\alpha$~Per,
and the Hyades with~\emph{Gaia} DR3, and then explore the
possible relation between the rotation of OCs and the cluster parameters
(e.g., age, mass, and radius).

The Pleiades, also known as the Seven Sisters, are easily visible to the naked eye 
in the constellation of Taurus~\citep{lynga1981}. 
The distance to the Pleiades has been estimated by many 
methods, such as the trigonometric parallax~\citep[e.g.,][]{gatewood2000,
soderblom2005}, 
fitting a color-magnitude diagram~\citep[e.g.,][]{anDeokkeun2007}, and 
adoption of the eclipsing binaries as standard candles~\citep[e.g.,][]{southworth2005}.
Currently, the distance to the Pleiades has been agreed to be the mean value
of~130--140 pc~\citep{melis2014,gaia2016}.
The age of the Pleiades cluster is a subject of debate,
with estimates ranging from 70 to 80 Myr based on comparisons of the 
color-magnitude diagram of the cluster with theoretical isochrones 
of stellar evolution~\citep[e.g.,][]{mermilliod1981,vandenberg1984},
to 120–130 Myr derived from the lithium-depletion boundary 
method~\citep[e.g.,][]{basri1996,dahm2015}, and 132 Myr inferred from 
the white dwarf in the cluster~\citep{lodieu2019}.
In addition, the Pleiades contain more than 1000 member stars 
that share common proper motions of
($\mu_{\alpha^{*}}$, $\mu_{\delta}$)~$\sim$~(19.5, --45.5)
mas yr$^{-1}$, a high Galactic latitude~($b$ $\sim$ 24$^{\circ}$), and 
low reddening along the line of sight of~$E(B - V)$ = 0.03 
mag~\citep[see][]{jones1981,robichon1999,vanLeeuwen2009}.
This cluster has a tidal radius of 11.6--19.5 pc, a core radius of 0.9--3.0 
pc, and a cluster mass of 721--870~${\rm M}_{\odot}$~\citep{pinfield1998,
raboud1998,converse2008,converse2010}.

The cluster $\alpha$~Per, also known as Melotte 20 or Collinder 
39~\citep{melotte1915,collinder1931}, is located in the northern constellation 
of Perseus. It is composed of  
several blue B-type stars~\citep{zuckerman2012}. 
The distance to this cluster is estimated to be between 170 and 190 pc~\citep[][]{robichon1999,vanLeeuwen2009,majaess2011,babusiaux2018}.
The trigonometric parallax and infrared color-magnitude diagram fitting 
methods give consistent distances for $\alpha$~Per, making it a benchmark 
on the cosmic distance ladder. 
In a comprehensive study, \citet{lodieu2019}~derived a mean distance of  
177.7 $\pm$ 0.8 pc for the cluster.
Furthermore, the $\alpha$~Per has an estimated age of approximately 50--70 
Myr~\citep{prosser1996,makarov2006}.
Its extinction along the line of sight is estimated to be $A_{V}$ = 0.30 mag, 
with a possible differential extinction~\citep{prosser1992}.
The $\alpha$~Per is located at a low Galactic latitude of~\citep[$b$ $\sim$ 
--7$^{\circ}$, e.g.,][]{makarov2006,vanLeeuwen2009}, has a proper motion
of ($\mu_{\alpha^{*}}$, $\mu_{\delta}$)~$\sim$~(23.1, --25.8)~mas 
yr$^{-1}$~\citep{lodieu2019}, a core radius of 2.3 $\pm$ 0.3 pc, and a tidal radius 
of $\sim$9.5 pc~\citep{lodieu2019}.

The Hyades, regarded as a sister cluster of the Pleiades, is the 
nearest open cluster and one of the best-studied star 
clusters~\citep{messier1781}. 
Using the $Hipparcos$ dataset, \citet{perryman1998}~deduced the cluster 
distance to be 46.3 $\pm$ 0.3 pc, similar to the results obtained by using the
$Hubble$ Space Telescope measurements~\citep{mcArthur2011} and infrared 
color-magnitude diagram fitting~\citep{majaess2011}.
Based on the \emph{Gaia}'s trigonometric parallax, the distance of the cluster has 
been updated to 47.0 $\pm$ 0.2 pc~\citep{lodieu2019}.
The age of the Hyades cluster has been estimated by different methods,
such as the theoretical isochrone fitting method~\citep[625 $\pm$ 50 
Myr,][]{maeder1981,mermilliod1981}, the cooling age of white dwarf 
members~\citep[648 $\pm$ 45 Myr,][]{deGennaro2009}, and the method of 
the lithium-depletion boundary~\citep[650 $\pm$ 70 Myr,][]{lodieu2018}.
The cluster consists of a roughly spherical group of hundreds of member stars 
presenting the proper motions of 70--140~mas yr$^{-1}$ on the
sky~\citep{perryman1998,vanLeeuwen2009,lodieu2019}.
The cluster has a core radius of 2.5--3.0 pc and a tidal radius of 10 
pc~\citep[see][]{perryman1998,roser2011}. 
The reddening of the Hyades is low, with $E(B - V)$ $\le$ 0.001 
mag~\citep{taylor2006}.

With the high-quality data from \emph{Gaia} DR3, the focus of this work is to 
reveal the internal kinematics of the Pleiades, $\alpha$~Per, and the Hyades 
clusters in detail. 
We first update the catalog of member stars of the three 
OCs based on~\emph{Gaia} DR3, and then we extract those members with astrometric 
parameters and radial velocities. 
Subsequently, we update the method used in~\citetalias{hao2022b} to 
investigate the rotation of the Pleiades, 
$\alpha$~Per, and the Hyades.
When a cluster does rotate, we determine its rotation axis, derive the 
rotational velocities, and analyze the rotational 
characteristics of the member stars.
Finally, we discuss the possible relations between the 
internal rotation of OCs and the cluster parameters.

\section{Data}
\label{data}

To study the internal kinematics of OCs, we adopt the \emph{Gaia} DR3 dataset. 
The astrometric and photometric parameters in \emph{Gaia} DR3 are 
inherited from~\emph{Gaia} EDR3, which contains 
celestial positions (R.A., Decl.), parallaxes ($\varpi$), 
proper motions ($\mu_{\alpha^{*}}$, $\mu_{\delta}$), and three photometric 
passbands ($G$, $G_{\rm BP}$, and $G_{\rm RP}$) of more than 
a billion stellar sources.
In addition, in comparison to \emph{Gaia} DR2,
more than 33 million objects with new determinations of the 
mean radial velocities are also included in \emph{Gaia} DR3.
In this work, we adopt the radial velocity measurements provided by~{\it Gaia} DR3
to ensure the sample homogeneity and avoid the potentially large systematic errors.
% 

%%%%%%%%%%%%%%%%%%%%%%%%%%%%%%%%%%%%%%%%%%%% Table. 1
\setlength{\tabcolsep}{15.0mm}
\begin{table}[ht]
	\centering
	\caption{Number of the member stars selected in different (sub) samples or different steps
	 for the Pleiades, $\alpha$ Per, and the Hyades.}
	\begin{tabular}{cccc} 
		\hline \hline 
		 &    Pleiades      &     $\alpha$ Per     &    Hyades    \\ \hline 
		Initial sample\tablenotemark{\footnotesize a} &     2281           &  3162    &   1764       \\  
		Step 1           &     66              &  182                      &     87        \\  
		Step 2           &     75              &  256                     &    133       \\    
		Step 3           &     77              &  894                     &    907       \\  
		Step 4           &     265            &  596                     &    45         \\  
		Sample\tablenotemark{\footnotesize b}          &     1798           &  1234   &    592       \\    \hline  
		Step 5           &     226            &  137                      &    215        \\  
		Step 6           &     1                &    0                       &       1         \\  
		Subsample\tablenotemark{\footnotesize c}     &     225            &  137     &     214       \\    \hline  
	\end{tabular}
	\tablecomments{For details of Steps 1--6 see Section~\ref{data}. 
	\tablenotetext{\footnotesize a}{The membership lists are from~\citet{lodieu2019} for
	the Pleiades and $\alpha$ Per, and from~\citet{lodieu2019b} for the Hyades.} 
	\tablenotetext{\footnotesize b}{After subtracting the member stars processed in ``Steps 1--4'' from 
	the ``Initial sample'', the samples contain the stars with reliable astrometric 
	parameters provided by {\it Gaia} DR3.}
	\tablenotetext{\footnotesize c}{After subtracting the member stars processed in ``Step 6'' from those of ``Step 5'', the subsamples contain the stars with radial velocities provided by {\it Gaia} DR3.}
	}
	\label{table:numbers}
\end{table}
%%%%%%%%%%%%%%%%%%%%%%%%%%%%%%%%%%%%%%%%%%%%%%%

The Pleiades, $\alpha$~Per, and the Hyades are three member-rich OCs 
in the solar neighborhood that span a wide range of ages.
We initiate the study by adopting the membership lists of the Pleiades, Hyades, 
and $\alpha$~Per 
clusters provided by~\citet{lodieu2019,lodieu2019b}.
\citet{lodieu2019} combined the \emph{Gaia} DR2 dataset with multiple large-scale 
public surveys from the optical to mid-infrared wavelengths to provide the 
revised membership lists for the Pleiades and $\alpha$~Per clusters.
The selection criteria for the member stars in~\citet{lodieu2019} are initially designed 
to be as inclusive as possible and become more selective in the later process.
Finally, they obtained 2281 member stars for the Pleiades and 3162 member stars 
for the $\alpha$~Per. 
With the same method, \citet{lodieu2019b}~compiled a list of member stars for the 
Hyades cluster, which includes a total of 1764 member stars.
We use \emph{Gaia} DR3 to update the aforementioned three membership lists. 
The details are described below. 
The numbers of selected member stars in each step are given 
in Table \ref{table:numbers}. 

\begin{itemize}
\item Step 1. Selection of the unmatched members after cross-matching the member 
stars with the~{\it Gaia} DR3 database through \emph{Gaia}'s~\texttt{source\_id}.

\item Step 2. Removal of the members that lack astrometric~($ \varpi$, $\mu_{\alpha^{*}}$, 
$\mu_{\delta}$) or photometric ($G$, $G_{\rm BP}$, and ~$G_{\rm RP}$) parameters.

\item Step 3. Rejection of those members that lie beyond three times of the tidal radius ($ r_{\rm td} $) 
from the cluster center. The distances of the members to the cluster center are taken
from~\citet{lodieu2019,lodieu2019b}.

\item Step 4. Selection of the members that have large parallax and/or proper motion 
uncertainties based on the exclusion criteria of the typical uncertainties of the faint 
stars ($G$ = 20) observed by \emph{Gaia}, i.e., e$_{\varpi}$ $\geq$ 0.5 mas, and
e$_{\mu_{\alpha^{*}, \delta}}$ $\geq$ 0.6 mas~yr$^{-1}$~\citep{gaia2023}.
Member stars whose parallaxes and/or proper motions significantly deviate (three 
times the standard deviation) from the overall sample are also excluded.

\item Step 5. Selection of the members with radial velocity measurements and corresponding 
uncertainties smaller than~2~km~s$^{-1}$, which is the same as~\citetalias{hao2022b}.  
Members whose radial velocities significantly deviate from the overall sample are
also deleted.

\item Step 6. Elimination of the nonsingle members that have not been correctly handled in 
the \emph{Gaia}'s astrometric solution according to the criterion 
of~\texttt{ipd\_gof\_harmonic\_amplitude} 
$>$ 0.1 and~\texttt{ruwe} $>$ 1.4 \citep{gaia2021}.
\end{itemize}

\section{Methods}
\label{methods}

The basic steps we develope to study the 3D rotation of OCs are described 
in detail in~\citetalias{hao2022b}.
Here, we only give a brief description.
First, we use the astrometric parameters of the cluster members to 
determine their coordinates ($x_{\rm g}$, $y_{\rm g}$, $z_{\rm g}$) 
and velocities ($v_{x_{\rm g}}$, $v_{y_{\rm g}}$, $v_{z_{\rm g}}$)
in the Galactic Cartesian coordinate 
system ($O_{\rm g}$--$X_{\rm g}$$Y_{\rm g}$$Z_{\rm g}$),
where the uncertainties are estimated with a Monte Carlo method by 
taking into account the observational errors.
Second, the coordinates ($x_{\rm c}$, $y_{\rm c}$, $z_{\rm c}$) and
velocities ($v_{x_{\rm c}}$, $v_{y_{\rm c}}$, $v_{z_{\rm c}}$)
of the members in the cluster Cartesian coordinate system 
($O_{\rm c}$--$X_{\rm c}$$Y_{\rm c}$$Z_{\rm c}$)
are converted from those in the 
$O_{\rm g}$--$X_{\rm g}$$Y_{\rm g}$$Z_{\rm g}$ system.
Third, based on the velocities ($v_{x_{\rm c}}$, $v_{y_{\rm c}}$, $v_{z_{\rm c}}$) 
of the cluster members, we calculate the 3D mean residual velocities and 
estimate their uncertainties with Monte Carlo simulations in order to 
investigate their dependence on the position 
angles (PA; i.e., $\alpha$, $\beta$, and $\gamma$ in Figure~\ref{fig:ccs2axis}),
as well as to check the rotation of the cluster and determine the rotation axis 
$\vec{l}$.
Afterward, the coordinates ($x_{\rm r}$, $y_{\rm r}$, $z_{\rm r}$) and
velocities ($v_{x_{\rm r}}$, $v_{y_{\rm r}}$, $v_{z_{\rm r}}$)
of the members in the cluster rotation Cartesian coordinate system 
($O_{\rm r}$--$X_{\rm r}$$Y_{\rm r}$$Z_{\rm r}$, Figure~\ref{fig:ccs2axis}) 
can be obtained according to the rotation axis, which can then be transformed 
into the cylindrical coordinate system $(r, \varphi, z)$ to reveal the internal 
rotation of an OC.

Based on the above steps, we have made some updates in this work.
First, the method is improved by analyzing the mean residual 
velocity of member stars within three times, one time, and half of the 
cluster's tidal radius ($r_{\rm td}$) as a function of PAs, respectively. 
This approach can enhance the reliability of the rotation representation 
within the cluster and better determine the rotation axis $\vec{l}$.
Another improvement addresses the difficulty of determining the cluster 
rotation axis in the $O_{\rm c}$--$X_{\rm c}$$Y_{\rm c}$$Z_{\rm c}$ 
system when the rotational signals (sinusoidal behavior) are not significant 
in all three dimensions simultaneously.
To fix this problem, we rotate the 
$O_{\rm c}$--$X_{\rm c}$$Y_{\rm c}$$Z_{\rm c}$ system around its origin 
($O_{\rm c}$) with a given angle.
Concretely speaking, with a set of PAs ($\alpha_{0}$, $\beta_{0}$, $\gamma_{0}$),
we transform the 3D coordinates and velocity components of cluster 
members from the $O_{\rm c}$--$X_{\rm c}$$Y_{\rm c}$$Z_{\rm c}$ system 
into the rotated Cartesian coordinate system 
(signed as $O_{\rm r}'$--$X_{\rm r}$$Y_{\rm r}$$Z_{\rm r}$).
Subsequently, we use the residual velocity method to determine the rotation in the
cluster and estimate the PAs 
($\alpha'$, $\beta'$, $\gamma'$) of the rotation axis $\vec{l}$ in the rotated system.
Then, the vectors of the rotation axis $\vec{l}$ can be converted from 
the $O_{\rm r}'$--$X_{\rm r}$$Y_{\rm r}$$Z_{\rm r}$ system back into the 
$O_{\rm c}$--$X_{\rm c}$$Y_{\rm c}$$Z_{\rm c}$ system via
\begin{equation}
  \begin{pmatrix}
   x_{\rm c} \\ y_{\rm c} \\ z_{\rm c}
  \end{pmatrix} = 
  \begin{pmatrix}
{\rm cos}~\alpha_{1} & {\rm cos}~\alpha_{2} & {\rm cos}~\alpha_{3} \\ {\rm cos}~\beta_{1} & {\rm cos}~\beta_{2} & {\rm cos}~\beta_{3} \\ {\rm cos}~\gamma_{1} & {\rm cos}~\gamma_{2} & {\rm cos}~\gamma_{3}
  \end{pmatrix}
  \begin{pmatrix}
   x_{\rm r} \\ y_{\rm r} \\ z_{\rm r}
  \end{pmatrix},
  \label{equ:ccs2rcc2}
\end{equation}
where $\alpha_{i}$, $\beta_{i}$, and $\gamma_{i}$ are the included angles
between the axes of the $O_{\rm c}$--$X_{\rm c}$$Y_{\rm c}$$Z_{\rm c}$ 
system and those of the $O_{\rm r}'$--$X_{\rm r}$$Y_{\rm r}$$Z_{\rm r}$ system.
When the vectors of $\vec{l}$ in the 
$O_{\rm c}$--$X_{\rm c}$$Y_{\rm c}$$Z_{\rm c}$ system 
are obtained, the PAs of $\vec{l}$ in the 
$O_{\rm c}$--$X_{\rm c}$$Y_{\rm c}$$Z_{\rm c}$ system 
($\alpha$, $\beta$, $\gamma$) can be determined.
Then, we continue to perform the aforementioned procedures to analyze 
the internal rotation of an OC.

In the same way as in~\citetalias{hao2022b}, to elucidate the rotational nature of 
the member stars of an OC,
we compare their root-mean-square (RMS) rotational 
velocities and the theoretical predictions from 
Newton’s theorems, with the following equation:
\begin{flalign}
\begin{split}
v_{\rm c} (r) = \left[G M_{\rm t} \cdot \frac{r^{2}}{(r^2~+~r_{\rm co}^{2})^{3/2}}\right]^{1/2}.
\label{equ:v-t}
\end{split}
\end{flalign}
Here, $G$ = $4.3 \times 10^{-3}~{\rm pc~M}_{\odot}^{-1}~({\rm km~s}^{-1})^{2}$ 
is the gravitational constant, $r_{\rm co}$ is the core radius, and $M_{\rm t}$ is the 
cluster mass inside the tidal radius.

\section{Results}
\label{results}

\subsection{Rotation in Praesepe with Gaia DR3}

In~\citetalias{hao2022b}, we studied the rotational properties of the 
Praesepe cluster based on~\emph{Gaia} EDR3.
In comparison to~EDR3, \emph{Gaia} DR3 provides high-precision radial velocity 
measurements for more cluster members.
We reanalyze the Praesepe rotation with the new data of cluster members
and the amended method described above.
We find that the features of the mean residual velocities of the member stars 
within 3 $r_{\rm td}$, 1 $r_{\rm td}$, and 0.5 $r_{\rm td}$ of the Praesepe 
all confirm the 3D rotation in the cluster.
In addition, the rotational characteristics of member stars within the cluster tidal radius 
are almost consistent with those in~\citetalias{hao2022b} and can be well 
described by Newton's theorems. 
Moreover,, the new dataset still shows no expansion or contraction in the
Praesepe.
As shown in~Table~\ref{table:pa_praesepe}, by comparison,
the consistent parameters confirm
that the results obtained in~\citetalias{hao2022b} are reliable, and the new 
data-analyzing method updated in this work is feasible.
%

%%%%%%%%%%%%%%%%%%%%%%%%%%%%%%%%%%%%%%%%%%%% Table. 3
\setlength{\tabcolsep}{8.0mm}
\renewcommand\arraystretch{1.0}
\begin{table}[ht]
\centering
\caption{Comparison of the rotational parameters of the Praesepe for the results obtained
in this work and those in~\citetalias{hao2022b}.}
\begin{tabular}{c|cc} 
\hline  \hline 
  &  This work &  \citetalias{hao2022b} \\  \hline 
Position angles  &    $\alpha$, $\beta$,  $\gamma$ &  $\alpha$, $\beta$,  $\gamma$  \\
0.5 $r_{\rm td}$   &  $119^{\circ}$ $\pm$ $10^{\circ}$, $133^{\circ}$ $\pm$ $12^{\circ}$,
$214^{\circ}$ $\pm$ $8^{\circ}$  &  
-- \\ 
1.0 $r_{\rm td}$    &  $128^{\circ}$ $\pm$ $4^{\circ}$, $132^{\circ}$ $\pm$ $7^{\circ}$,
$212^{\circ}$ $\pm$ $5^{\circ}$  &     
$121^{\circ}$ $\pm$ $5^{\circ}$, $118^{\circ}$ $\pm$ $17^{\circ}$,
$211^{\circ}$ $\pm$ $8^{\circ}$  \\  
3.0 $r_{\rm td}$   &  $136^{\circ}$ $\pm$ $6^{\circ}$, $148^{\circ}$ $\pm$ $5^{\circ}$, 
$210^{\circ}$ $\pm$ $5^{\circ}$  &  
$140^{\circ}$ $\pm$ $3^{\circ}$, $152^{\circ}$ $\pm$ $7^{\circ}$, 
$210^{\circ}$ $\pm$ $2^{\circ}$ \\     \hline  
$\vec{l}$ vs. GP &    $34^{\circ}$ $\pm$ $15^{\circ}$ &  $41^{\circ}$ $\pm$ $12^{\circ}$  \\  \hline  
Rotation velocity &    0.12 $\pm$ 0.05~km~s$^{-1}$ &  0.16 $\pm$ 0.05~km~s$^{-1}$  \\  \hline
Core radius &    2.0 $\pm$ 0.4~pc &  1.6 $\pm$ 0.5~pc  \\  \hline  
Tidal mass  &    609 $\pm$ 135~${\rm M}_{\odot}$ &  537 $\pm$ 146~${\rm M}_{\odot}$  \\  \hline  
 \end{tabular}
 \tablecomments{$r_{\rm td}$: tidal radius of the Praesepe. GP: Galactic plane.
 $\vec{l}$ vs. GP: the included angle between $\vec{l}$ and the Galactic plane.}
 \label{table:pa_praesepe}
\end{table}
%%%%%%%%%%%%%%%%%%%%%%%%%%%%%%%%%%%%%%%%%%%%%%%

%

\subsection{The Pleiades cluster}

%%%%%%%%%%%%%%%%%%%%%%%%%%%%%%%%%%%%%%% Figure 2
\begin{figure}[ht]
	\begin{center}
		\includegraphics[width=1.00\textwidth]{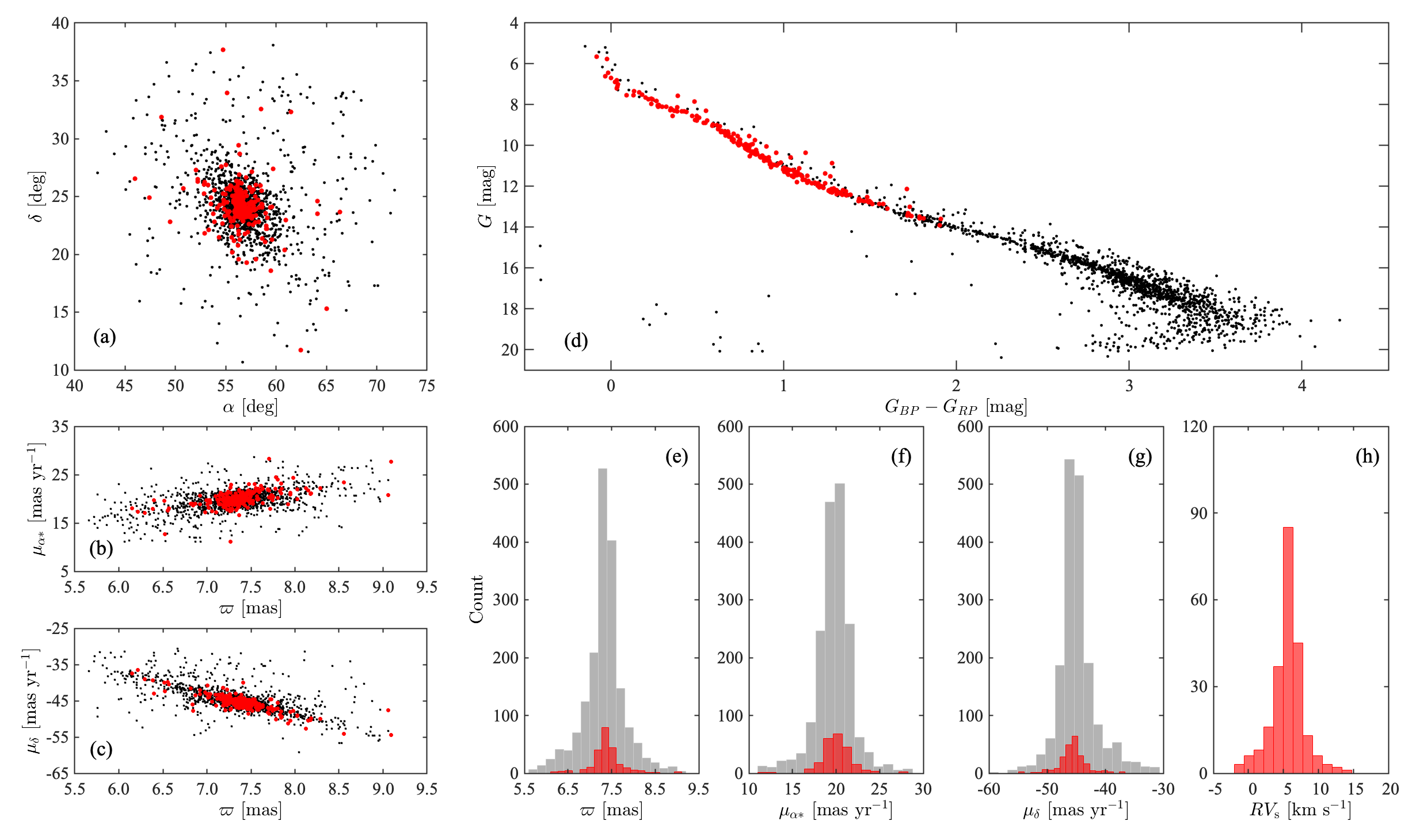}
		\caption{Multidimensional distributions of all the member stars (gray) and those with radial 
		velocity measurements (red) for the Pleiades cluster. 
			Panel~(a): the distribution in the sky; 
			Panels~(b) and (c): the relationship between parallax and proper motions; 
			Panel~(d): the color-magnitude diagram; 
			Panels~(e), (f), (g) and (h): the histograms of parallax, proper motions and radial 
			velocity, respectively.
		}
		\label{fig:distribution_pleiades}
	\end{center}
\end{figure}
%%%%%%%%%%%%%%%%%%%%%%%%%%%%%%%%%%%%%%%

Figure~\ref{fig:distribution_pleiades}~shows the multidimensional distributions of  
astrometric parameters (R.A., Decl., $\varpi$, $\mu_{\alpha^{*}}$, $\mu_{\delta }$) 
and color-magnitude diagram of all the member stars (grey) and those with radial 
velocities (red) for the Pleiades.
The observed parameters of the member stars can be 
approximated by a multidimensional normal distribution.
In this section, we describe the rotational properties of the Pleiades cluster and explore 
the rotational characteristics of the cluster members.
%

%%%%%%%%%%%%%%%%%%%%%%%%%%%%%%%%%%%%%%%%%%%%% Figure. 3
\begin{figure*}[ht]
\centering
  \includegraphics[width=0.32\textwidth]{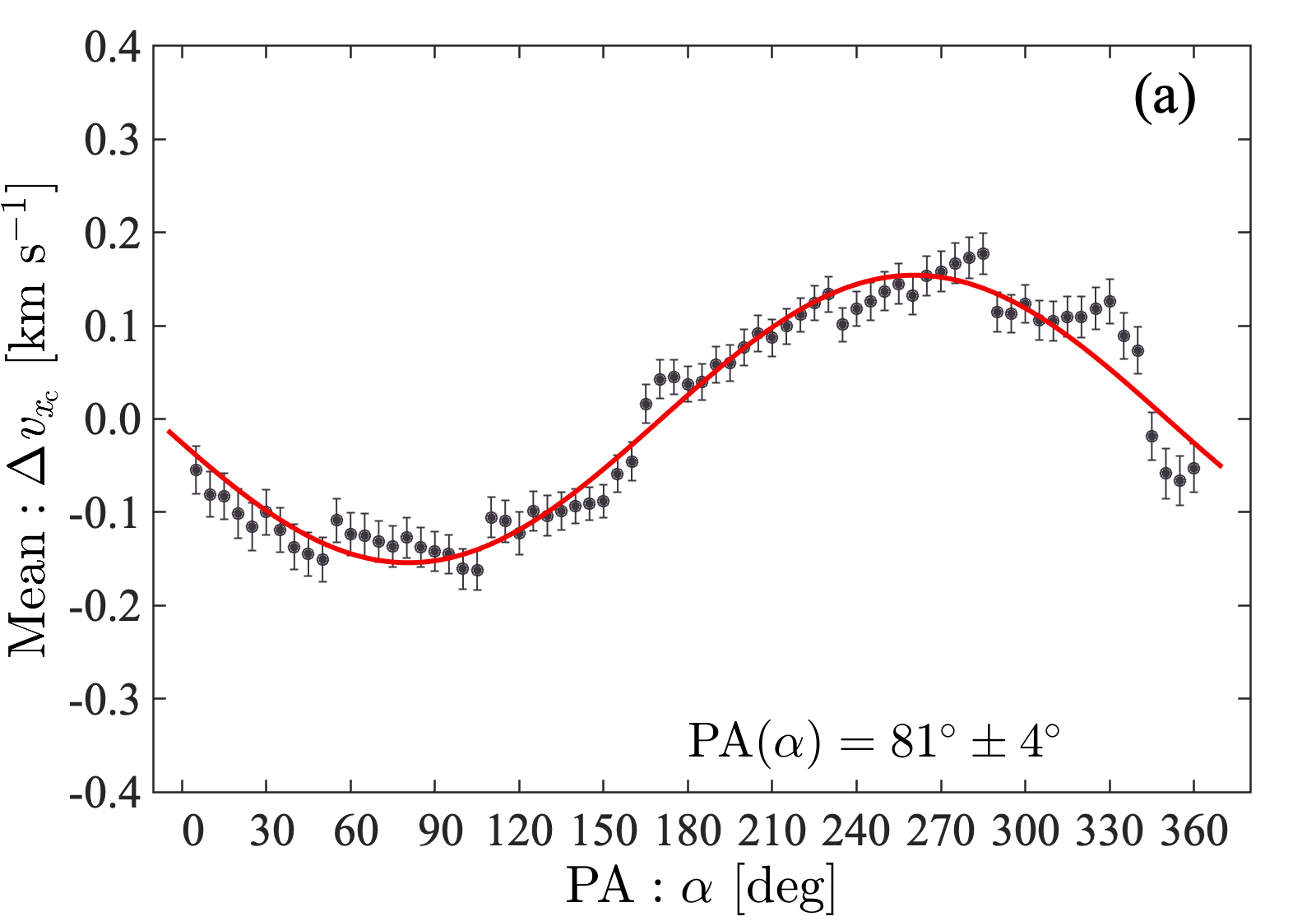}
  \includegraphics[width=0.32\textwidth]{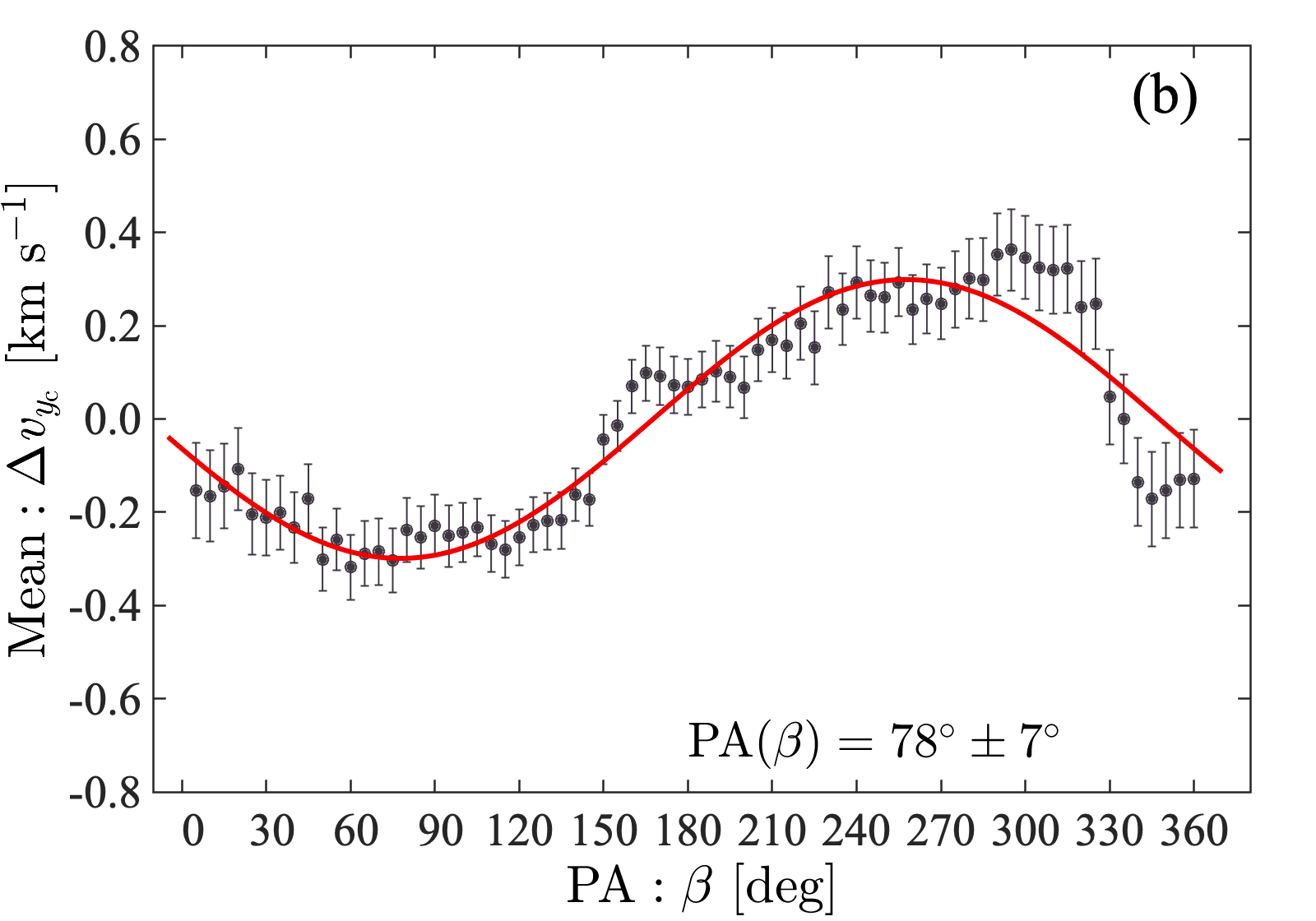}
  \includegraphics[width=0.32\textwidth]{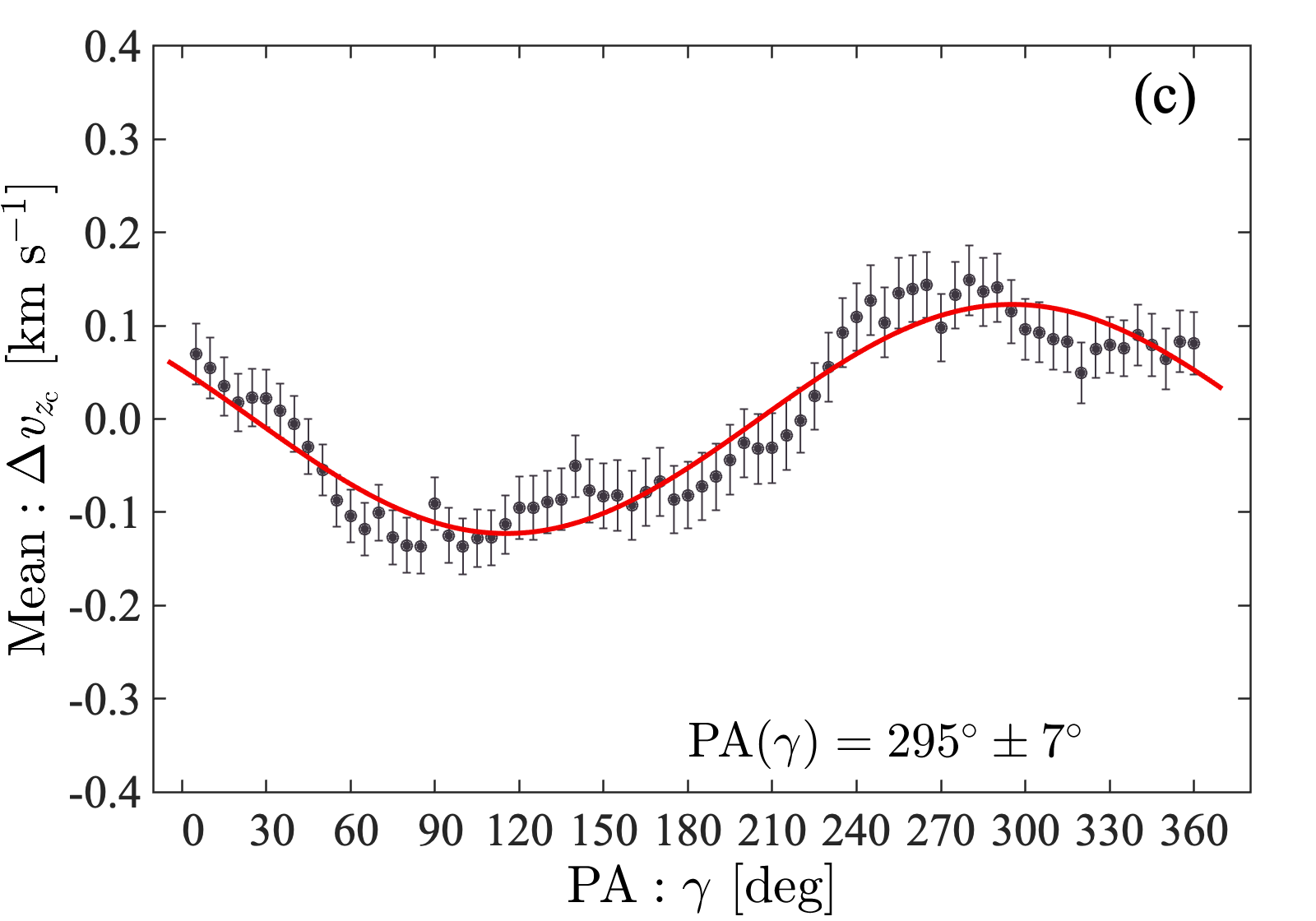}
  \includegraphics[width=0.32\textwidth]{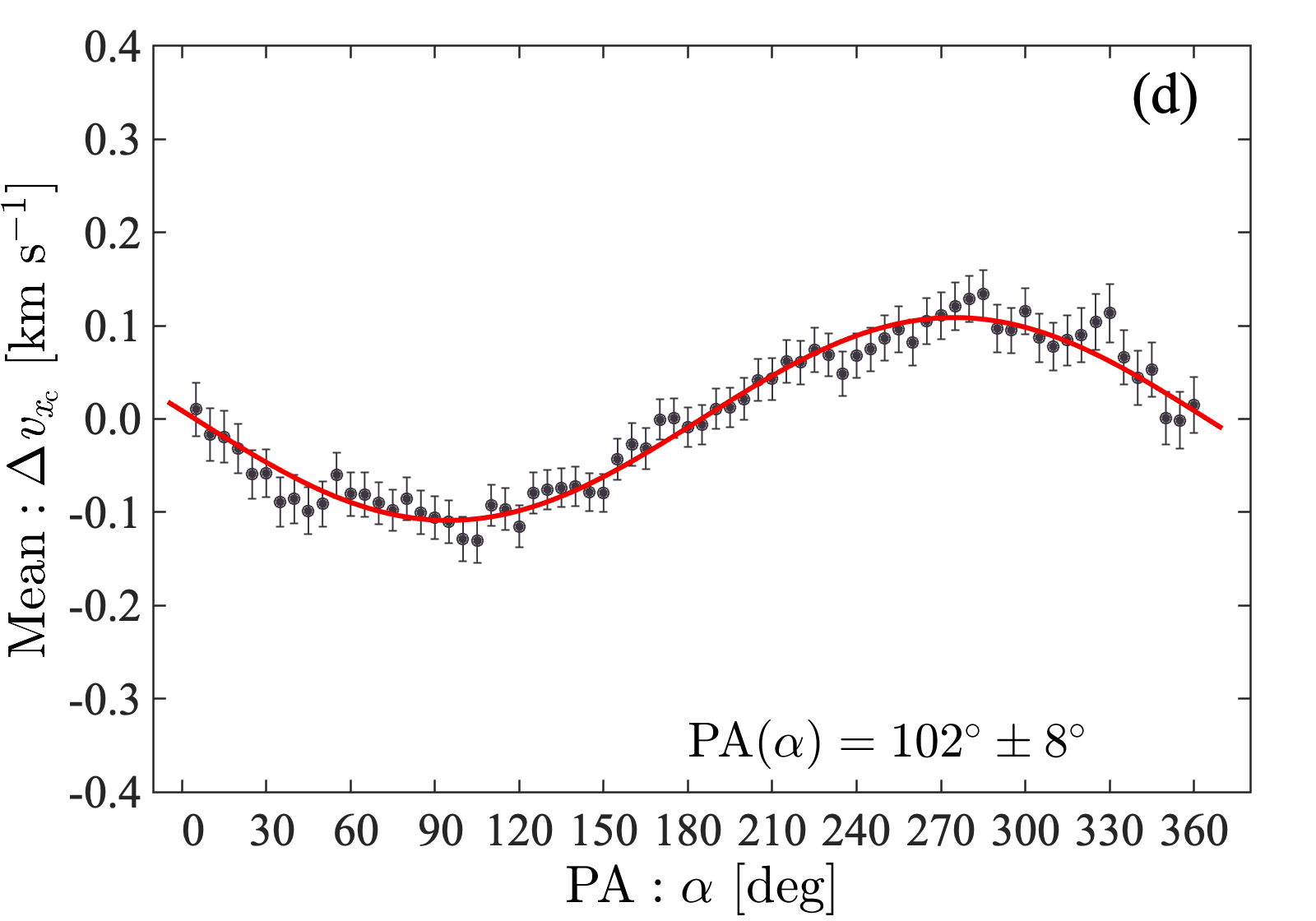}
  \includegraphics[width=0.32\textwidth]{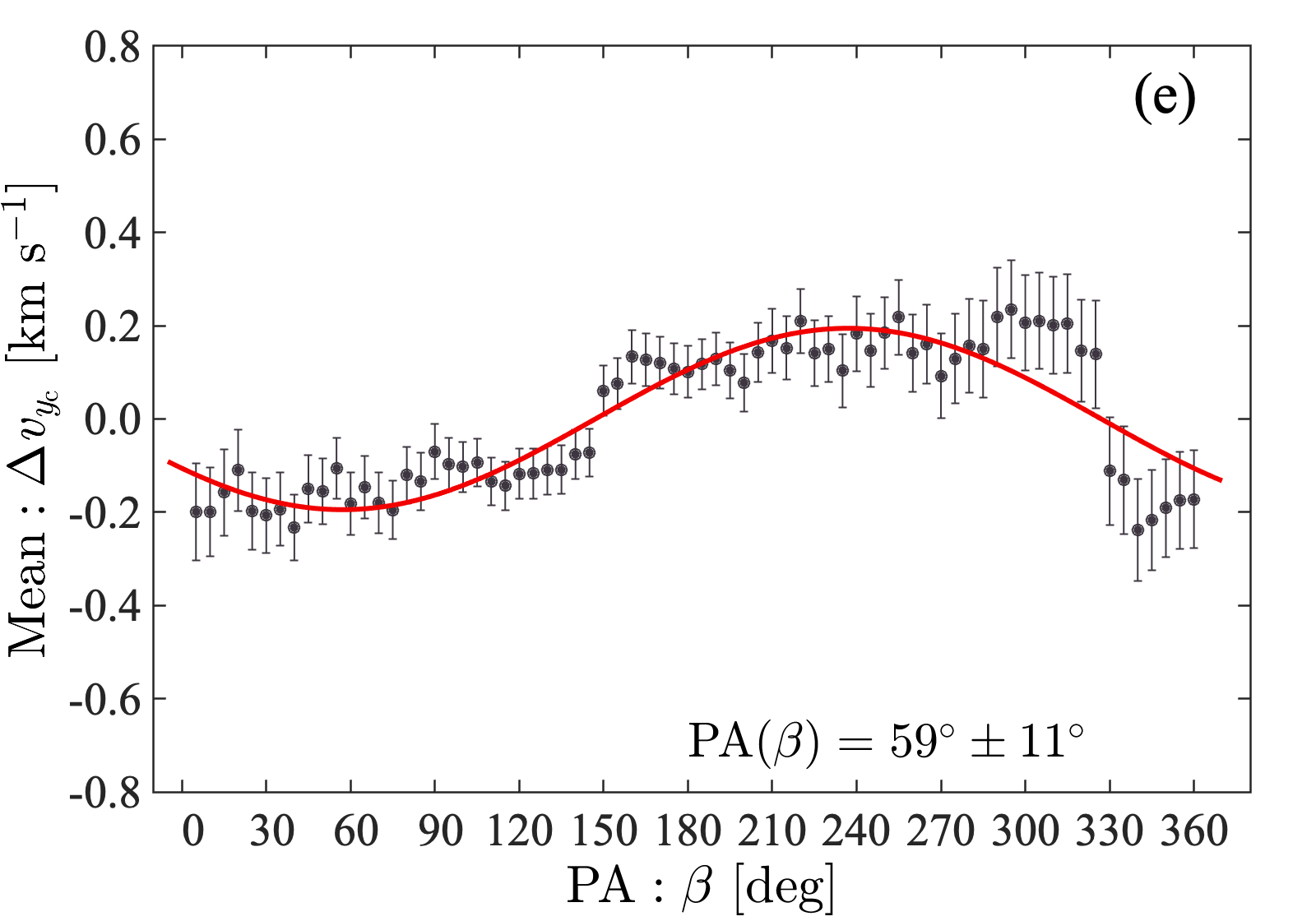}
  \includegraphics[width=0.32\textwidth]{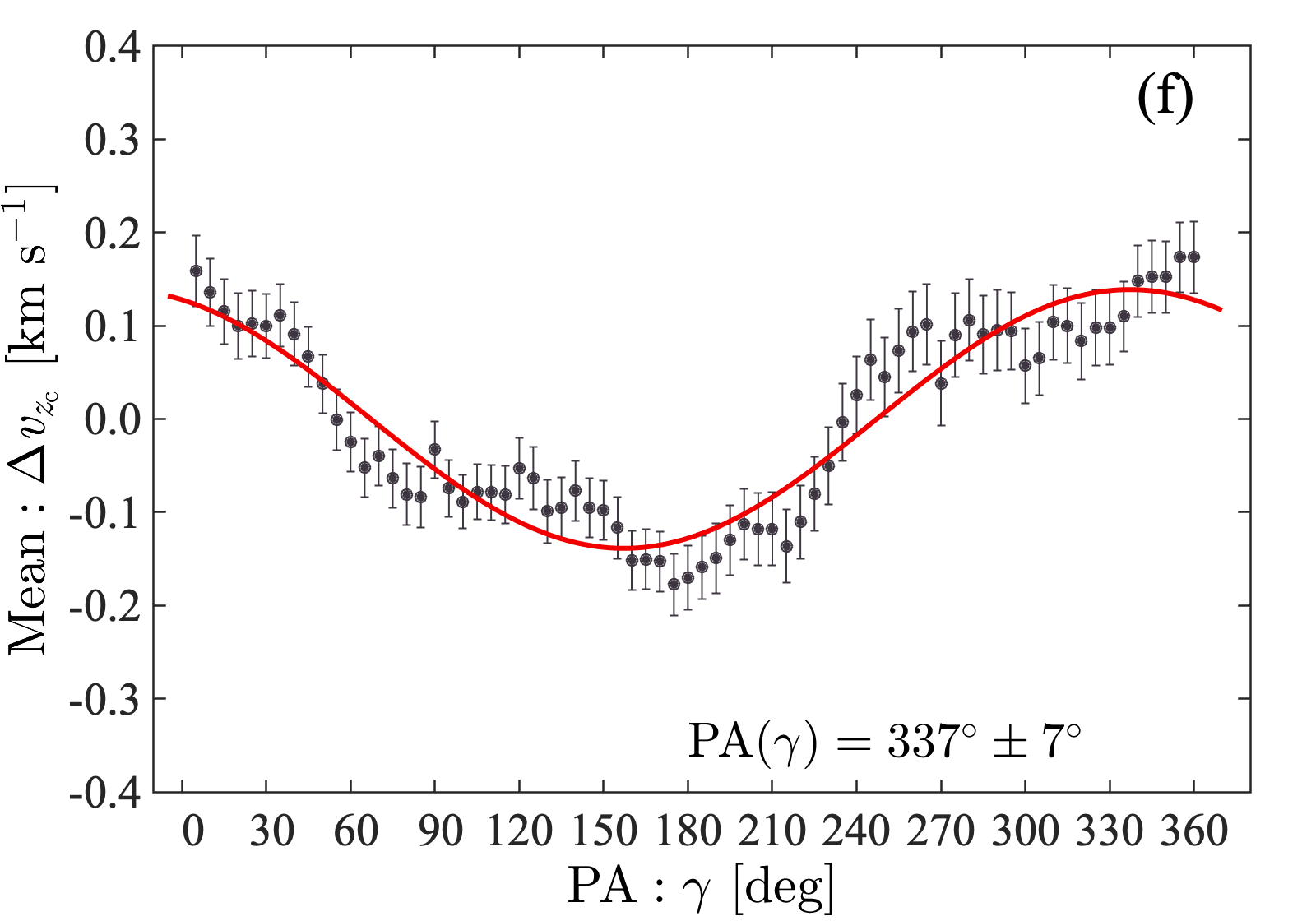}
  \includegraphics[width=0.32\textwidth]{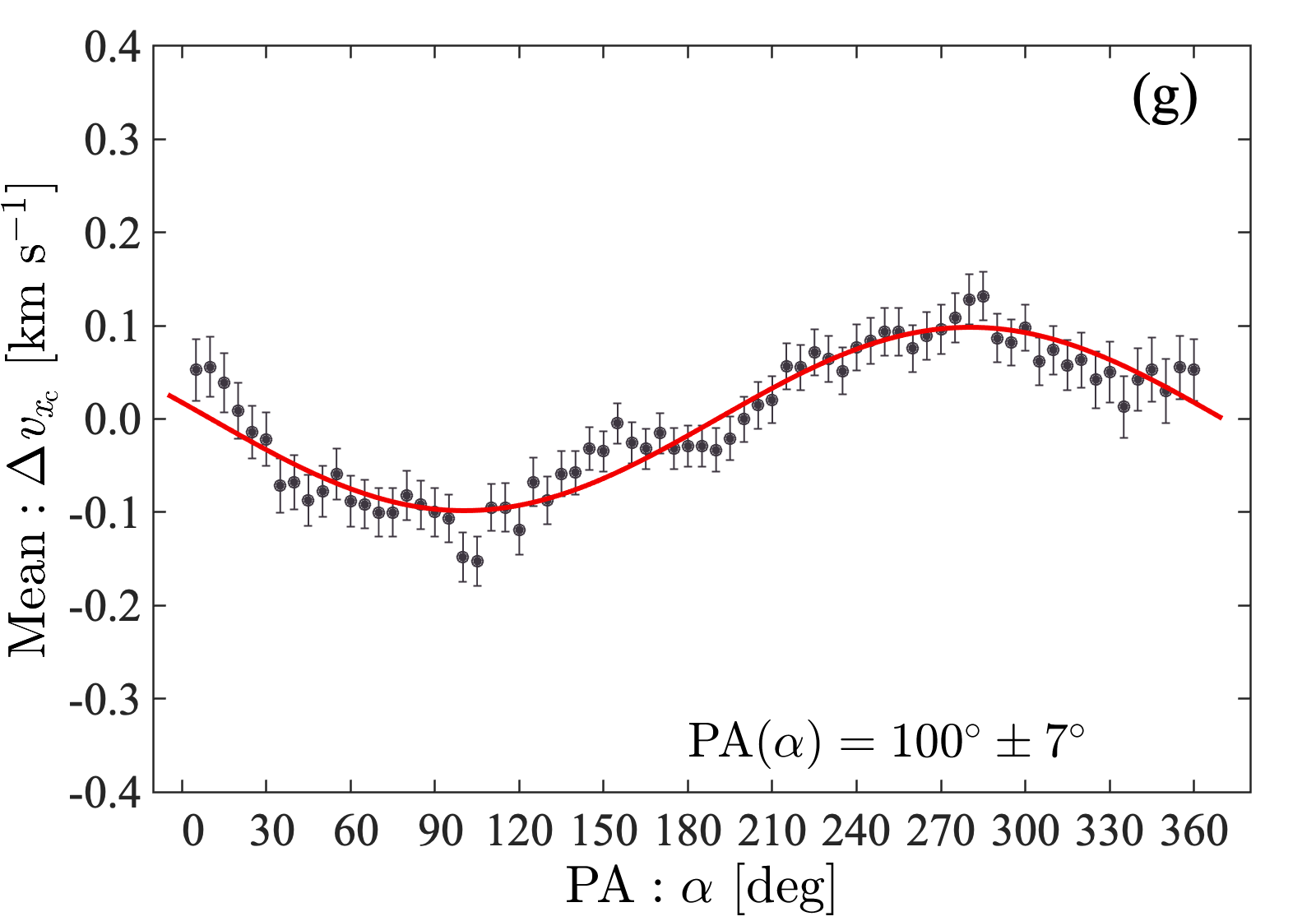}
  \includegraphics[width=0.32\textwidth]{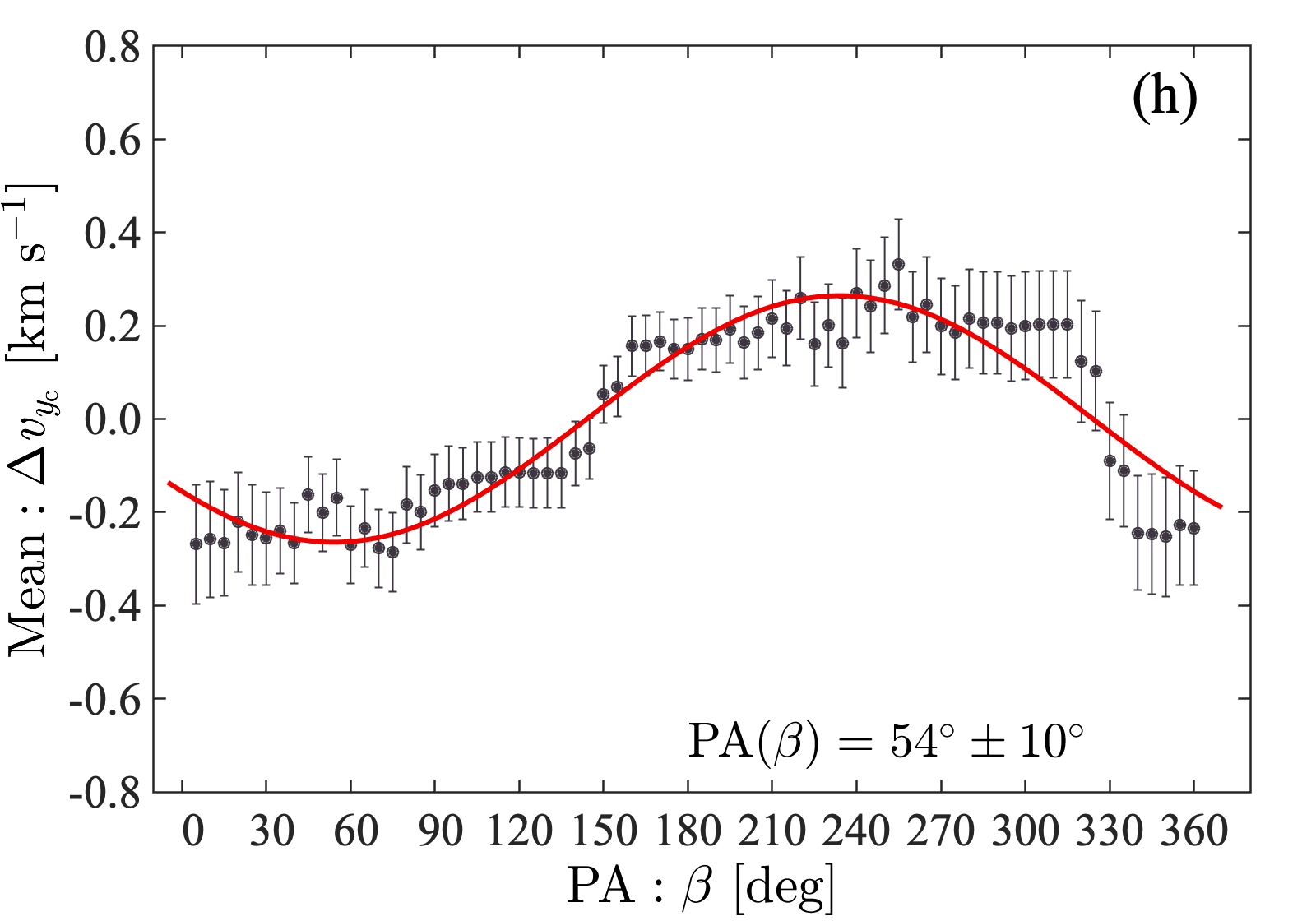}
  \includegraphics[width=0.32\textwidth]{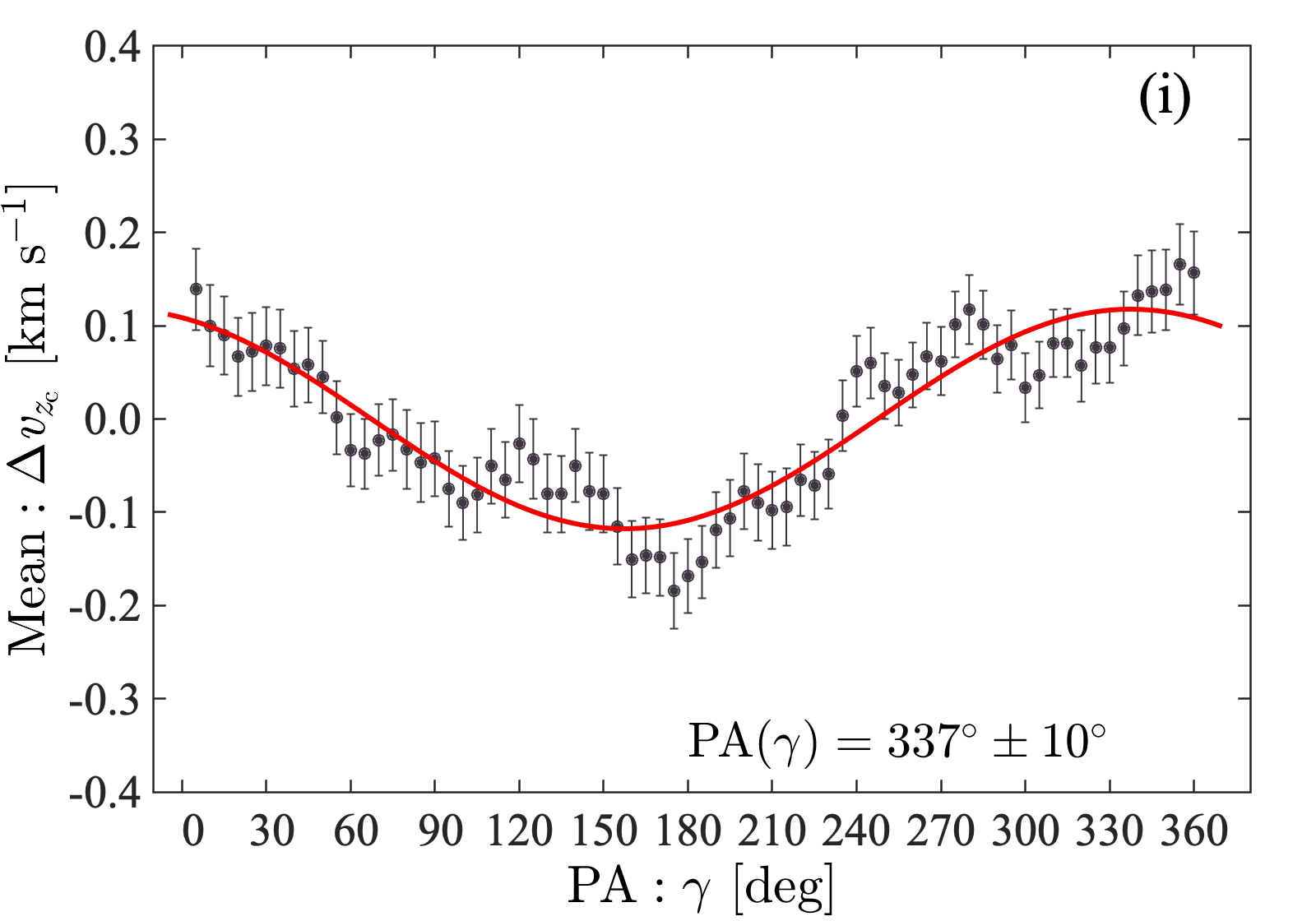}
\caption{Mean residual velocities as a function of the position angles
    (PAs) for the Pleiades. The left, middle, and right panels show the 
    $v_{x_{\rm c}}$ vs. $ \alpha $, $v_{y_{\rm c}}$ vs. $ \beta$, and 
    $v_{z_{\rm c}}$ vs. $ \gamma $ plots, respectively. 
    The results for the member stars within three times of the tidal radius (Panels a, b, c),
    within a tidal radius (Panels d, e, f), and within half a tidal radius (Panels g, h, i) 
    are shown from top to bottom, respectively.
    The error bars (gray) and the best-fitting sine function (red) are also shown in each panel.}
\label{fig:pa_pleiades}
\end{figure*}
%%%%%%%%%%%%%%%%%%%%%%%%%%%%%%%%%%%%%%%%%%%%%

\subsubsection{Rotation in the Pleiades}
\label{sec:Pleiades}
For the 225 member stars with radial velocities (see Table~\ref{table:numbers}), 
the median uncertainties of parallax, proper 
motion~$\mu_{\alpha^{*}}$~and~$\mu_{\delta}$~are~0.02 mas, 0.02~mas~yr$^{-1}$, 
and~0.02~mas~yr$^{-1}$, respectively. 
The median uncertainty of radial velocity is~0.6~km s$^{ -1}$. 
Based on these member stars, the means and standard deviations 
of the astrometric parameters of the Pleiades cluster are determined as: 
(R.A., Decl.) = ($56.58^{\circ}$ $\pm$ $2.51^{\circ}$, $24.35^{\circ}$ $\pm$ $2.47^{\circ}$), 
$\varpi$ = $7.38$ $\pm $ $0.37$ mas, 
proper motions ($\mu_{\alpha^{*}}$, $\mu_{\delta}$) = 
($20.04$ $\pm$ $1.80$, $-45.40$ $\pm$ $2.59$ ) mas yr$^{-1}$, 
and radial velocity RV = $5.57$ $\pm$ $0.77$ km s$^{-1}$. 
These values are consistent with the previous determinations~\citep{liu1991,mermilliod1997,
gao2019,babusiaux2018,lodieu2019}.
With these parameters, we derive the 3D coordinates and velocities of the
Pleiades and its member stars in the Galactic Cartesian coordinate system 
($O_{\rm g}$--$X_{\rm g}$$Y_{\rm g}$$Z_{\rm g}$), and then transform 
them to the $O_{\rm c}$--$X_{\rm c}$$Y_{\rm c}$$Z_{\rm c}$ 
system.
After that, the mean residual velocities and uncertainties of the member stars 
as a function of PAs are calculated. 
We assign an $r_{\rm td}$ of 11.6 pc to the Pleiades cluster, 
which comes from~\citet{lodieu2019}. 
Figure~\ref{fig:pa_pleiades}~(a--c), (d--f), and (g--i)~present the 
mean residual velocities as functions of PAs for the member stars within 
3 $r_{\rm td}$, 1 $r_{\rm td}$, and 0.5 $r_{\rm td}$, respectively. 
The error bars in the figures 
are the uncertainties of the mean residual velocities. 
As shown in Figure~\ref{fig:pa_pleiades}, the clear sinusoidal 
behaviors indicate the presence of 3D rotation in the Pleiades cluster.
%

%%%%%%%%%%%%%%%%%%%%%%%%%%%%%%%%%%%%%%%%%%%% Table. 3
\setlength{\tabcolsep}{10.0mm}
\begin{table}[ht]
\centering
\caption{Best-fit PAs of the Pleiades cluster.}
\begin{tabular}{c|ccc} 
\hline  \hline 
           &    $\alpha$      &    $\beta$    &    $\gamma     $  \\ \hline 
3.0 $r_{\rm td}$   &  $81^{\circ}$ $\pm$ $4^{\circ}$   &  $78^{\circ}$ $\pm$ $7^{\circ}$  
          &  $295^{\circ}$ $\pm$ $7^{\circ}$     \\  
1.0 $r_{\rm td}$      &  $102^{\circ}$ $\pm$ $8^{\circ}$  &  $59^{\circ}$ $\pm$ $11^{\circ}$    
          &  $337^{\circ}$ $\pm$ $7^{\circ}$     \\  
0.5 $r_{\rm td}$   &  $100^{\circ}$ $\pm$ $7^{\circ}$   &  $54^{\circ}$ $\pm$ $10^{\circ}$    
          &  $337^{\circ}$ $\pm$ $10^{\circ}$     \\    \hline  
 \end{tabular}
 \tablecomments{$r_{\rm td}$: tidal radius of the Pleiades.}
 \label{table:pa_pleiades}
\end{table}
%%%%%%%%%%%%%%%%%%%%%%%%%%%%%%%%%%%%%%%%%%%%%%%

Using the residual velocity method, the PAs ($\alpha$, $\beta$, $\gamma$) of the 
rotation axis $\vec{l}$ of the Pleiades cluster in the 
$O_{\rm c}$--$X_{\rm c}$$Y_{\rm c}$$Z_{\rm c}$ system can be determined.
The best-fitting results are shown in Table~\ref{table:pa_pleiades} and
Figure~\ref{fig:pa_pleiades}, 
where the errors come from the Monte Carlo simulations by
considering the uncertainties of the mean residual velocities.
The PAs obtained by the member stars within 1 $r_{\rm td}$ are consistent with 
those fitted by using the members within 0.5 $r_{\rm td}$. They satisfy the 
relation of tan $\alpha$ $\cdot$ tan $\gamma$ = tan $\beta$
considering the uncertainties.
Nevertheless, the PAs obtained by the member stars of the entire sample are 
different from the above results, although the three angles also satisfy 
tan $\alpha$ $\cdot$ tan $\gamma$ = tan $\beta$ considering the uncertainties.
This phenomenon aligns with the findings in~\citetalias{hao2022b} for the Praesepe 
cluster, which may be due to the fact that the member 
stars located outside the cluster's tidal radius are not only dominated by the 
OC itself, but also by the tidal force of the Milky Way.

We employ the best-fitting PAs for the member stars within 1 $r_{\rm td}$ 
to determine the rotation axis $\vec{l}$ of the Pleiades cluster. 
The angle between $\vec{l}$ and the Galactic plane is estimated 
to be~$62^\circ\pm13^\circ$, where the errors come from the uncertainties
of the obtained PAs.
Considering the relationship~of tan $\alpha$ $\cdot$ tan $\gamma$ = tan $\beta$, 
we adopt 
($\alpha$, $\beta$, $\gamma$) = ($102^{ \circ}$, $63^{\circ}$, $337^{\circ}$) 
to derive the vectors of the $X_{\rm r}$, $Y_{\rm r}$, and $Z_{\rm r}$ axes in the 
$O_{\rm c}$--$X_{\rm c}$$Y_{\rm c}$$Z_{\rm c}$ system, as described in
Section~\ref{methods}.
Then, the 3D coordinates and velocity components of the member 
stars are transformed from the $O_{\rm c}$--$X_{\rm c}$$Y_{\rm c}$$Z_{\rm c}$ 
system to the $O_{\rm r}$--$X_{\rm r}$$Y_{\rm r}$$Z_{\rm r}$ system,
and eventually converted to the 3D coordinates 
($r$, $\varphi$, $z$) and velocity components ($v_ {r}$, $v_{\varphi}$, $v_{z}$)
in the cylindrical coordinate system.
Based on the rotational velocities ($v_{\varphi}$) and the corresponding 
uncertainties of the member stars within 
1 $r_{\rm td}$, the mean rotation velocity of the Pleiades cluster is estimated 
to be~0.24 $\pm$ 0.04~km~s$^{-1}$, which is in general agreement with the 
results illustrated in Figure~\ref{fig:pa_pleiades}.
%
%

%%%%%%%%%%%%%%%%%%%%%%%%%%%%%%%%%%%%%%%%%%%%% Figure. 4
\begin{figure}[ht]
\centering
\includegraphics[scale=0.18]{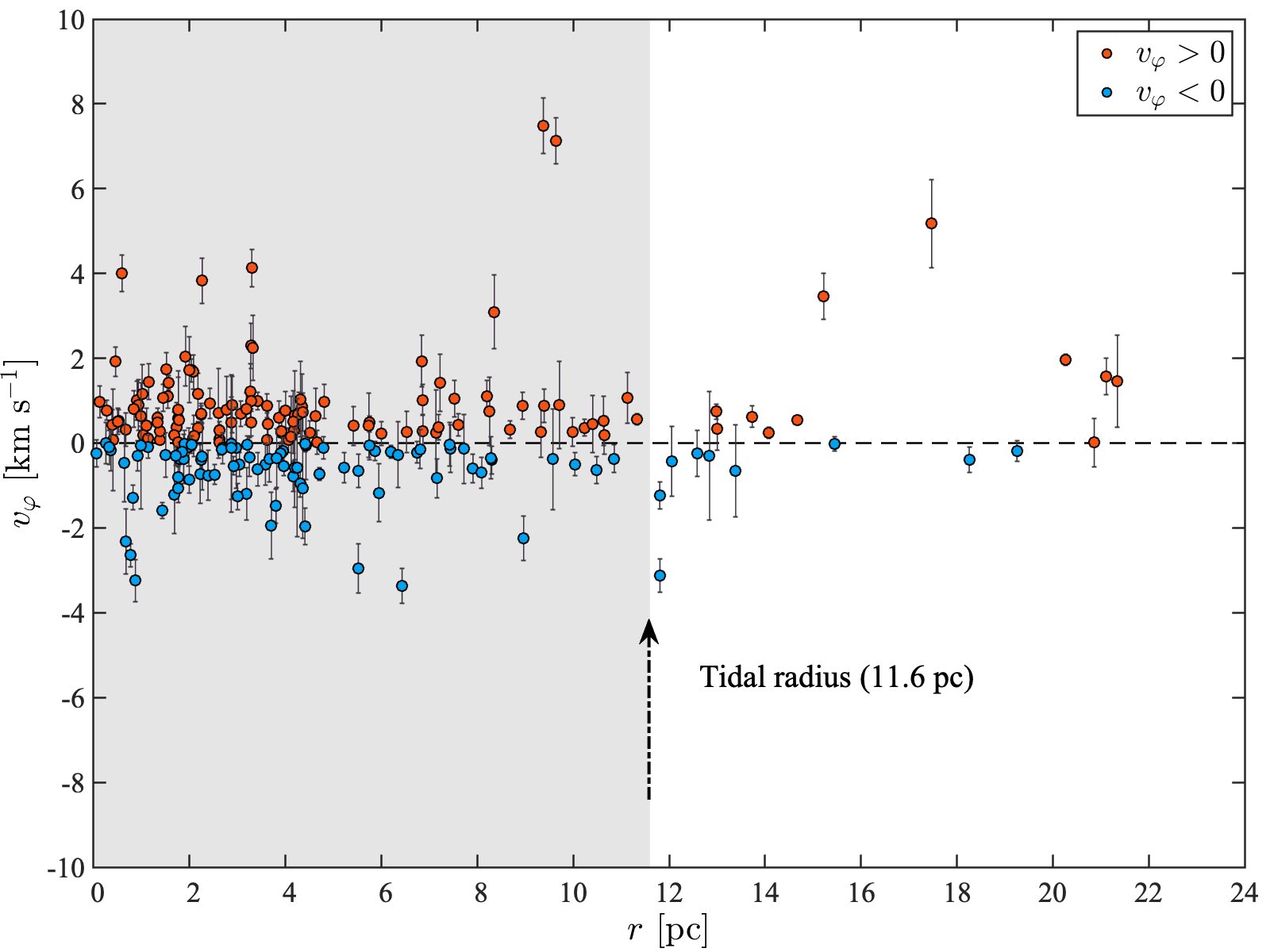}
\caption{Rotational velocity as a function of the
	distance to the cluster center for the member stars within
	twice the tidal radius of the Pleiades.}
\label{fig:rotation_pleiades}
\end{figure}
%%%%%%%%%%%%%%%%%%%%%%%%%%%%%%%%%%%%%%%%%%%%%

\subsubsection{Rotational properties of the member stars of Pleiades}
\label{sec:rotation-Pleiades}

%%%%%%%%%%%%%%%%%%%%%%%%%%%%%%%%%%%%%%%%%%%% Table. 4
\setlength{\tabcolsep}{8.0mm}
\begin{table}[ht]
\centering 
\caption{RMS rotational velocities of the member stars within
    the tidal radius of the Pleiades.}
\begin{tabular}{cc|cccc} 
  \hline  \hline 
$R_{i}$  &  $R_{o}$ & $R_{m}$  & $N$  & RMS $v_{\varphi}$  &  $\epsilon_{{\rm RMS} \, v_{\varphi}}$   \\ %\hline
      {[pc]}  &  {[pc]}    &  {[pc]}  &  & {[km~s$^{-1}$]} & {[km~s$^{-1}$]}  \\
       (1)    &   (2)   &  (3)     & (4)  & (5) & (6) \\
\hline
 0.0 &  2.0    &  1.2  &  54  &  0.79  &  0.06   \\  
 2.0 &  4.0    &  3.0  &  56  &  0.86  &  0.06   \\  
 4.0 &  6.0    &  4.7  &  34  &  0.70  &  0.10   \\  
 6.0 &  8.0    &  7.1  &  18  &  0.73  &  0.11   \\  
 8.0 & 10.0    &  8.9  &  12  &  0.66  &  0.12   \\  
10.0 & 11.6    &  10.6  &  9  &  0.57  &  0.13   \\    \hline  
 \end{tabular}
 \tablecomments{Columns 1--2: the inner and outer distances of each bin;
   Columns 3--6: the average distance, the number of the member stars, 
   RMS rotational velocity, and its uncertainty derived from the
     stars in each bin.}
 \label{table:rms_pleiades}
\end{table}
%%%%%%%%%%%%%%%%%%%%%%%%%%%%%%%%%%%%%%%%%%%%

%%%%%%%%%%%%%%%%%%%%%%%%%%%%%%%%%%%%%%%%%%%%% Figure. 5
\begin{figure}[ht]
\centering
\includegraphics[scale=0.18]{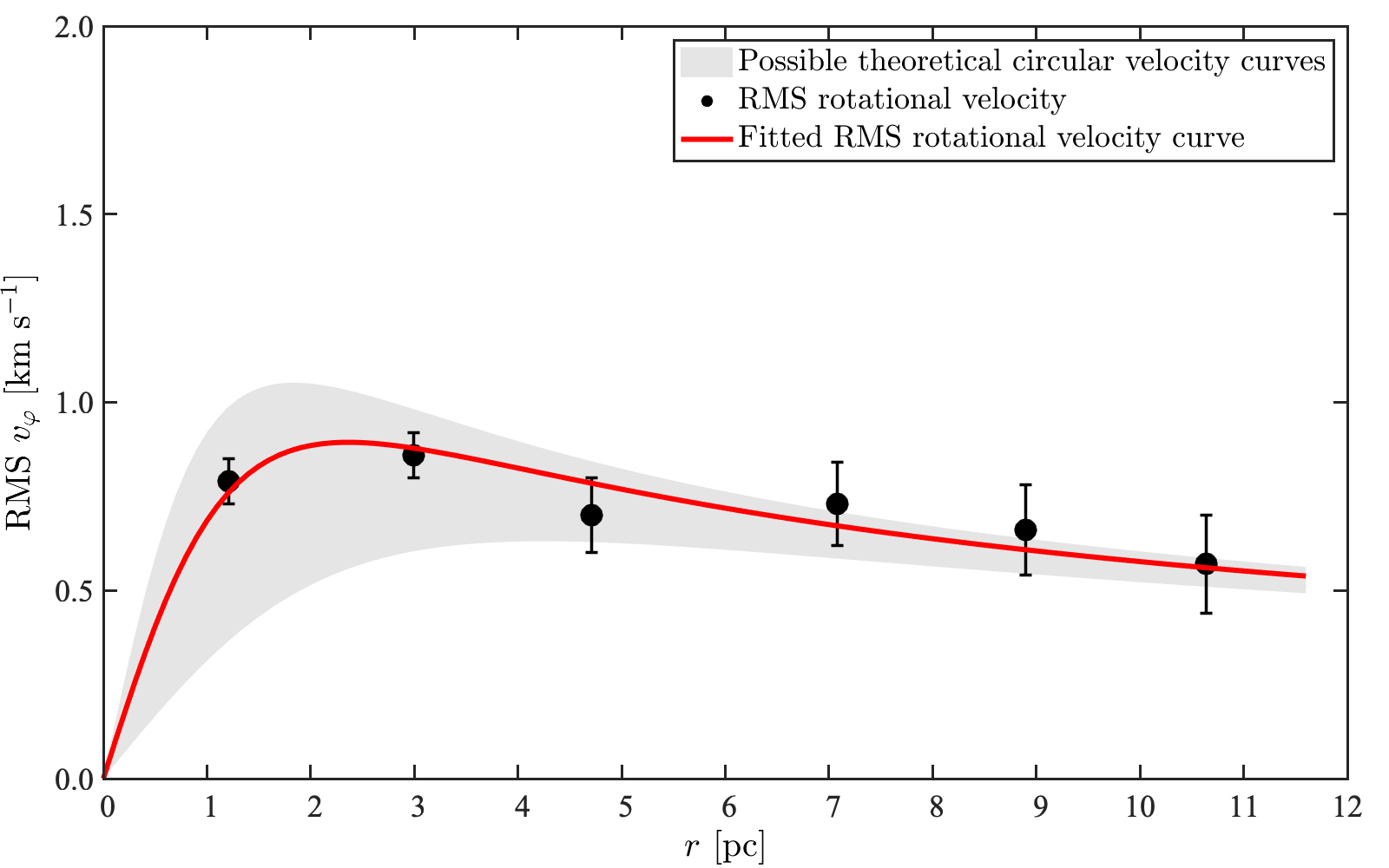}
\caption{RMS rotational velocity as the function of distance to the center of 
the Pleiades cluster. The black dots denote the observed RMS rotational
    velocities of the member stars. The uncertainties are indicated by the error bars. 
    The gray shadowed area shows the theoretical values calculated by
    Equation~(\ref{equ:v-t}) after considering the possible ranges of the
    tidal mass and the core radius of the Pleiades. The red line is the
    best-fitting curve for the observed RMS rotational velocities.}
\label{fig:rms_pleiades}
\end{figure}
%%%%%%%%%%%%%%%%%%%%%%%%%%%%%%%%%%%%%%%%%%%%%

Figure~\ref{fig:rotation_pleiades} presents the rotational velocities
of the member stars of the Pleiades as a function of their distances from the
cluster center.
Within the tidal radius of the Pleiades (11.6 pc), the rotation velocities of 
the member stars in the inner region are slightly higher than those
in the outer region.
Similar to the Preasepe cluster~\citepalias{hao2022b} and 
globular clusters~\citep[e.g.,][]{lanzoni2018,leanza2022}, 
not all member stars rotate in the same direction. 
There are $\sim$1.3 times as many member stars with positive rotational 
velocities as rotate with negative rotational velocities.
This situation may be attributed to the presence of mutual interference of 
stellar feedback during the formation of the cluster progenitor.
In Figure~\ref{fig:rotation_pleiades},
the rotation of the member stars located close to or outside the cluster's tidal 
radius seem to be disorganized, possibly because these member stars are 
partially affected by the Galactic tidal force and are not purely
governed by the gravitational force of the cluster itself.
In addition, within the tidal radius of the Pleiades, several stars exhibit 
peculiar rotational velocities that deviate from most of the cluster members, which 
may be passing stars and not the genuine 
members of the Pleiades.

We compare the observations with the theoretical predictions to address
whether Newton’s theorems can describe the rotation of the member stars within 
the tidal radius of the Pleiades cluster.
Firstly, we calculate the skewness of the astrometric parameters of the member 
stars within $r_{\rm td}$ to investigate the symmetry of their density distributions.
The skewness of the Galactic longitude, Galactic latitude, and parallax are~0.1, 
--0.2~and~0.1, respectively, which confirms the assumption of a spherically symmetric 
distribution of the stars in the Pleiades.
Further, we attempt to make a comparison between the 
RMS rotational velocities derived from the cluster members 
and the theoretical expectations of Newton’s theorems.
For the~196~member stars within the tidal radius of the Pleiades cluster, the 
median absolute value of rotational velocities~$v_{\varphi}$~is
$\sim$0.5~km~s$^{-1}$, and the standard deviation is~$\sim$1.0 km s$^{-1}$. 
To study the rotational characteristics displayed by the member stars of the
Pleiades, we divide them into several bins based on their 
distances from the cluster center. 
A few stars have peculiar rotational velocities that significantly deviate 
from those of most member stars. 
We identify 13 such kind of stars by checking the distribution 
of the rotational velocities of the member stars in each bin and eliminate them.
Then, we compute the RMS rotation velocity in each distance bin and estimate 
the corresponding uncertainties using the Monte 
Carlo simulations.
The results are presented in Table~\ref{table:rms_pleiades} and 
Figure~\ref{fig:rms_pleiades}. 

With Equation~(\ref{equ:v-t}), the theoretical rotational velocity profile provides 
a good fit to the RMS rotation velocities of the cluster members.
In addition, based on previous estimates of the tidal mass and core radius (see 
Section \ref{intro}), we calculate the possible theoretical circular velocity 
curves for the Pleiades cluster, as shown by the 
gray shaded area in Figure~\ref{fig:rms_pleiades}.
It is clear that the best-fit RMS rotational velocity curve is consistent	 with 
the theoretical predictions. 
Therefore, we argue that Newton's theorems is capable of characterizing the 
rotational properties of the member stars within the tidal radius of Pleiades.
Based on the result in Table~\ref{table:rms_pleiades} and Equation~(\ref{equ:v-t}), 
we estimate the core radius and the mass within $r_{\rm td}$ to be 
1.7 $\pm$ 0.3~pc and 806 $\pm$ 144~${\rm M}_{\odot}$, 
respectively, where the errors are estimated with the Monte Carlo simulations.
These two values agree well with previous results (see Section \ref{intro}).
%

%%%%%%%%%%%%%%%%%%%%%%%%%%%%%%%%%%%%%%%%%%%%% Figure. 6
\begin{figure}[ht]
\centering
\includegraphics[scale=0.18]{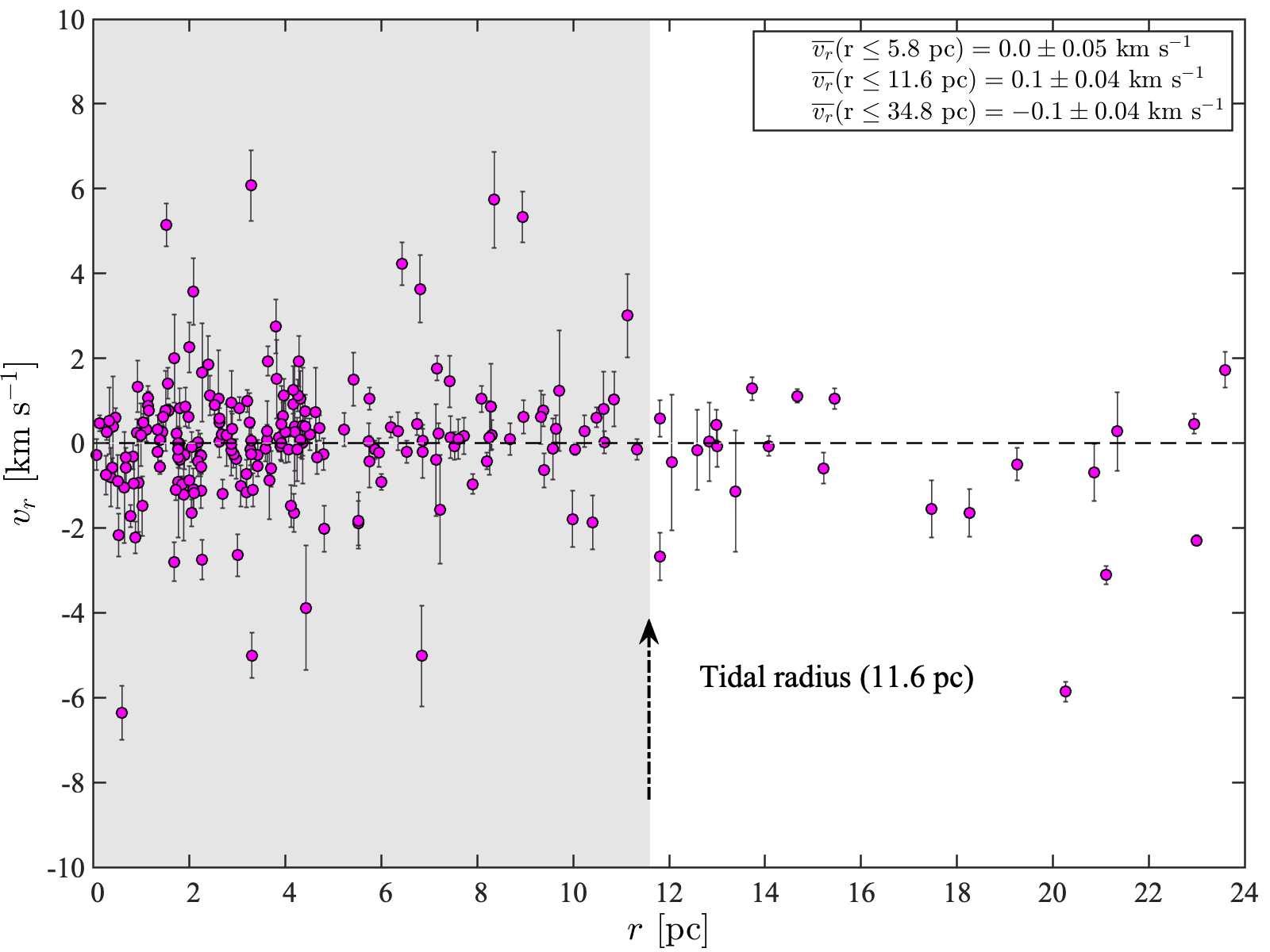}
\caption{Radial velocity as a function of the distance to the cluster center,
for the member stars of the Pleiades. The mean values of the radial components within 
0.5 $r_{\rm td}$, 1 $r_{\rm td}$, and 3 $r_{\rm td}$ are close to zero, as shown in the panel.}
\label{fig:radial_pleiades}
\end{figure}
%%%%%%%%%%%%%%%%%%%%%%%%%%%%%%%%%%%%%%%%%%%%%

The Pleiades are about 100 Myr old (see Section \ref{intro}).
Whether the Pleiades cluster is expanding or contracting can be examined 
by a statistical analysis of the radial components~$v_{r}$~of the member 
stars perpendicular to the cluster rotation axis. 
In~Figure~\ref{fig:radial_pleiades},
the mean radial components~$\overline{v_{r}}$~of the member stars within
0.5 $r_{\rm td}$, 1 $r_{\rm td}$, and 3 $r_{\rm td}$~are~0.0 $\pm$ 0.05 
km s$^{- 1}$, 
0.1 $\pm$ 0.04 km s$^{-1}$, and~--0.1 $\pm$ 0.04 km s$^{-1}$, respectively. 
Here, the errors are obtained with the Monte Carlo method by taking into 
account the uncertainties of~${v_{r}}$. 
These results indicate 
the absence of expansion or contraction in the Pleiades cluster, and this implies 
that the rotation of the member stars within $r_{\rm td}$ can exhibit 
closed-loop motions. 
However, as shown in~Figure~\ref{fig:radial_pleiades},
we note the presence of non-negligible radial component 
acceleration and dispersion of members in the Pleiades, although the dispersion
is partially influenced by the astrometric measurement uncertainties.  
Considering the absence of expansion or contraction in the Pleiades,
the system should possess additional mass to provide the force that support 
the radial components acceleration of the member stars. 
Hence, the derived mass of the Pleiades cluster represents a lower limit.
%

%%%%%%%%%%%%%%%%%%%%%%%%%%%%%%%%%%%%%%% Figure 7
\begin{figure}
	\begin{center}
		\includegraphics[width=1.00\textwidth]{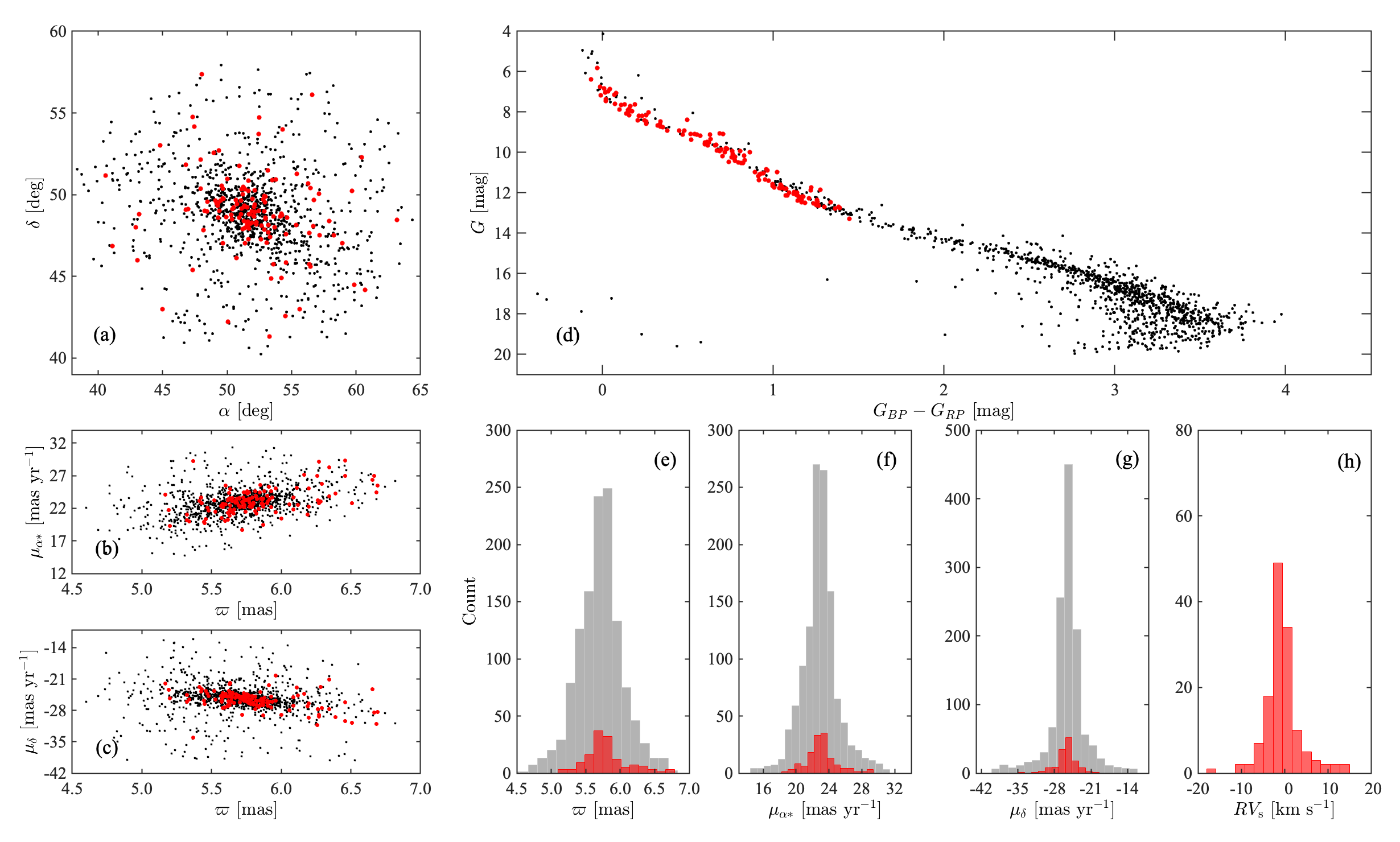}
		\caption{Same as Figure~\ref{fig:distribution_pleiades}, but for the $\alpha$~Per cluster.
		}
		\label{fig:distribution_alphaper}
	\end{center}
\end{figure}
%%%%%%%%%%%%%%%%%%%%%%%%%%%%%%%%%%%%%%%

\subsection{Cluster $\alpha$~Per}

Similar to Figure~\ref{fig:distribution_pleiades}, we present the multidimensional 
distributions and color-magnitude diagram for cluster~$\alpha$~Per  
in Figure~\ref{fig:distribution_alphaper}.
In this section, we show the rotational properties of cluster $\alpha$~Per, and analyze 
the rotational characteristics of the cluster members.

%%%%%%%%%%%%%%%%%%%%%%%%%%%%%%%%%%%%%%%%%%%%% Figure. 8
\begin{figure*}
\centering
  \includegraphics[width=0.32\textwidth]{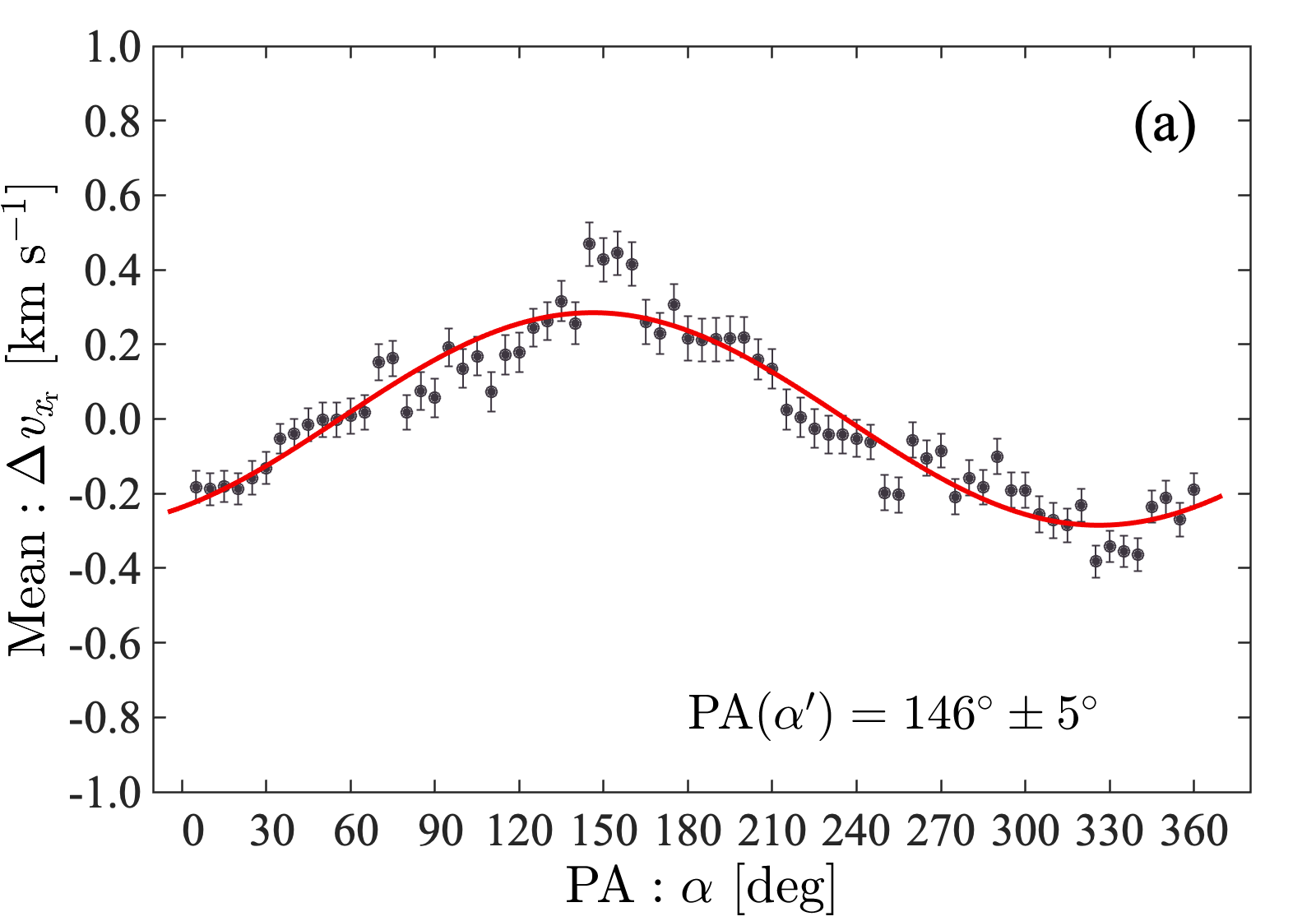}
  \includegraphics[width=0.32\textwidth]{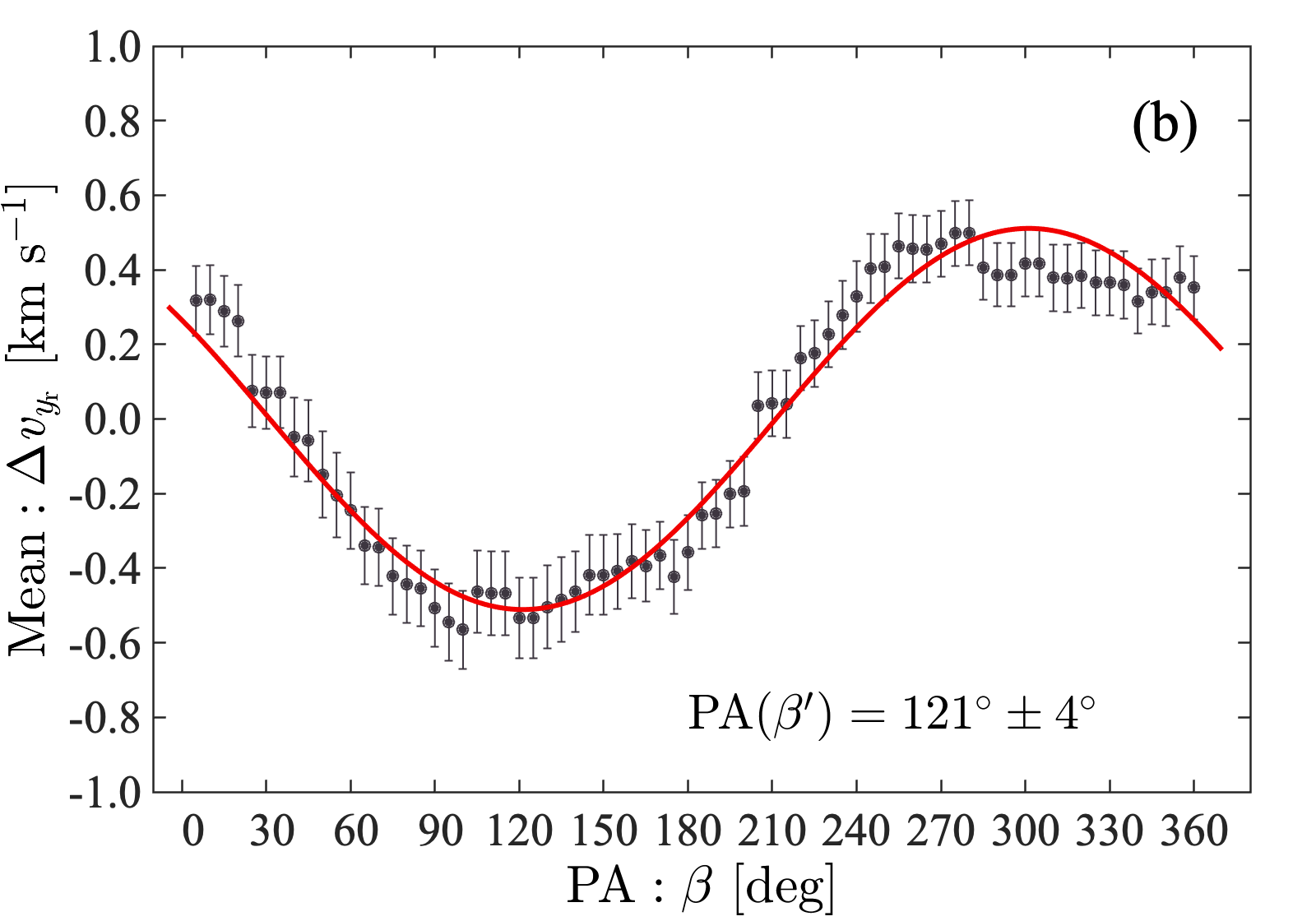}
  \includegraphics[width=0.32\textwidth]{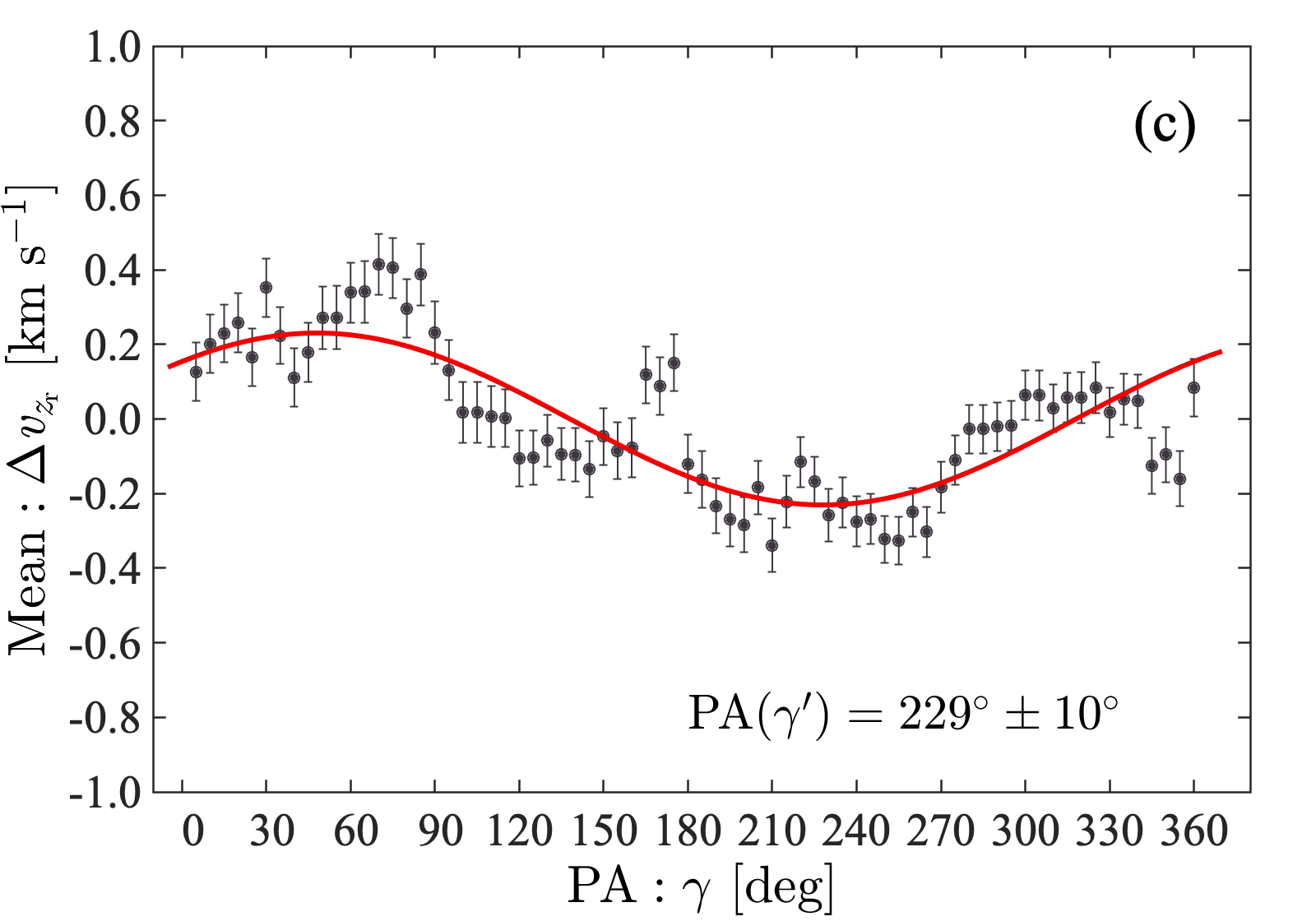}
  \includegraphics[width=0.32\textwidth]{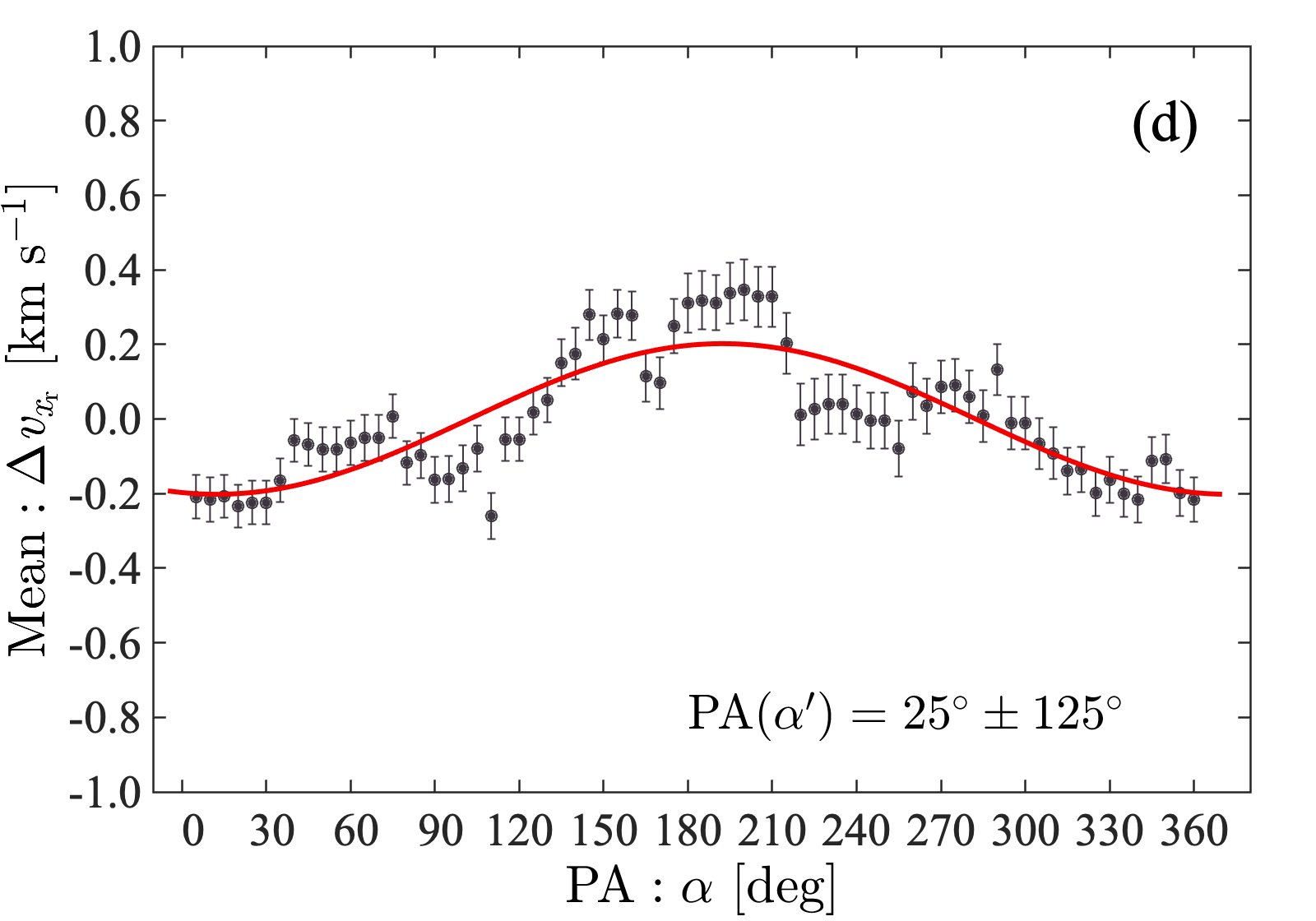}
  \includegraphics[width=0.32\textwidth]{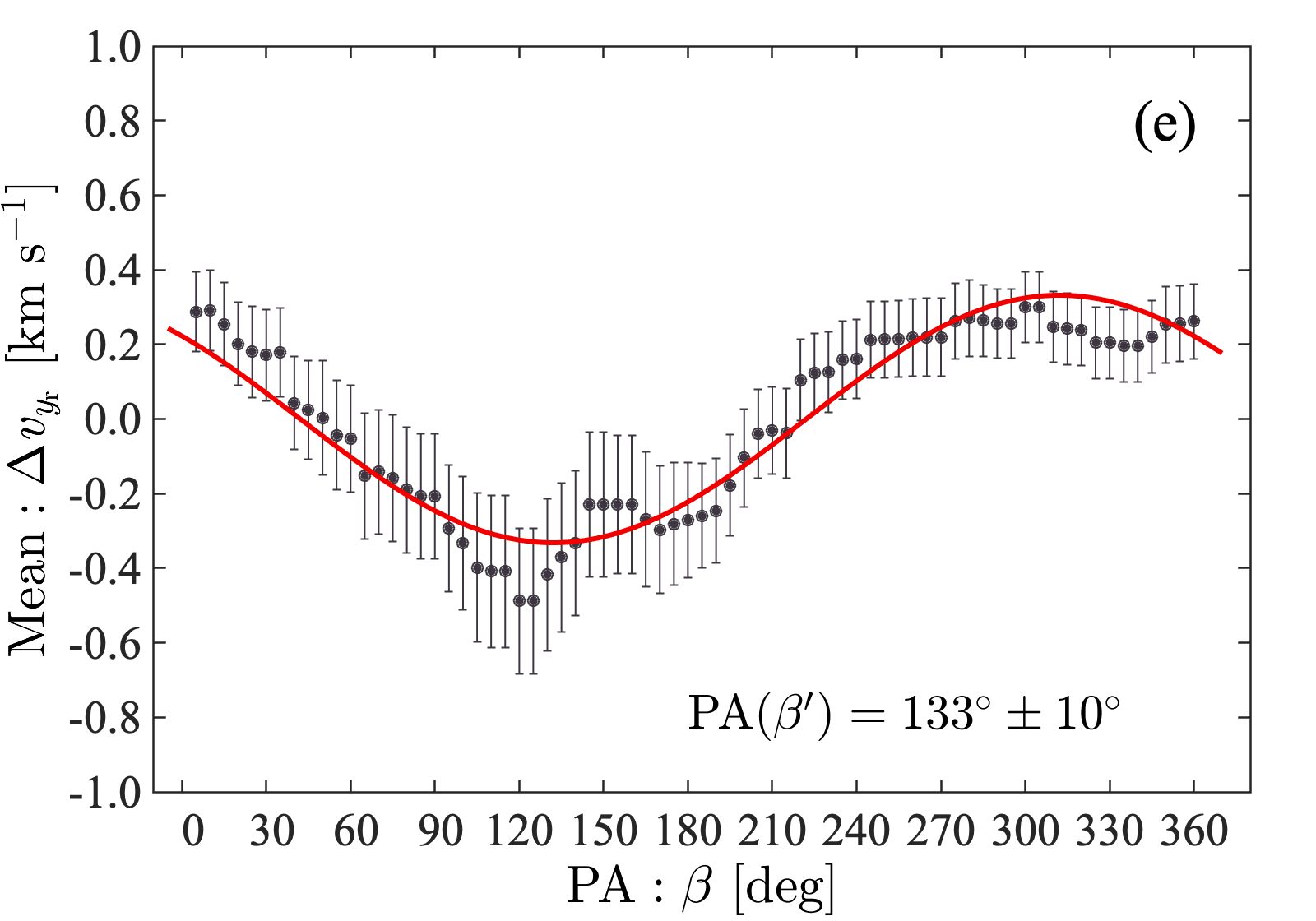}
  \includegraphics[width=0.32\textwidth]{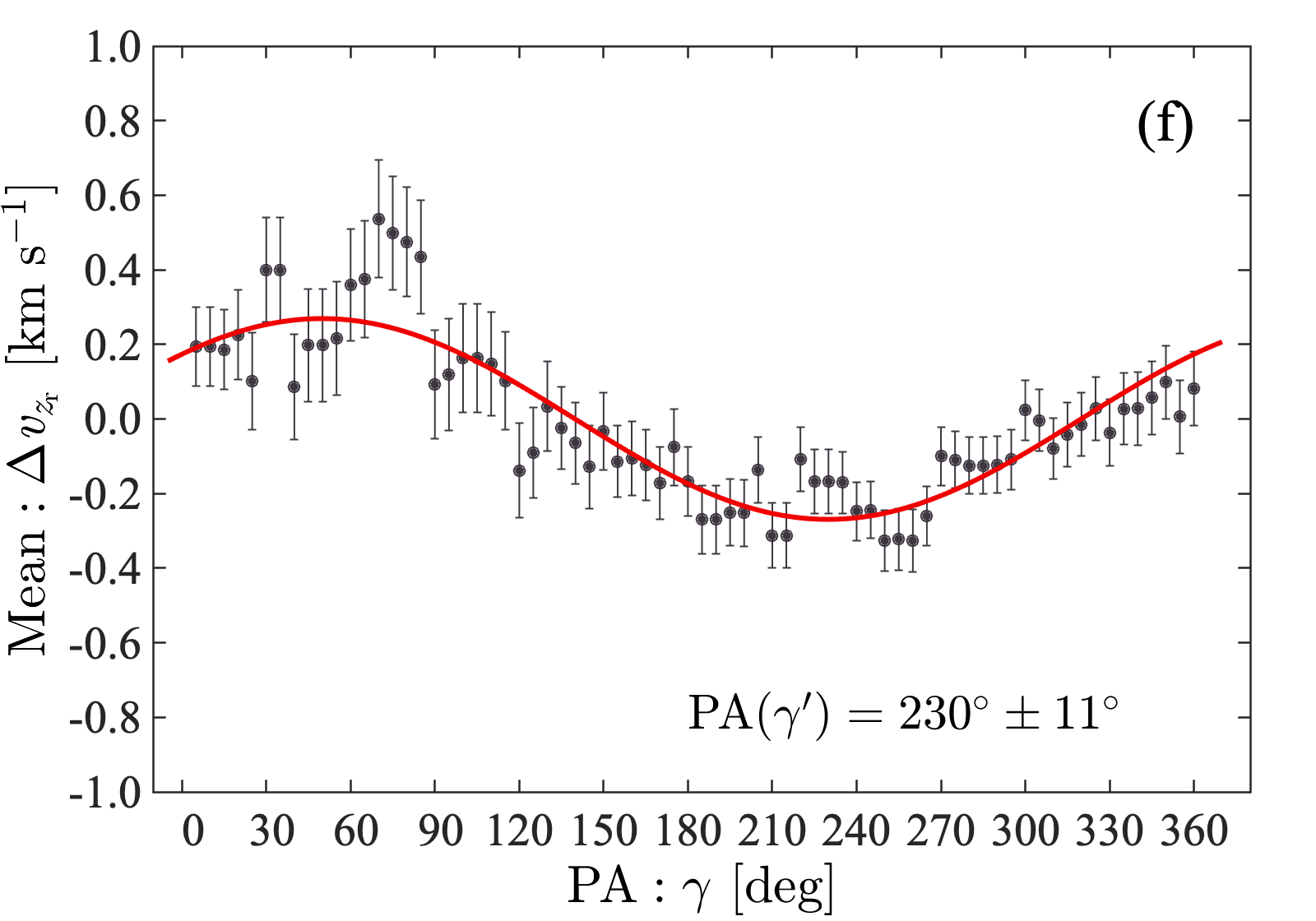}
  \includegraphics[width=0.32\textwidth]{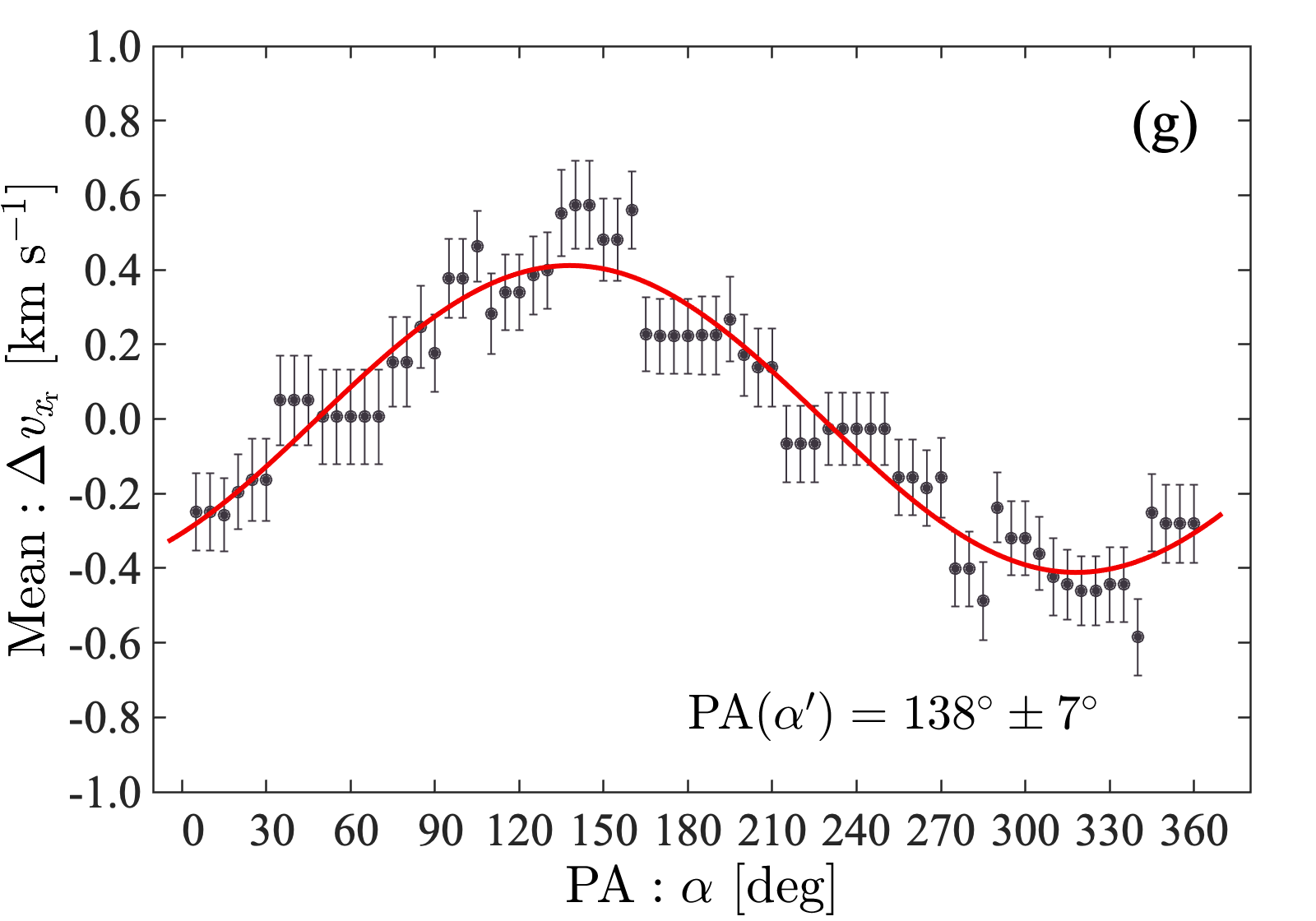}
  \includegraphics[width=0.32\textwidth]{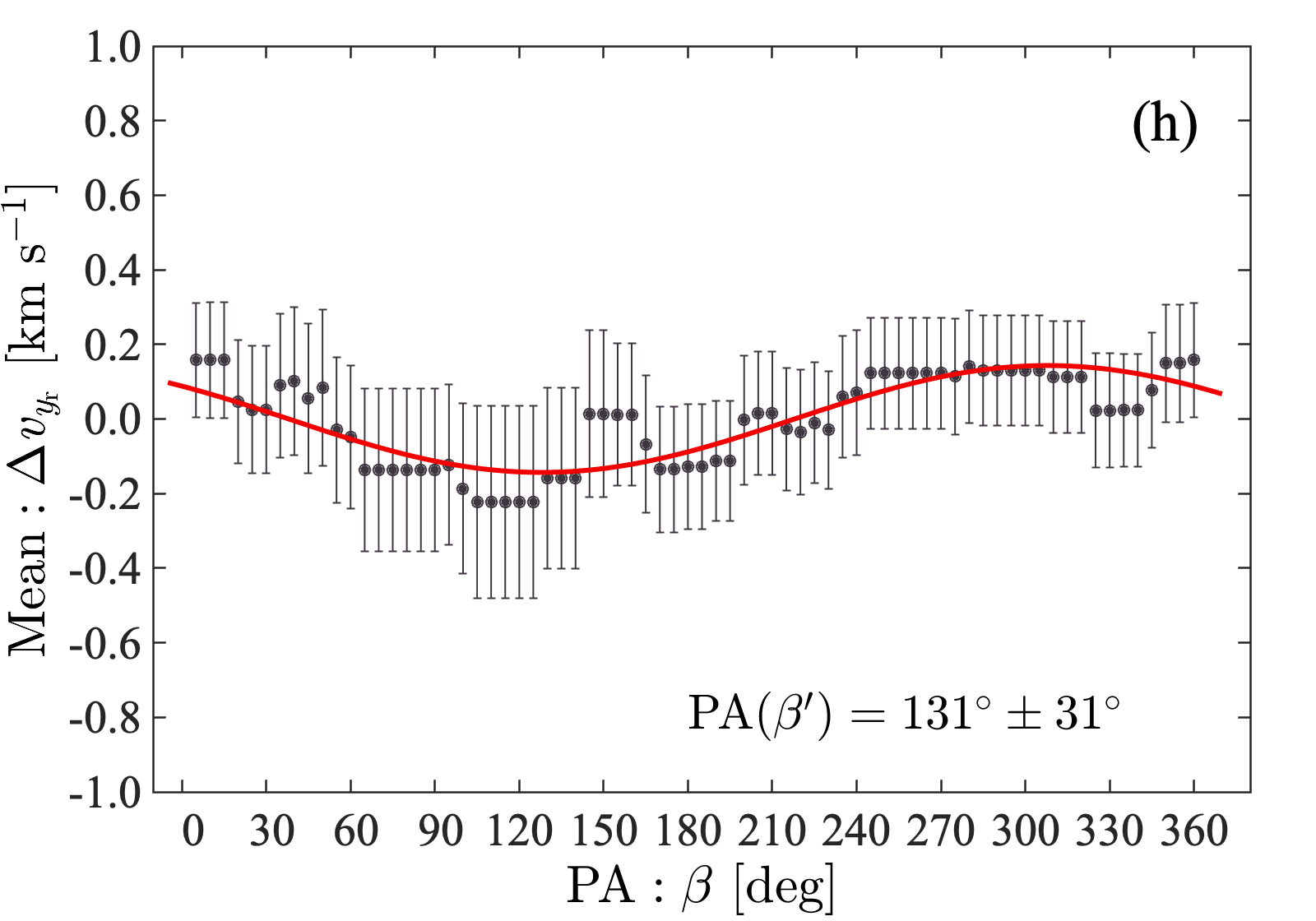}
  \includegraphics[width=0.32\textwidth]{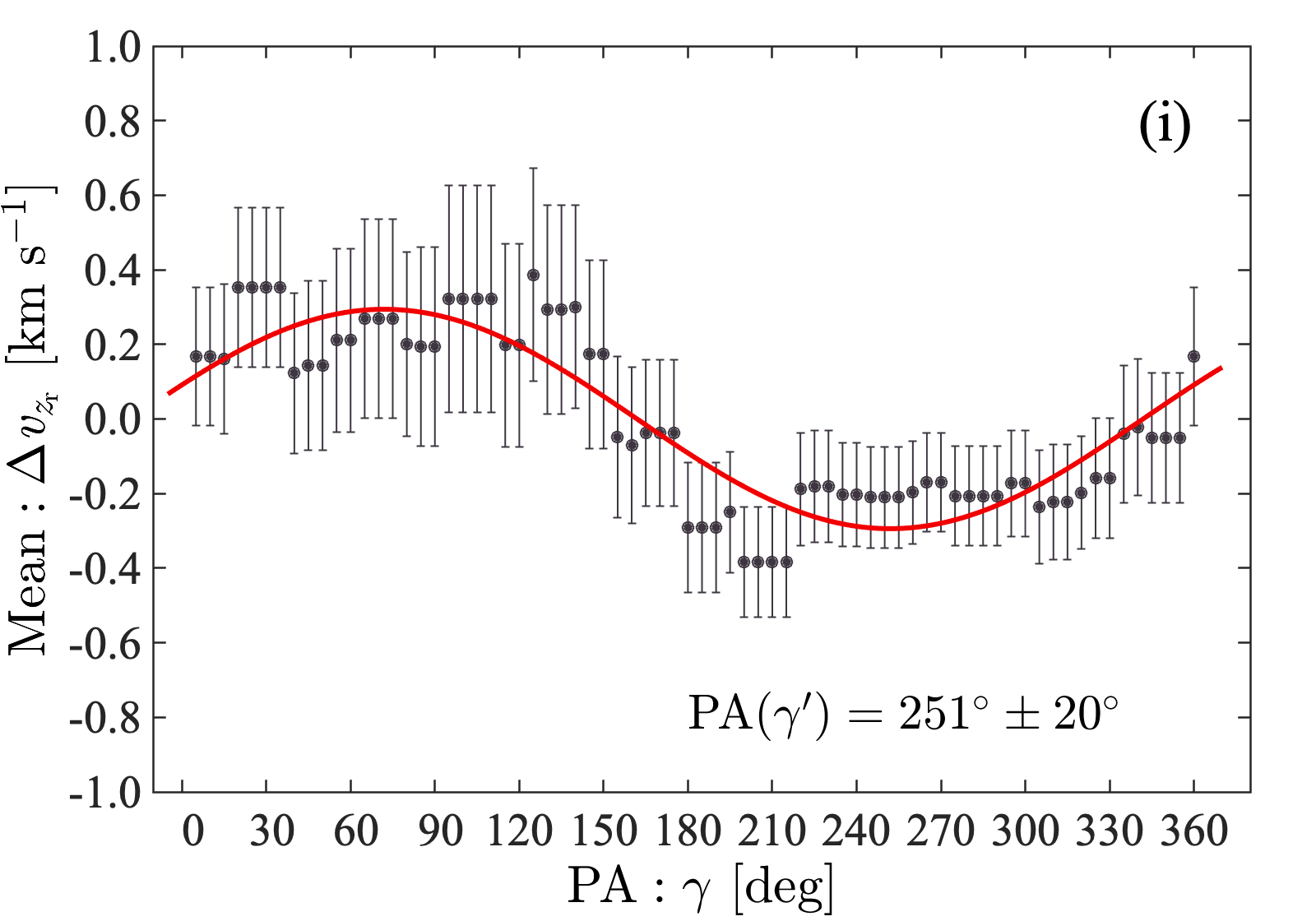}
\caption{Same as Figure~\ref{fig:pa_pleiades}, but for the $\alpha$~Per cluster. }
\label{fig:pa_alphaper}
\end{figure*}
%%%%%%%%%%%%%%%%%%%%%%%%%%%%%%%%%%%%%%%%%%%%%

\subsubsection{Rotation in $\alpha$~Per}
\label{sec:alphaper}
For the~137~member stars with radial velocities (see Table~\ref{table:numbers}), 
the median uncertainties of the parallax, proper 
motion~$\mu_{\alpha^{*}}$~and~$\mu_{\delta}$, and for the radial velocity 
are~0.02~mas, 0.02~mas~yr$^{-1}$, 0.02~mas~yr$^{-1}$, and~0.9~km s$^{ -1}$, 
respectively.
Based on the observational parameters of these member stars, 
the means and standard deviations of the fundamental astrometric parameters 
of the $\alpha$~Per cluster are determined as: 
(R.A., Decl.)=($51.87^{ \circ}$ $\pm$ $3.65^{\circ}$, $48.93^{\circ}$ $\pm$
$2.47^{\circ}$), $\varpi$ = $5.81$ $\pm$ $0.30$ mas, 
($\mu_{\alpha^{*}}$, $\mu_{\delta}$) = ($23.11$ $\pm$ $1.87$, 
$-25.76$ $\pm$ $1.97$) mas yr$^{-1}$, and
RV = $-0.43$ $\pm$ $4.10$ km s$^ {-1}$, which are in good agreement with 
previous determinations~\citep{stauffer1989,prosser1994,mermilliod2008,
babusiaux2018,lodieu2019}. 
Similar to the Pleiades, we obtain the 3D coordinates and motion of the $\alpha$~Per 
cluster in the $O_{\rm g}$--$X_{\rm g}$$Y_{\rm g}$$Z_{\rm g}$ system and the 
3D coordinates and motions of its member stars in the
$O_{\rm c}$--$X_{\rm c}$$Y_{\rm c}$$Z_{\rm c}$ system.

The direct usage of the residual velocity method in the 
$O_{\rm c}$--$X_{\rm c}$$Y_{\rm c}$$Z_{\rm c}$ system captures the rotation 
signals of the $\alpha$~Per cluster, but does not allow a good determination of 
the PAs ($\alpha$, $\beta$, $\gamma$) of the rotation axis $\vec{l}$.
In order to resolve this matter, as described in Section~\ref{methods}, we rotate the 
$O_{\rm c}$--$X_{\rm c}$$Y_{\rm c}$$Z_{\rm c}$ system around its origin
with a set of angles ($\alpha_{0}$, $\beta_{0}$, $\gamma_{0}$) =
($40^{\circ}$, $40^{\circ}$, $45^{\circ}$).
The 3D coordinates and velocity components of the cluster members are then 
transformed from the $O_{\rm c}$--$X_{\rm c}$$Y_{\rm c}$$Z_{\rm c}$ system 
to the rotated $O_{\rm r}'$--$X_{\rm r}$$Y_{\rm r}$$Z_{\rm r}$ system.
After this, in the same way as for the Pleiades, 
we estimate the residual velocities of the cluster
members and uncertainties to determine the PAs ($\alpha'$, $\beta'$, $\gamma'$) 
of the rotation axis $\vec{l}$ of the $\alpha$~Per in the rotated system.
Figure~\ref{fig:pa_alphaper} presents the mean residual velocities as functions 
of PAs ($\alpha'$, $\beta'$, $\gamma'$) for the member stars within 3 
$ r_{\rm td} $ (a--c), 1 $ r_{\rm td} $ (d--f), and 0.5 $ r_{\rm td} $ (g--i) of the 
$\alpha$~Per cluster, where $ r_{\rm td} =9.5$ pc is taken from \citet{lodieu2019}.
The sinusoidal relations between the mean residual velocities 
and PAs indicate the 3D rotation of the $\alpha$~Per cluster,
although the result in Figure~\ref{fig:pa_alphaper}(d) is not very significant.

The fitted PAs ($\alpha'$, $\beta'$, $\gamma'$) of the rotation axis 
$\vec{l}$ of the $\alpha$~Per cluster in the rotated system are listed in 
Table~\ref{table:pa_alphaper} and marked in Figures~\ref{fig:pa_alphaper}.
The PAs fitted by the members within 0.5 $ r_{\rm td} $
satisfy the relation of tan $\alpha'$ $\cdot$ tan $\gamma'$ = tan $\beta'$
considering the uncertainties.
The PA $\alpha'$ value cannot be well determined by the members within 
1 $ r_{\rm td} $, which suffer from large uncertainties.
Alternatively, PA $\alpha'$ = $138^{\circ}$ can be determined from the 
tan $\alpha'$ $\cdot$ tan $\gamma'$ = tan $\beta'$ relation, combined with 
the well-fit $\beta'$ and $\gamma'$ values.
In this way, the three PAs are compatible with those derived by the members 
within 0.5 $ r_{\rm td} $, and they can therefore be used to determine
the rotation axis $\vec{l}$.
Compared with the Pleiades cluster, the stronger rotation of the $\alpha$~Per 
cluster makes the PAs obtained by using the member stars within 3 $ r_{\rm td} $ 
almost consistent with the above results, considering the uncertainties.

The PAs obtained by using the members within 1 $ r_{\rm td} $ are 
employed to determine the rotation axis $\vec{l}$ of the $\alpha$~Per cluster
in the rotated system ($O_{\rm r}'$--$X_{\rm r}$$Y_{\rm r}$$Z_{\rm r}$),
i.e., ($\alpha'$, $\beta'$, $\gamma'$) = ($138^{ \circ}$, $133^{\circ}$, 
$230^{\circ}$).
Then, the PAs of ($\alpha$, $\beta$, $\gamma$) = ($68^{ \circ}$, $51^{\circ}$, 
$26^{\circ}$) for $\vec{l}$ in the 
$O_{\rm c}$--$X_{\rm c}$$Y_{\rm c}$$Z_{\rm c}$ system can be obtained by
using Equation~(\ref{equ:ccs2rcc2}). 
Thus, $\vec{l}$ is tilted at an angle 
of~$48^\circ\pm14^\circ$ with respect to the Galactic plane.
With the PAs,
we derive the vectors of the $X_{\rm r}$, $Y_{\rm r}$, and $Z_{\rm r}$ axes 
of the $O_{\rm r}$--$X_{\rm r}$$Y_{\rm r}$$Z_{\rm r}$ system in the 
$O_{\rm c}$--$X_{\rm c}$$Y_{\rm c}$$Z_{\rm c}$ system.
After transforming the 3D coordinates and velocity components of the cluster 
members from the $O_{\rm c}$--$X_{\rm c}$$Y_{\rm c}$$Z_{\rm c}$ system   
to those in the $O_{\rm r}$--$X_{\rm r}$$Y_{\rm r}$$Z_{\rm r}$ system, 
we also calculate the 3D coordinates ($r$, $\varphi$, $z$) and velocity 
components ($v_ {r}$, $v_{\varphi}$, $v_{z}$) of
the cluster members in the cylindrical coordinate system.
As for the Pleiades,
using the rotation velocities ($v_{\varphi}$) and uncertainties of the member stars, we
estimate the mean rotation velocity of the $\alpha$~Per cluster within its
tidal radius to be~0.43 $\pm$ 0.08~km~s$^{-1}$, which is in consistent 
with the results shown in Figures~\ref{fig:pa_alphaper}.
%

%%%%%%%%%%%%%%%%%%%%%%%%%%%%%%%%%%%%%%%%%%%% Table. 5
\setlength{\tabcolsep}{10.0mm}
\begin{table}
\centering
\caption{Best-fit PAs of the $\alpha$~Per cluster.}
\begin{tabular}{c|ccc} 
\hline  \hline 
           &    $\alpha'$      &    $\beta'$    &    $\gamma'$  \\ \hline 
3.0 $r_{\rm td}$   &  $146^{\circ}$ $\pm$ $5^{\circ}$   &  $121^{\circ}$ $\pm$ $4^{\circ}$  
          &  $229^{\circ}$ $\pm$ $10^{\circ}$     \\  
1.0 $r_{\rm td}$      &  $25^{\circ}$ $\pm$ $125^{\circ}$  &  $133^{\circ}$ $\pm$ $10^{\circ}$    
          &  $230^{\circ}$ $\pm$ $11^{\circ}$     \\  
0.5 $r_{\rm td}$   &  $138^{\circ}$ $\pm$ $7^{\circ}$   &  $131^{\circ}$ $\pm$ $31^{\circ}$    
          &  $251^{\circ}$ $\pm$ $20^{\circ}$     \\    \hline  
 \end{tabular}
 \tablecomments{$r_{\rm td}$: tidal radius of the $\alpha$~Per cluster.}
 \label{table:pa_alphaper}
\end{table}
%%%%%%%%%%%%%%%%%%%%%%%%%%%%%%%%%%%%%%%%%%%%%%%

\subsubsection{Rotational properties of the stars in $\alpha$~Per}
\label{sec:rotation-alphaper}

%%%%%%%%%%%%%%%%%%%%%%%%%%%%%%%%%%%%%%%%%%%%% Figure. 9
\begin{figure}[ht]
\centering
\includegraphics[scale=0.18]{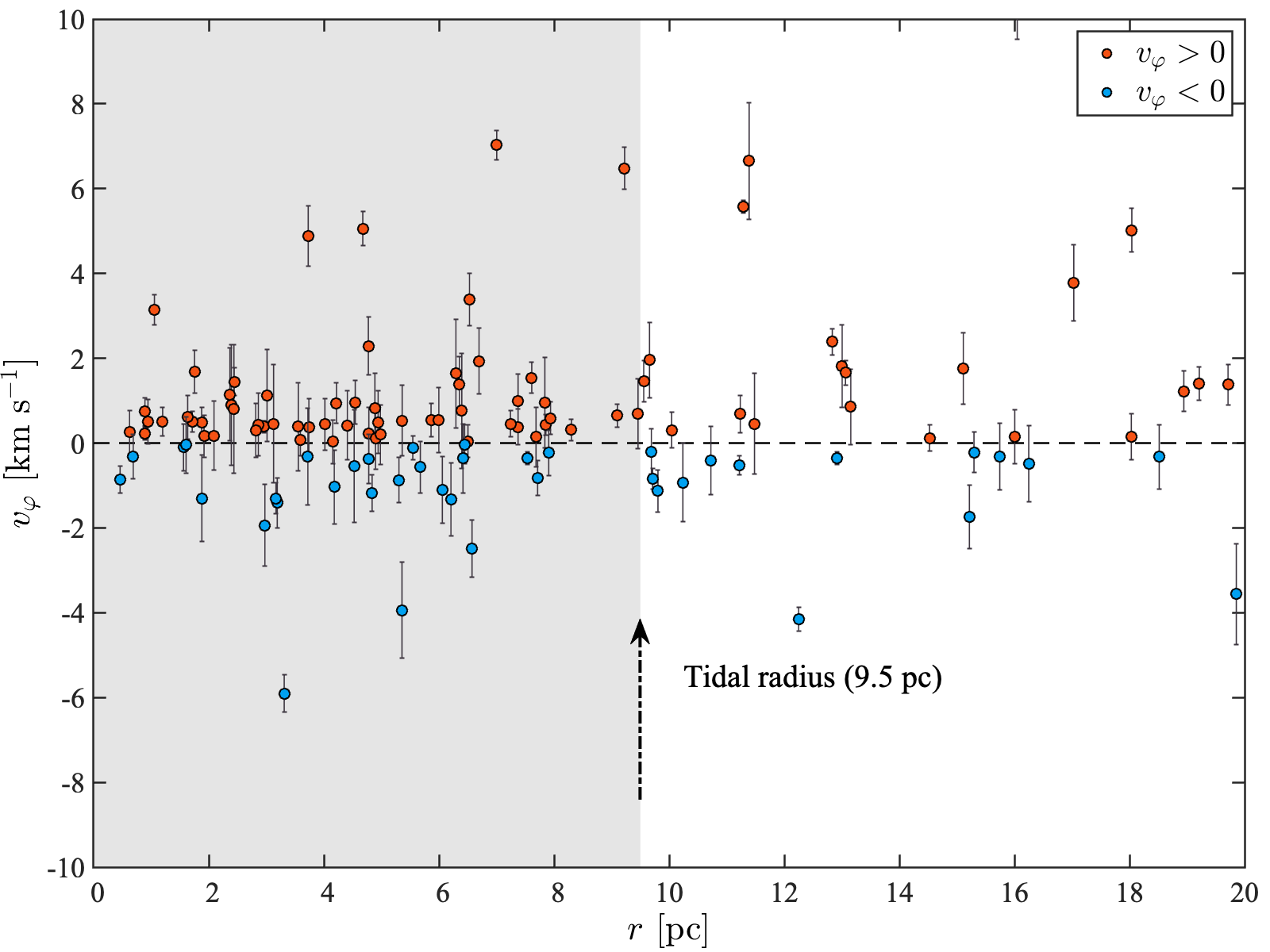}
\caption{Rotational velocity as a function of 
distance to the cluster center, for the member stars within
two times of the tidal radius of the $\alpha$~Per.}
\label{fig:rotation_alphaper}
\end{figure}
%%%%%%%%%%%%%%%%%%%%%%%%%%%%%%%%%%%%%%%%%%%%%

%%%%%%%%%%%%%%%%%%%%%%%%%%%%%%%%%%%%%%%%%%%%% Figure. 10
\begin{figure}[ht]
\centering
\includegraphics[scale=0.18]{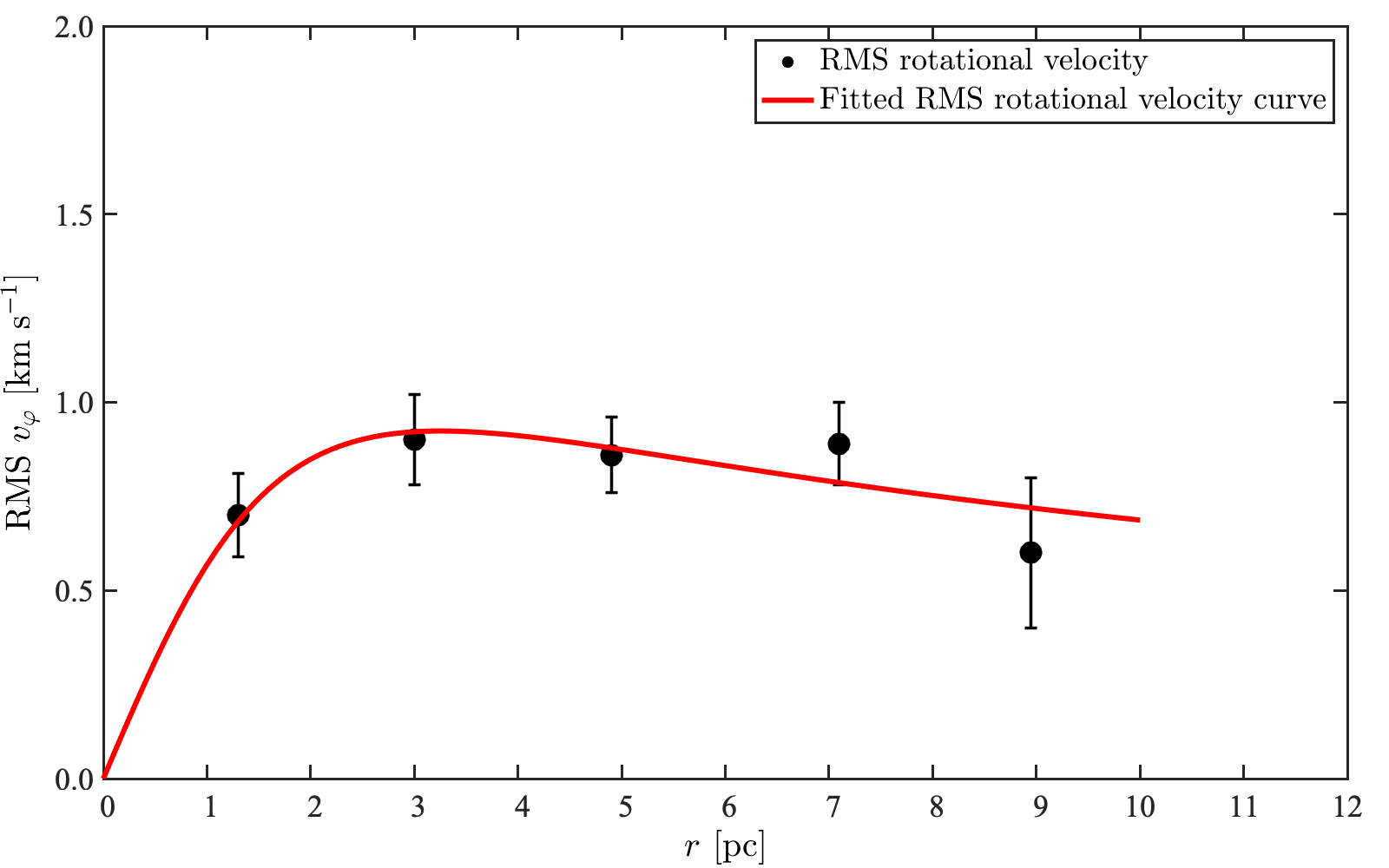}
\caption{RMS rotational velocity as a function of distance to the center 
of the $\alpha$~Per cluster. The black dots denote the observed RMS rotational
velocities of the member stars, with uncertainties indicated by error bars. The red line is the
best-fitting curve for the observed RMS rotational velocities according to the Newton's theorems.}
\label{fig:rms_alphaper}
\end{figure}
%%%%%%%%%%%%%%%%%%%%%%%%%%%%%%%%%%%%%%%%%%%%%

%%%%%%%%%%%%%%%%%%%%%%%%%%%%%%%%%%%%%%%%%%%%% Table. 6
\setlength{\tabcolsep}{8.0mm}
\begin{table}[ht]
\centering 
\caption{RMS rotational velocities of the member stars within
    the tidal radius of the $\alpha$~Per cluster.}
\begin{tabular}{cc|cccc} 
  \hline  \hline 
$R_{i}$  &  $R_{o}$ & $R_{m}$  & $N$  & RMS $v_{\varphi}$  &  $\epsilon_{{\rm RMS} \, v_{\varphi}}$   \\ %\hline
      {[pc]}  &  {[pc]}    &  {[pc]}  &  & {[km~s$^{-1}$]} & {[km~s$^{-1}$]}  \\
       (1)    &   (2)   &  (3)     & (4)  & (5) & (6) \\
\hline
 0.0 &  2.0    &  1.3  &  15  &  0.70  &  0.11   \\  
 2.0 &  4.0    &  3.0  &  17  &  0.90  &  0.12   \\  
 4.0 &  6.0    &  4.9  &  21  &  0.89  &  0.10   \\  
 6.0 &  8.0    &  7.1  &  21  &  0.86  &  0.11   \\  
 8.0 &  9.5    &  8.9  &  3  &  0.58  &  0.24   \\  \hline  
 \end{tabular}
 \tablecomments{The same format as Table~\ref{table:rms_pleiades},
but for the $\alpha$~Per cluster.}
 \label{table:rms_alphaper}
\end{table}
%%%%%%%%%%%%%%%%%%%%%%%%%%%%%%%%%%%%%%%%%%%%%

The rotational velocities of the member stars as a 
function of their distances to the cluster center are presented in 
Figure~\ref{fig:rotation_alphaper}.
Similar to the Preasepe \citepalias{hao2022b} and the Pleiades (this paper),
not all of the member stars of the $\alpha$~Per cluster rotate in the same direction. 
The ratio of the number of members with positive $v_{\varphi}$ values to those 
with negative values is $\sim$2.3, which is higher than that of
the Pleiades.
Due to the tidal force of the Milky Way, the rotation of the member stars located 
beyond the cluster tidal radius (9.5 pc) seems to be disordered,
as shown in Figure~\ref{fig:rotation_alphaper}.
Moreover, in a few stars, the peculiar rotational velocities deviate from those of
most cluster members within the tidal radius of the $\alpha$~Per.

Understanding the rotational patterns of the member stars within the tidal radius 
of the $\alpha$ Per cluster is an intriguing issue. To this end, we compare our 
observations with the theoretical predictions of Newton’s theorems, following 
the same analysis procedure as for the Pleiades.
Within the tidal radius of the $\alpha$~Per, the skewness of the Galactic longitudes, 
Galactic latitudes, and parallaxes of member stars are~0.2, 
0.2,~and~--0.4, respectively, indicating that the assumption of the spherically symmetric 
distribution in the cluster is reasonable.
Next, we try to use Equation~(\ref{equ:v-t}) to describe the RMS rotational velocity 
derived from the member stars of the $\alpha$~Per cluster.
Within the tidal radius of the $\alpha$~Per cluster, there are~85~member stars in the
sample. 
The median absolute value of their rotation velocities~$v_{\varphi}$~is 
$\sim$0.6~km~s$^{-1}$, with a standard deviation of~$\sim$1.4 km s$^{-1}$. 
For these member stars, we study their rotational characteristics by dividing 
them into several bins according to their distances from the cluster center.
Eight stars with peculiar rotational velocities deviating from the 
overall velocity are eliminated.
Similar to the Pleiades, using the remaining~77~member stars, we calculate the 
RMS rotation velocity and uncertainty in each distance bin.
The results are given in Table~\ref{table:rms_alphaper} and 
Figure~\ref{fig:rms_alphaper}. 
The RMS rotational velocities of the member stars can be well fit by
the velocity profile of circular motion expected by Newton's theorems,
suggesting that the rotation of the member stars within the tidal radius of the 
$\alpha$~Per can be elucidated by Newton's theorems. 
The core radius and the mass within the tidal radius of the $\alpha$~Per cluster 
is 2.3 $\pm$ 0.5~pc and 1186 $\pm$ 312~${\rm M}_{\odot}$, respectively, derived by 
fitting the RMS velocities listed in Table~\ref{table:rms_alphaper} with 
Equation~(\ref{equ:v-t}), with the same technique for the Pleiades.
We find that the best-fitting core radius matches the result
of 2.3 $\pm$ 0.3 pc reported by~\citet{lodieu2019}, but is significantly smaller 
than the early estimate~\citep[$\sim$4 pc, see][]{artyukhina1972,kharchenko2005}.
%

%%%%%%%%%%%%%%%%%%%%%%%%%%%%%%%%%%%%%%%%%%%%% Figure. 11
\begin{figure}[ht]
\centering
\includegraphics[scale=0.18]{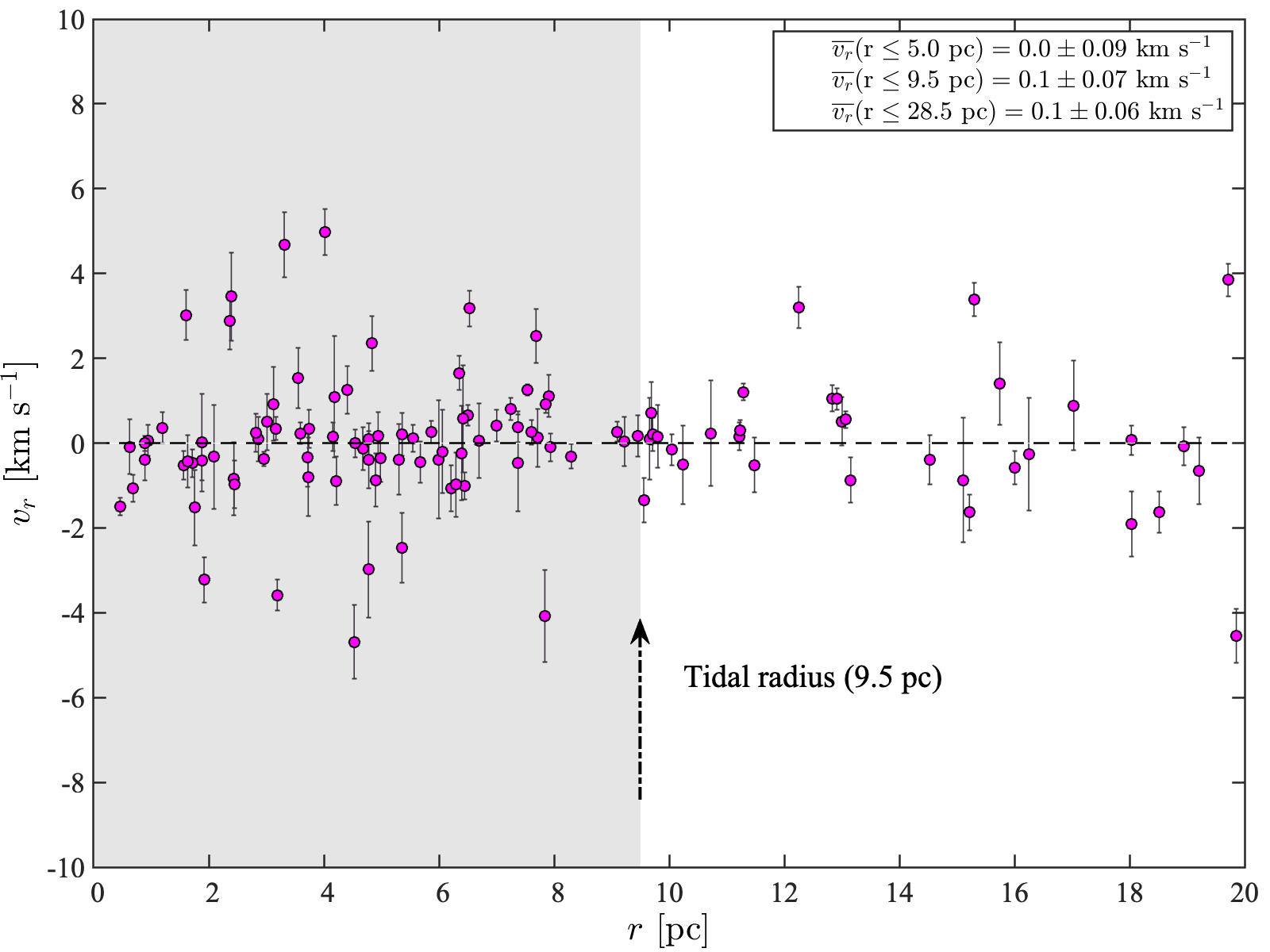}
\caption{Same as Figure~\ref{fig:radial_pleiades}, but for the $\alpha$~Per cluster.}
\label{fig:radial_alphaper}
\end{figure}
%%%%%%%%%%%%%%%%%%%%%%%%%%%%%%%%%%%%%%%%%%%%%

The age of the $\alpha$~Per cluster has been estimated to be approximately 
50--70 Myr (see Section \ref{intro}).
As presented in Figure~\ref{fig:radial_alphaper},
the mean radial components~$\overline{v_{r}}$~of the member stars within~0.5 $r_{\rm td}$, 
1 $r_{\rm td}$, and 3 $r_{\rm td}$~are~0.0 $\pm$ 0.09, 
0.1 $\pm$ 0.07, and~0.1 $\pm$ 0.06 km s$^{-1}$, respectively,
using the same approach as for the Pleiades.
In this context, we posit that the $\alpha$~Per cluster shows no obvious signs of 
expansion or contraction, suggesting that the rotation of the member stars within the 
tidal radius exhibits closed-loop motions.
Additionally, ass for the Pleiades, the above estimated mass of the $\alpha$~Per 
is expected to be a lower limit, since the non-negligible radial component acceleration 
and dispersion (Figure~\ref{fig:radial_alphaper}) of the cluster members imply that the 
cluster ought to have additional mass.

\subsection{The Hyades cluster}
\begin{figure}
	\begin{center}
		\includegraphics[width=1.00\textwidth]{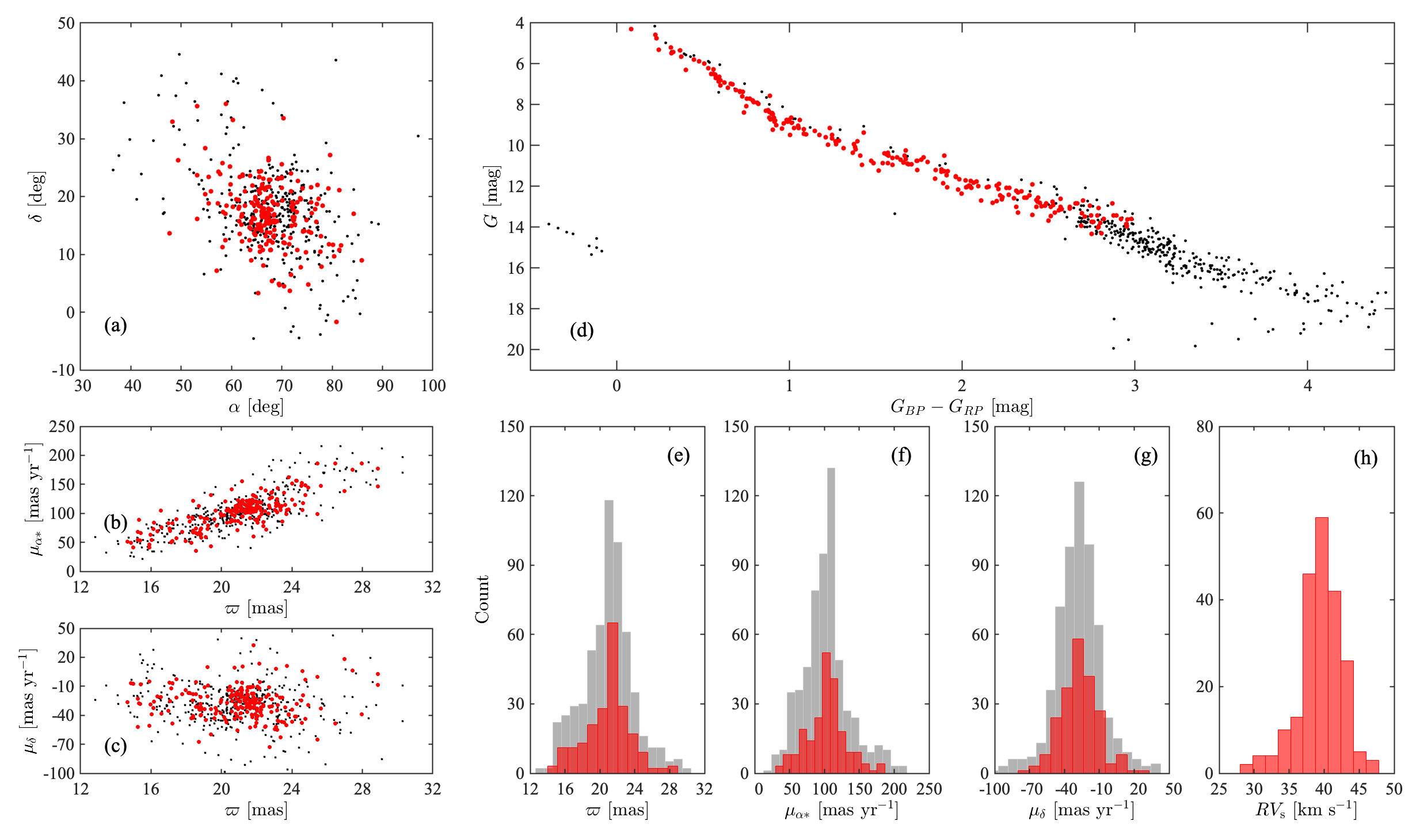}
		\caption{Same as Figure~\ref{fig:distribution_pleiades}, but for the Hyades cluster.
		}
		\label{fig:distribution_hyades}
	\end{center}
\end{figure}

In Figure~\ref{fig:distribution_hyades}, we display the multidimensional distributions 
and color-magnitude diagram for the Hyades cluster.
In this section, we picture the rotation in the Hyades cluster, and investigate 
the rotational characteristics of the member stars.
%

%%%%%%%%%%%%%%%%%%%%%%%%%%%%%%%%%%%%%%% Figure. 12
%%%%%%%%%%%%%%%%%%%%%%%%%%%%%%%%%%%%%%%

\subsubsection{Rotation in Hyades}
\label{sec:hyades}
For the subsample of~214~member stars with radial velocities (see 
Table~\ref{table:numbers}), the median uncertainties of parallax, proper 
motion~$\mu_{\alpha^{*}}$~and~$\mu_{\delta}$~are~0.02 mas, 0.03~mas~yr$^{-1}$, 
and~0.02~mas~yr$^{-1}$, respectively. 
The median uncertainty of the radial velocity is~0.3~km s$^{-1}$. 
From these member stars, we estimate the means and standard deviations of 
the basic parameters of the Hyades cluster to be: 
(R.A., Decl.)=($67.41^{\circ}$ $\pm$ $6.50^{\circ}$, 
$17.18^{\circ}$ $\pm$ $5.69^{\circ}$), 
$\varpi$ = $20.99$ $\pm$ $2.61$ mas, 
($\mu_{\alpha^{*}}$, $\mu_{\delta}$) = 
($102.68$ $\pm$ $27.97$, $-27.25$ $\pm$ $16.30$) mas yr$^{-1}$, 
and RV = $39.39$ $\pm$ $3.10$ km s$^{-1}$, 
in concordance with previous results~\citep{perryman1998,
vanLeeuwen2009,majaess2011,lodieu2019b}. 
With these parameters, we calculate the center, 3D motions, and corresponding 
uncertainties of the Hyades cluster in the Galactic 
$O_{\rm g}$--$X_{\rm g}$$Y_{\rm g}$$Z_{\rm g}$ system.
For each member of the Hyades, 
we first calculate its 3D coordinates, 3D motions, and 
corresponding uncertainties in the Galactic coordinate system, and we then transform 
them into the $O_{\rm c}$--$X_{\rm c}$$Y_{\rm c}$$Z_{\rm c}$ system. 

\begin{figure*}[ht]
\centering
  \includegraphics[width=0.32\textwidth]{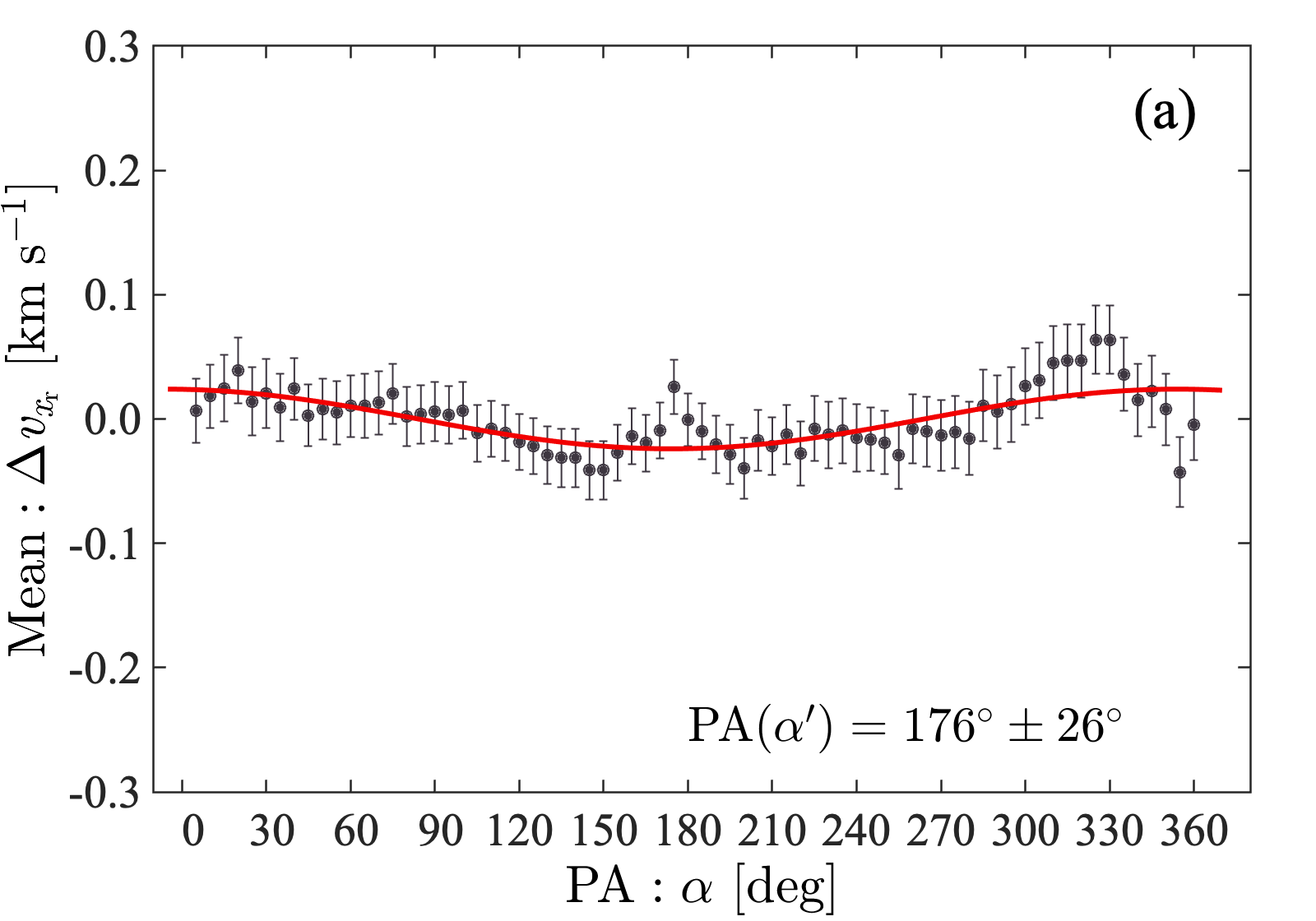}
  \includegraphics[width=0.32\textwidth]{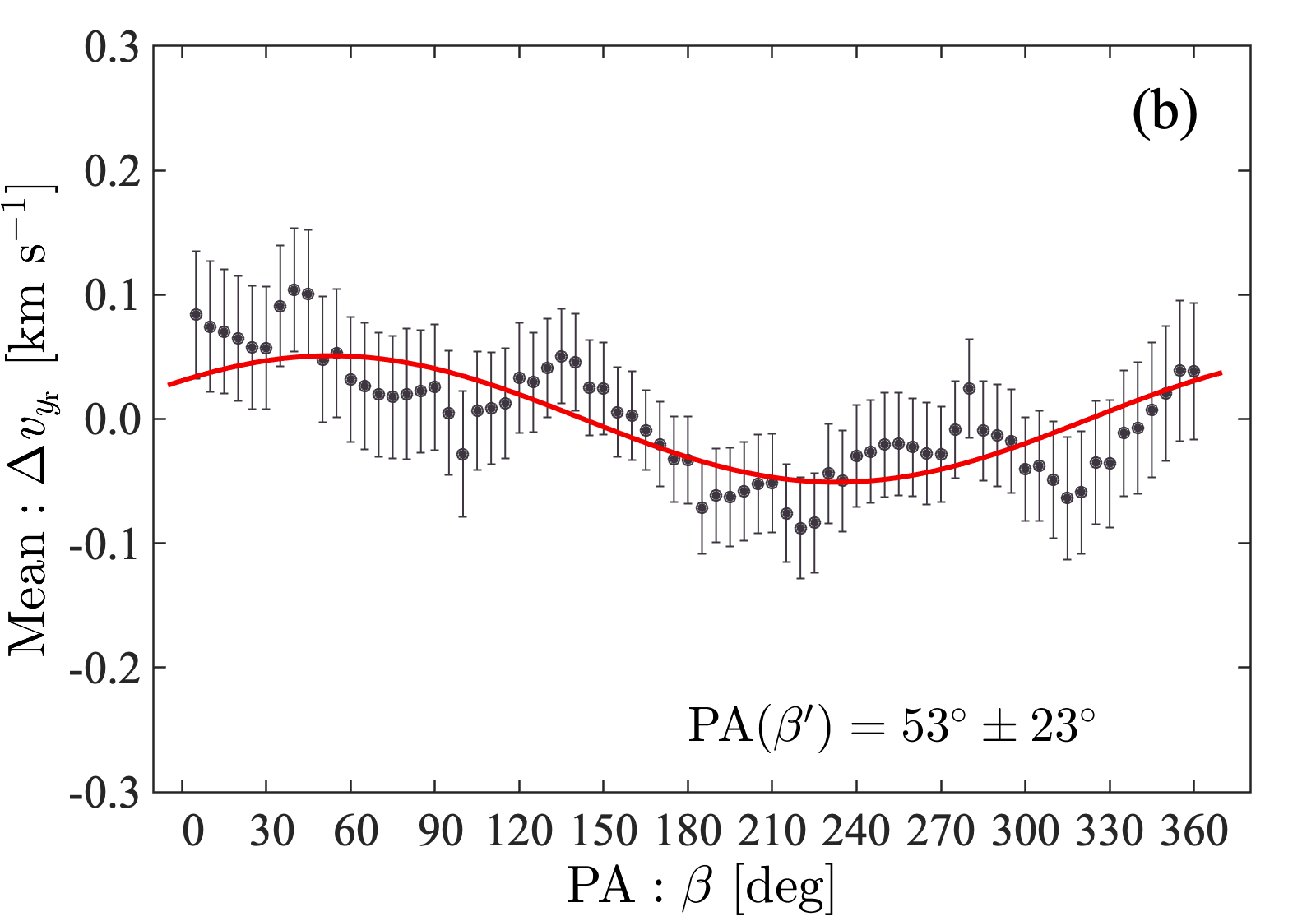}
  \includegraphics[width=0.32\textwidth]{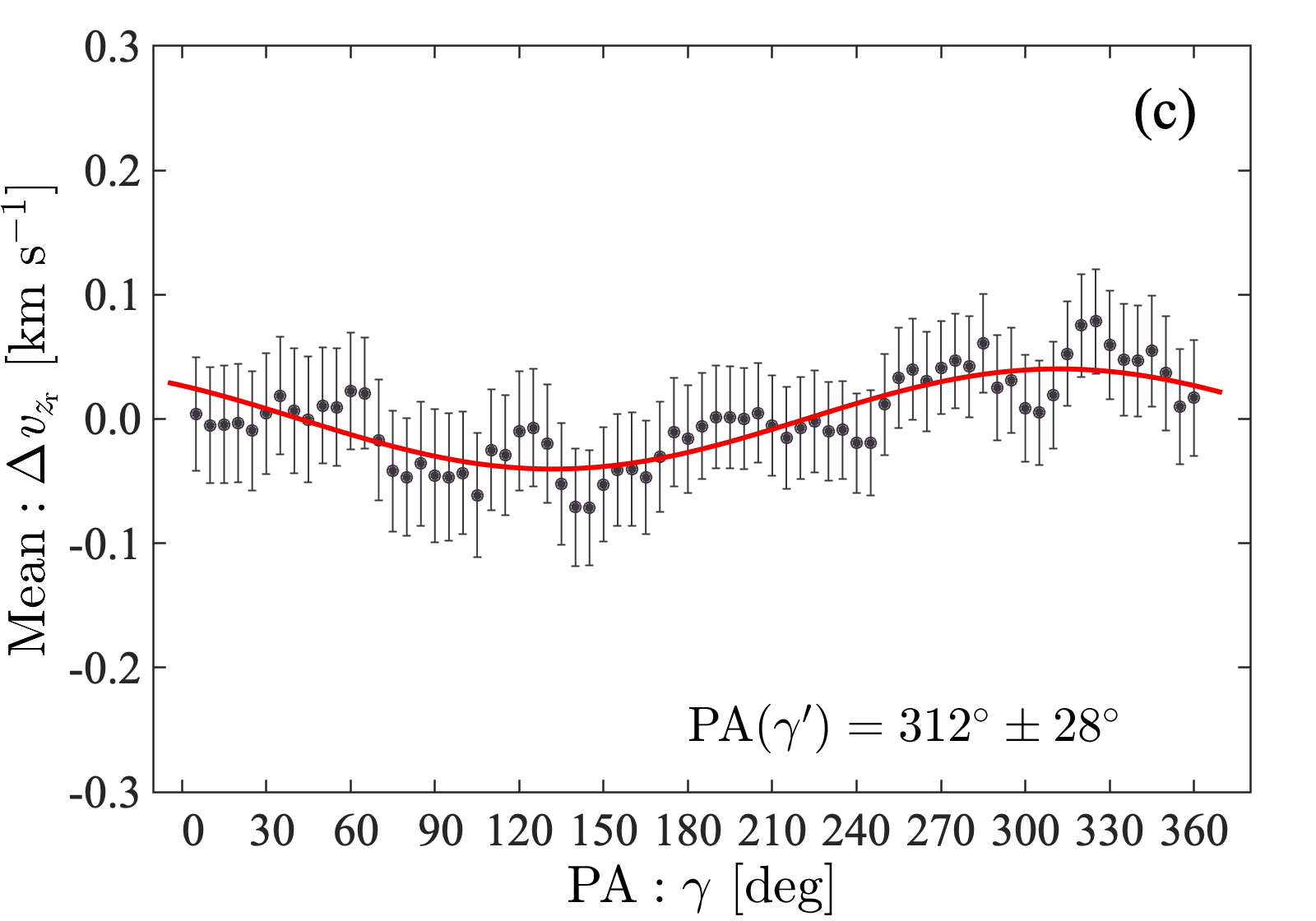}
  \includegraphics[width=0.32\textwidth]{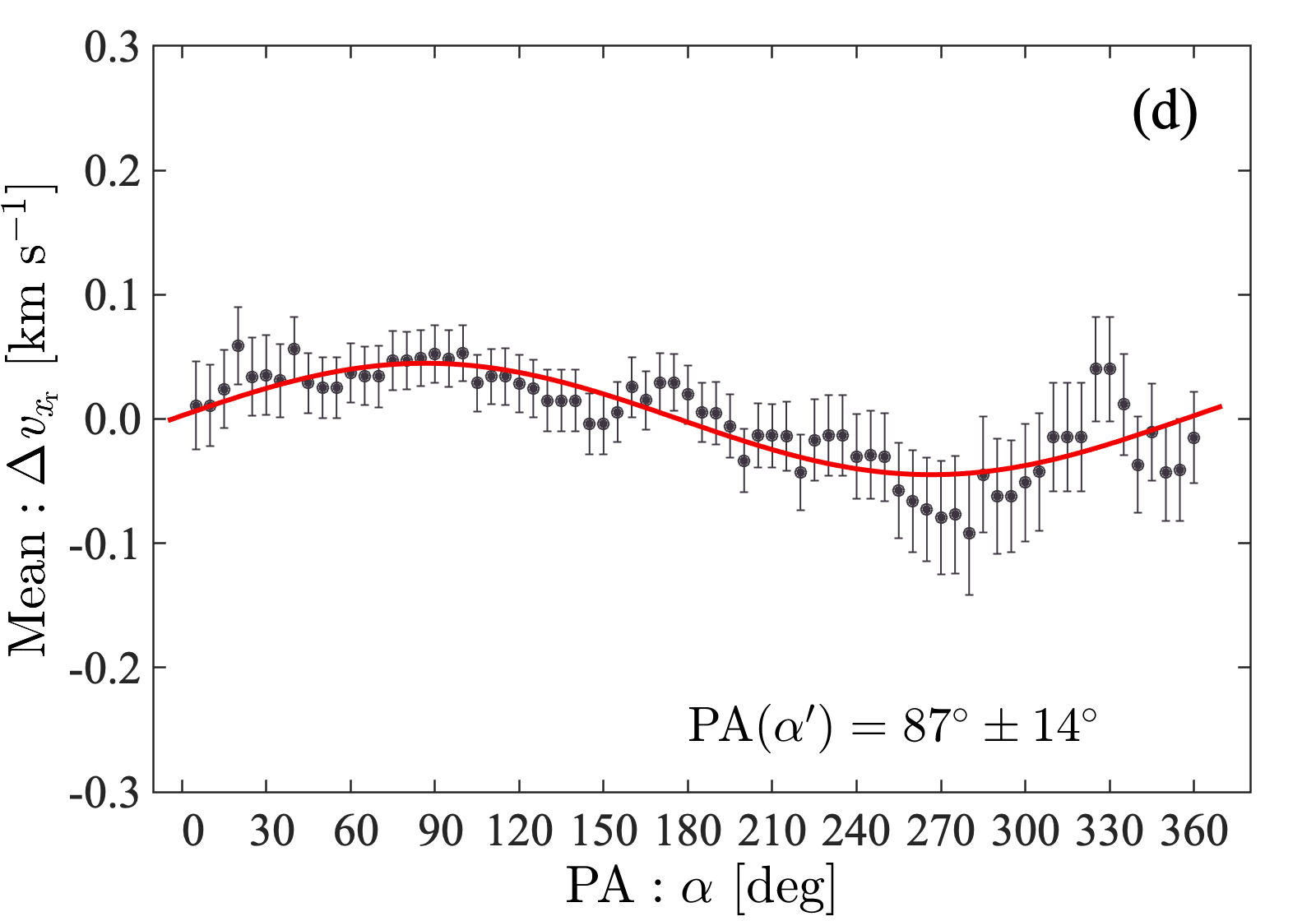}
  \includegraphics[width=0.32\textwidth]{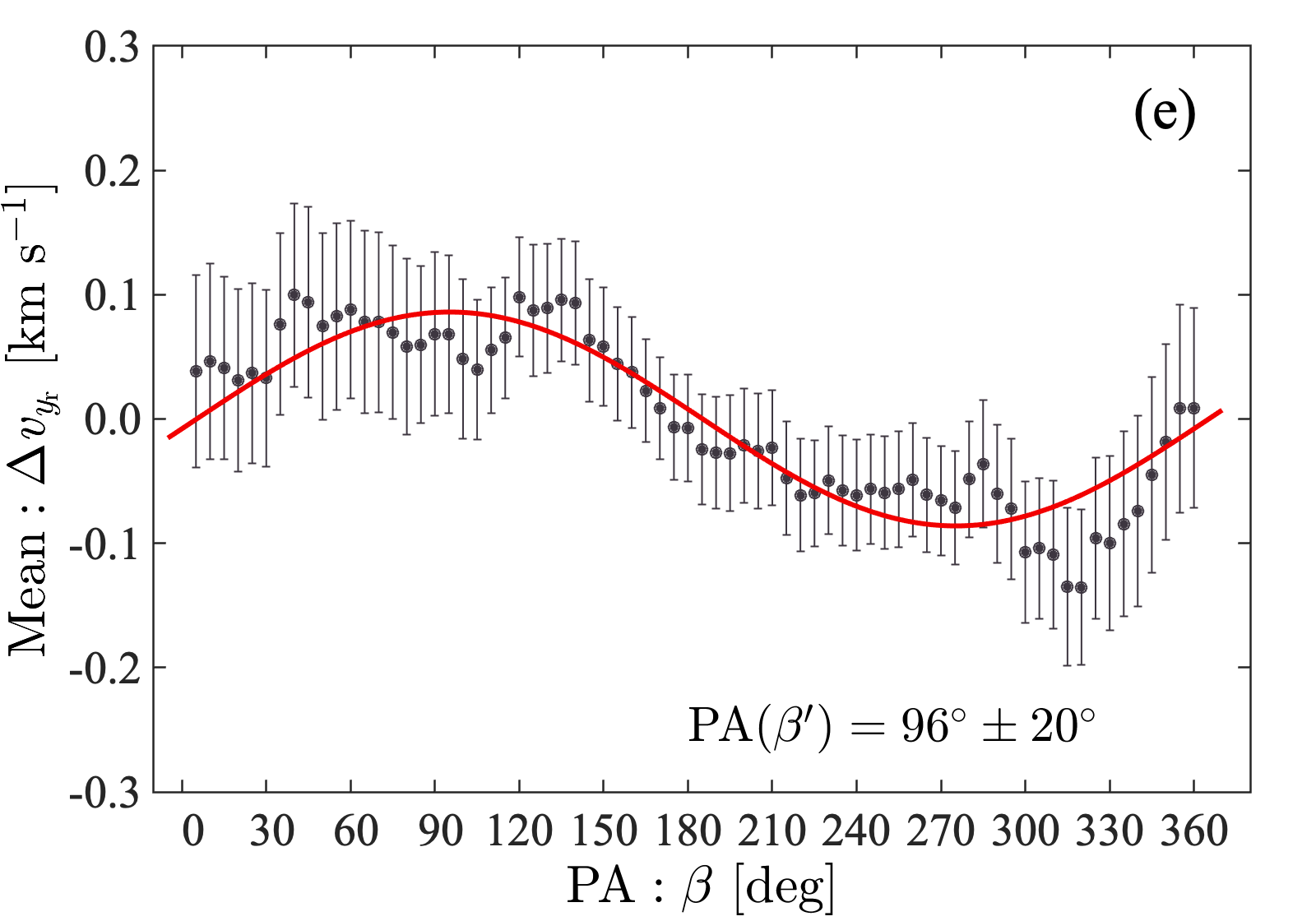}
  \includegraphics[width=0.32\textwidth]{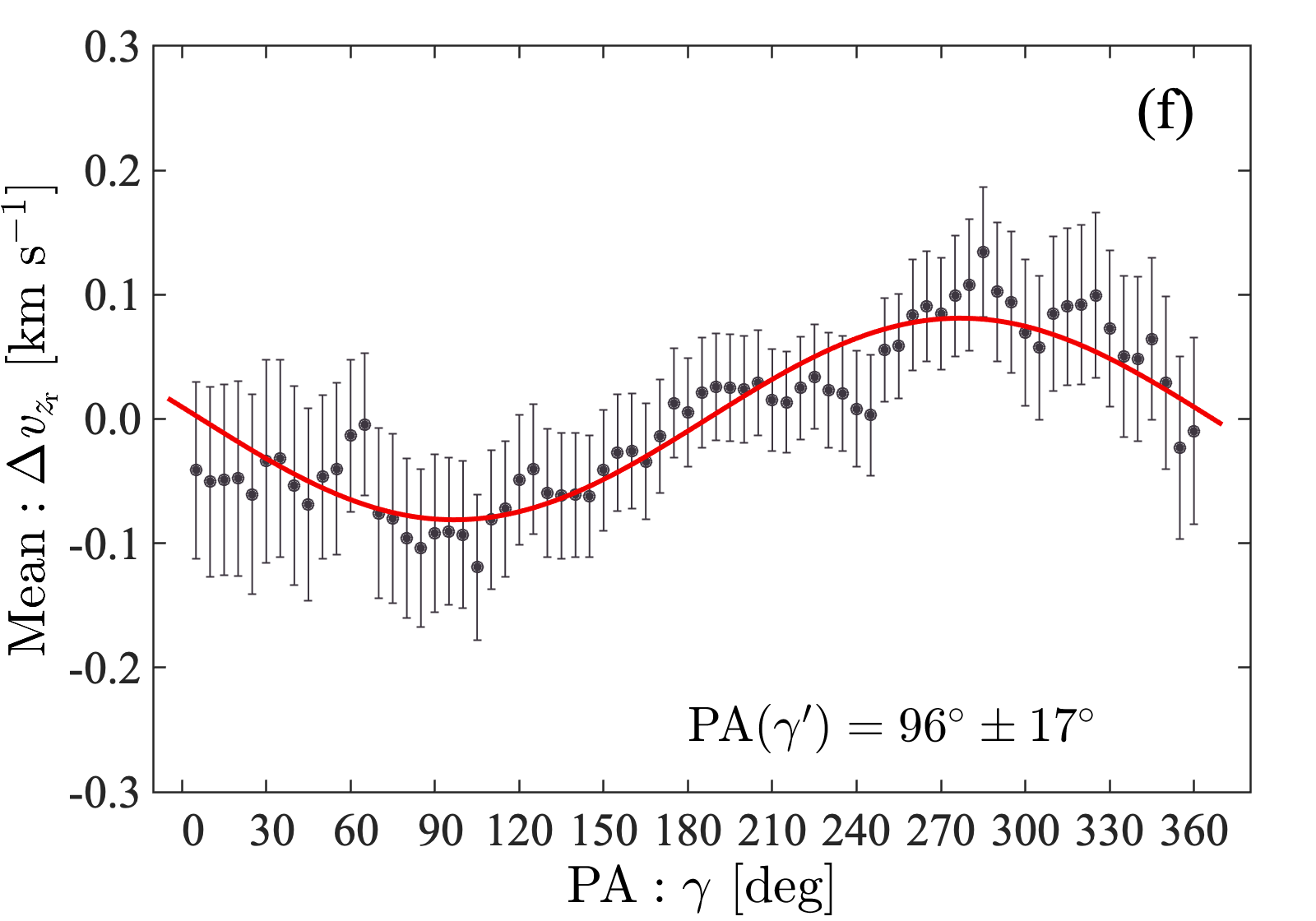}
  \includegraphics[width=0.32\textwidth]{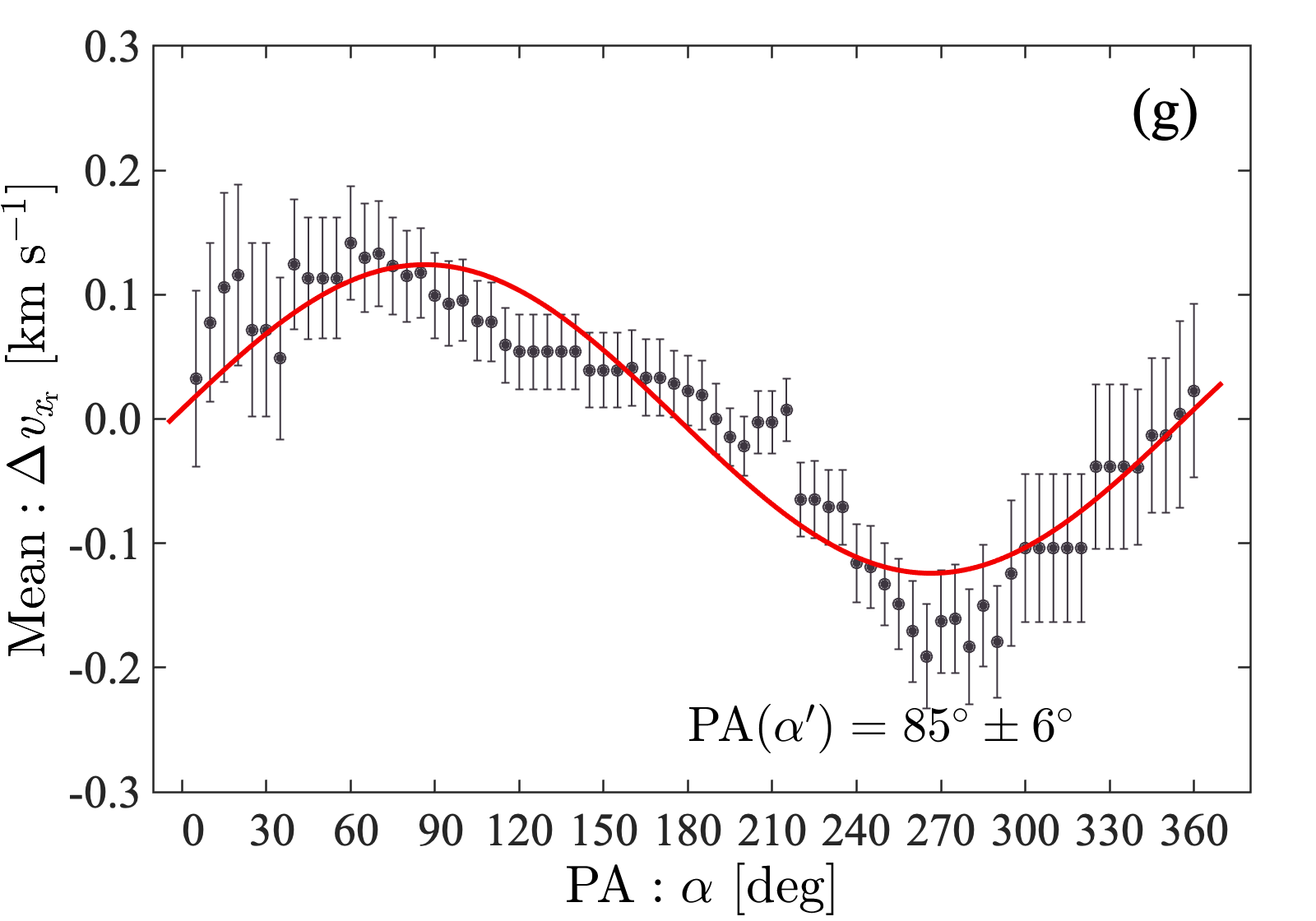}
  \includegraphics[width=0.32\textwidth]{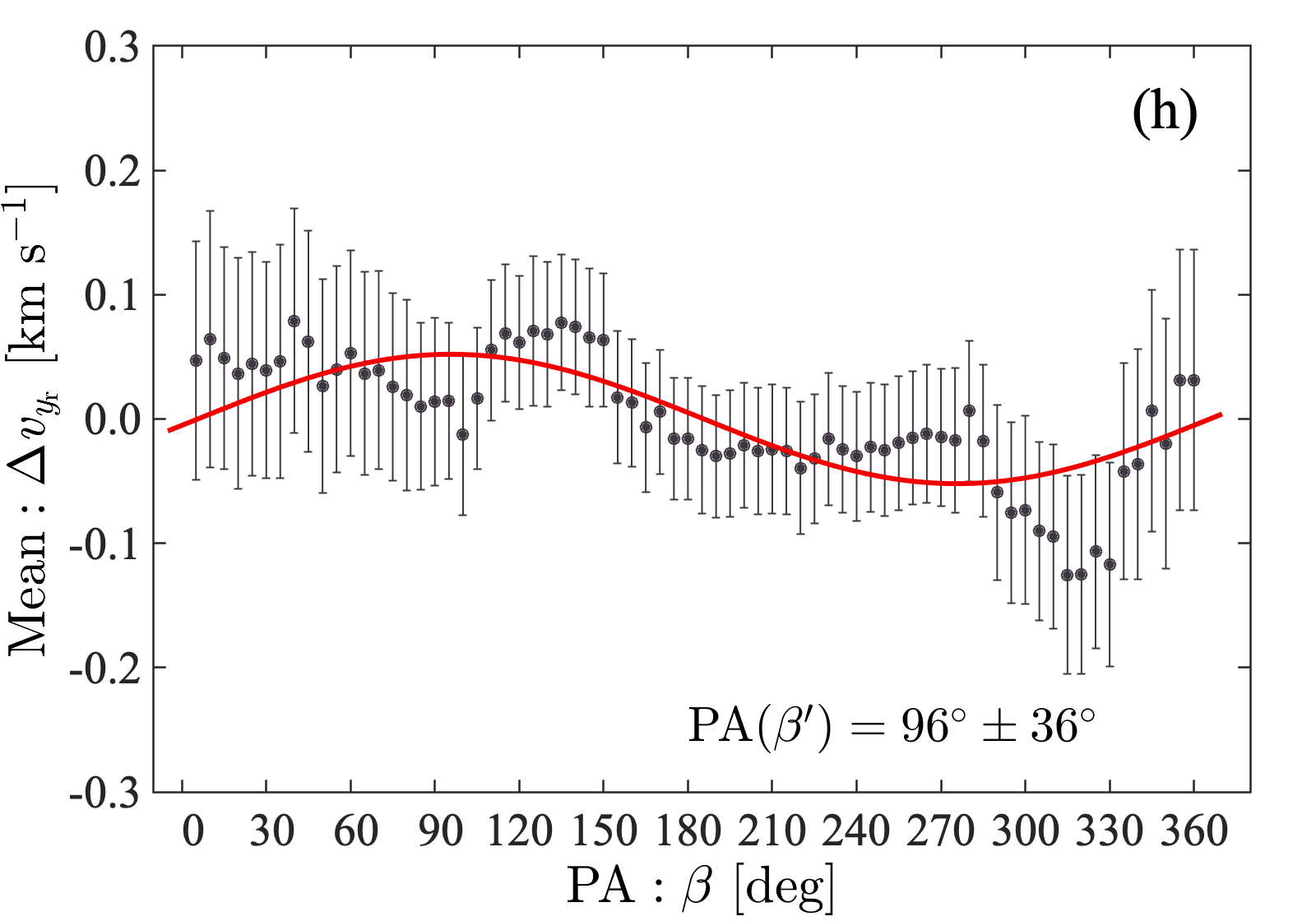}
  \includegraphics[width=0.32\textwidth]{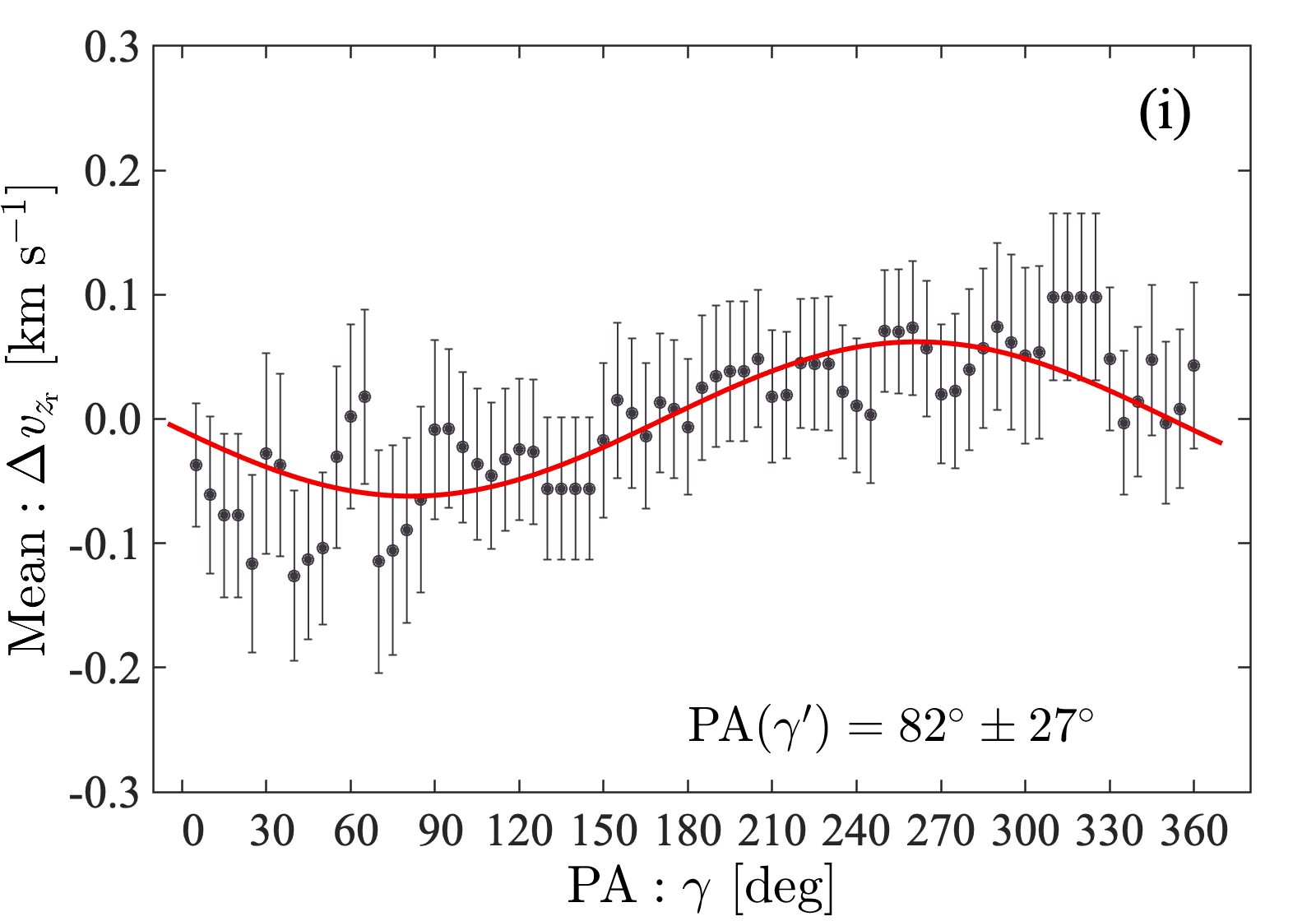}
\caption{Same as Figure~\ref{fig:pa_pleiades}, but for the Hyades cluster. }
\label{fig:pa_hyades}
\end{figure*}

As in the case of the $\alpha$~Per, we can determine the rotation signals in the Hyades
by using the residual velocity method in the 
$O_{\rm c}$--$X_{\rm c}$$Y_{\rm c}$$Z_{\rm c}$ system, but we can not pin down the 
PAs ($\alpha$, $\beta$, $\gamma$) of the rotation axis $\vec{l}$.
Using a set of PAs ($\alpha_{0}$, $\beta_{0}$, $\gamma_{0}$) =
($10^{\circ}$, $10^{\circ}$, $45^{\circ}$),
the 3D coordinates and velocity components of the cluster members are transformed
from the $O_{\rm c}$--$X_{\rm c}$$Y_{\rm c}$$Z_{\rm c}$ system to the rotated
$O_{\rm r}'$--$X_{\rm r}$$Y_{\rm r}$$Z_{\rm r}$ system.
We then determine the PAs of the rotation axis $\vec{l}$ ($\alpha'$, $\beta'$, 
$\gamma'$) in the rotated system by using the same method as 
for the aforementioned clusters.
In Figure~\ref{fig:pa_hyades}, we display the mean residual velocities 
as functions of PAs, using the member stars within 3 $ r_{\rm td} $ (a--c), 
1 $ r_{\rm td} $ (d--f), and 0.5 $ r_{\rm td} $ (g--i) of the Hyades cluster.
According to~\citet{lodieu2019b}, the Hyades cluster has a tidal radius of~9.0 pc.
The sinusoidal behavior of the mean residual velocities 
as a function of PAs is an excellent indication of the 3D rotation of the Hyades 
cluster, although the result in Figure~\ref{fig:pa_hyades}(a) is not very significant.
%

%%%%%%%%%%%%%%%%%%%%%%%%%%%%%%%%%%%%%%%%%%%%% Figure. 13
%%%%%%%%%%%%%%%%%%%%%%%%%%%%%%%%%%%%%%%%%%%%%

Table \ref{table:pa_hyades} and~Figure \ref{fig:pa_hyades} summarize the fitted 
PAs ($\alpha'$, $\beta'$, $\gamma'$) of the rotation axis $\vec{l}$ of the 
Hyades in the rotated system.
Like the Pleiades, the PAs obtained by using the member stars within the tidal radius 
of the Hyades are concordant with those obtained by using the members within half 
of the tidal radius, and they all satisfy the relation of 
tan $\alpha$ $\cdot$ tan $\gamma$ = tan $\beta$ considering the uncertainties.
In contrast, the PAs derived by using the member stars within three times of the
tidal radius are different from the above results, although they also satisfy 
tan $\alpha$ $\cdot$ tan $\gamma$ = tan $\beta$ considering the uncertainties.
This phenomenon has also been shown for the Praesepe and Pleiades clusters.

For the Hyades cluster, the PAs obtained by the members within the tidal radius 
are exploited to determine the rotation axis $\vec{l}$.
Considering the relation tan $\alpha'$ $\cdot$ tan $\gamma'$ = tan $\beta'$, 
we adopt the PAs of ($\alpha'$, $\beta'$, $\gamma'$) = 
($87^{ \circ}$, $90^{\circ}$, $96^{\circ}$) 
to obtain the vectors of the rotation axis $\vec{l}$ in the 
$O_{\rm c}$--$X_{\rm c}$$Y_{\rm c}$$Z_{\rm c}$ system, and we then use
Equation~(\ref{equ:ccs2rcc2}) to determine the PAs of the rotation axis $\vec{l}$ in 
the $O_{\rm c}$--$X_{\rm c}$$Y_{\rm c}$$Z_{\rm c}$ system to be 
($\alpha$, $\beta$, $\gamma$) 
= ($10^{ \circ}$, $11^{\circ}$, $48^{\circ}$).
We find that the rotation axis of the Hyades cluster is nearly parallel to the Galactic 
plane, with an estimated angle of~$7^\circ\pm7^\circ$.
The vectors of the $X_{\rm r}$, $Y_{\rm r}$, and $Z_{\rm r}$ axes 
of the $O_{\rm r}$--$X_{\rm r}$$Y_{\rm r}$$Z_{\rm r}$ system in the 
$O_{\rm c}$--$X_{\rm c}$$Y_{\rm c}$$Z_{\rm c}$ system can be given by 
the PAs ($\alpha$, $\beta$, $\gamma$).
Subsequently, we transform the 3D coordinates and velocity components of 
the cluster members from the $O_{\rm c}$--$X_{\rm c}$$Y_{\rm c}$$Z_{\rm c}$ 
system to the $O_{\rm r}$--$X_{\rm r}$$Y_{\rm r}$$Z_{\rm r}$ system, 
and we finally obtain the 3D coordinates 
($r$, $\varphi$, $z$) and velocity components ($v_ {r}$, $v_{\varphi}$, $v_{z}$)
of the cluster members in the cylindrical coordinate system.
Using the same technique as the previous two clusters,
averaging the rotation velocities ($v_{\varphi}$) of the member stars yields a 
rotation velocity of 0.09 $\pm$ 0.03~km~s$^{-1}$ for the Hyades cluster, in 
agreement with the results shown in Figures~\ref{fig:pa_hyades}.
%

%%%%%%%%%%%%%%%%%%%%%%%%%%%%%%%%%%%%%%%%%%%% Table. 7
\setlength{\tabcolsep}{10.0mm}
\begin{table}[ht]
\centering
\caption{Best-fit PAs of the Hyades cluster.}
\begin{tabular}{c|ccc} 
\hline \hline 
           &    $\alpha'$      &    $\beta'$    &    $\gamma'$  \\ \hline 
3.0 $r_{\rm td}$   &  $176^{\circ}$ $\pm$ $26^{\circ}$   &  $53^{\circ}$ $\pm$ $23^{\circ}$  
          &  $312^{\circ}$ $\pm$ $28^{\circ}$     \\  
1.0 $r_{\rm td}$      &  $87^{\circ}$ $\pm$ $14^{\circ}$  &  $96^{\circ}$ $\pm$ $20^{\circ}$    
          &  $96^{\circ}$ $\pm$ $17^{\circ}$     \\  
0.5 $r_{\rm td}$   &  $85^{\circ}$ $\pm$ $6^{\circ}$   &  $96^{\circ}$ $\pm$ $36^{\circ}$    
          &  $82^{\circ}$ $\pm$ $27^{\circ}$     \\    \hline  
 \end{tabular}
 \tablecomments{$r_{\rm td}$: tidal radius of the Hyades cluster.}
 \label{table:pa_hyades}
\end{table}
%%%%%%%%%%%%%%%%%%%%%%%%%%%%%%%%%%%%%%%%%%%%%%%

\subsubsection{Rotational properties of stars in Hyades}
\label{sec:rotation-hyades}

\begin{figure}[ht]
\centering
\includegraphics[scale=0.18]{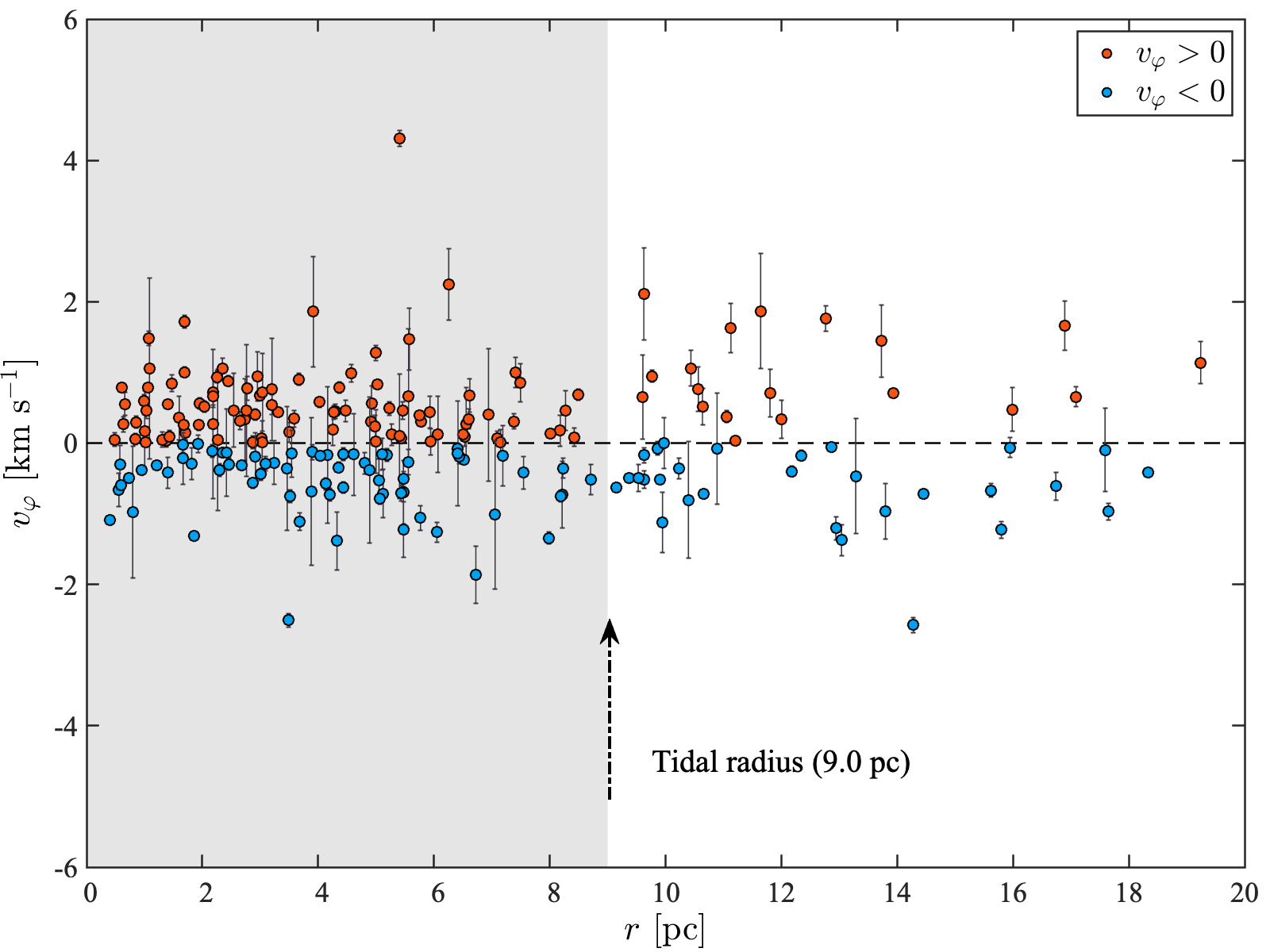}
\caption{Rotational velocity as a function of 
	distance to the cluster center for the member stars within
	twice of the tidal radius of the Hyades.}
\label{fig:rotation_hyades}
\end{figure}

Figure~\ref{fig:rotation_hyades} shows the variation in the rotational velocities
of the member stars of the Hyades cluster vary with their distance to the
cluster center.
Similar to the properties of the former two clusters, 
not all member stars rotate in the same direction. 
The number of the member stars with positive rotational velocities is 1.4 times 
of those with negative rotational velocities.
Compared with the Pleiades and the $\alpha$~Per cluster, there are fewer member stars have 
peculiar rotational velocities that deviate from that of most cluster members within the tidal 
radius of the Hyades.
In addition, the rotation of the member stars located near or beyond the tidal radius does 
not appear to be disorganized, as seen from Figure~\ref{fig:rotation_hyades}.
%

%%%%%%%%%%%%%%%%%%%%%%%%%%%%%%%%%%%%%%%%%%%%% Figure. 14
%%%%%%%%%%%%%%%%%%%%%%%%%%%%%%%%%%%%%%%%%%%%%

%
In the same way as for the Pleiades and $\alpha$~Per,
the skewness of the astrometric parameters of the member stars
within the tidal radius of the Hyades cluster are calculated to assess the 
symmetry of the cluster density distribution.
The skewness of the Galactic longitude, Galactic latitude, and parallax 
are~0.1, 0.1,~and~0.0, respectively, indicating that it makes sense for us
to assume a spherically symmetric distribution in the Hyades. 
%

%%%%%%%%%%%%%%%%%%%%%%%%%%%%%%%%%%%%%%%%%%%%% Figure. 15
\begin{figure}[ht]
\centering
\includegraphics[scale=0.18]{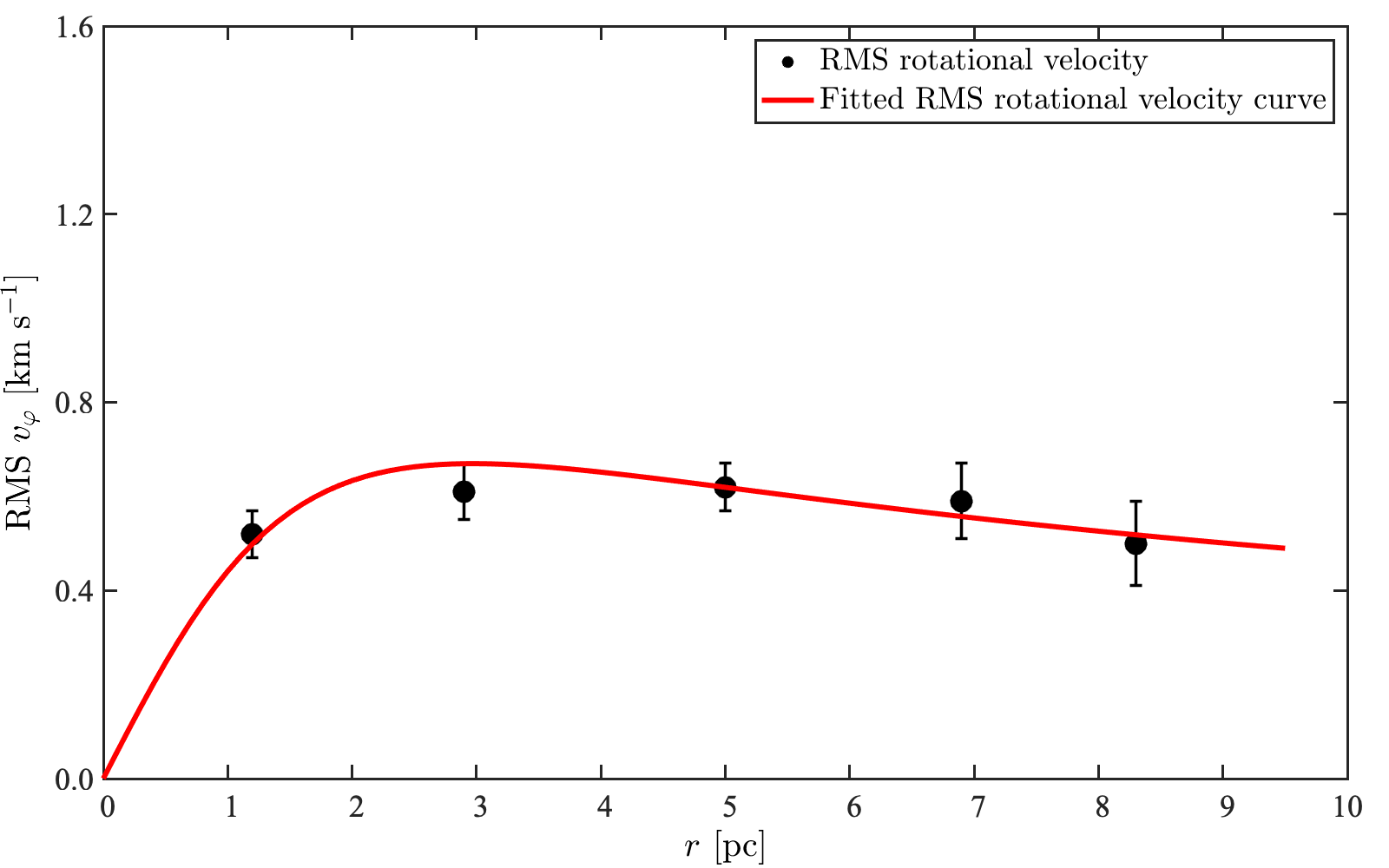}
\caption{Same as Figure \ref{fig:rms_alphaper}, but for the Hyades cluster.}
\label{fig:rms_hyades}
\end{figure}
%%%%%%%%%%%%%%%%%%%%%%%%%%%%%%%%%%%%%%%%%%%%%

%%%%%%%%%%%%%%%%%%%%%%%%%%%%%%%%%%%%%%%%%%%%% Table. 8
\setlength{\tabcolsep}{8.0mm}
\begin{table}[ht]
\centering 
\caption{RMS rotational velocities of the members stars within 
the tidal radius of the Hyades cluster.}
\begin{tabular}{cc|cccc} 
  \hline \hline 
$R_{i}$  &  $R_{o}$ & $R_{m}$  & $N$  & RMS $v_{\varphi}$  &  $\epsilon_{{\rm RMS} \, v_{\varphi}}$   \\ %\hline
      {[pc]}  &  {[pc]}    &  {[pc]}  &  & {[km~s$^{-1}$]} & {[km~s$^{-1}$]}  \\
       (1)    &   (2)   &  (3)     & (4)  & (5) & (6) \\
\hline
 0.0 &  2.0    &  1.2  &  36  &  0.52  &  0.05   \\  
 2.0 &  4.0    &  2.9  &  44  &  0.61  &  0.06   \\  
 4.0 &  6.0    &  5.0  &  46  &  0.62  &  0.05   \\  
 6.0 &  8.0    &  6.9  &  21  &  0.59  &  0.08   \\  
 8.0 &  9.5    &  8.3  &  9  &  0.50  &  0.09   \\  \hline  
 \end{tabular}
 \tablecomments{The same format as Table~\ref{table:rms_pleiades},
but for the Hyades cluster.}
 \label{table:rms_hyades}
\end{table}
%%%%%%%%%%%%%%%%%%%%%%%%%%%%%%%%%%%%%%%%%%%%%

%%%%%%%%%%%%%%%%%%%%%%%%%%%%%%%%%%%%%%%%%%%%% Figure. 16
\begin{figure}[ht]
\centering
\includegraphics[scale=0.18]{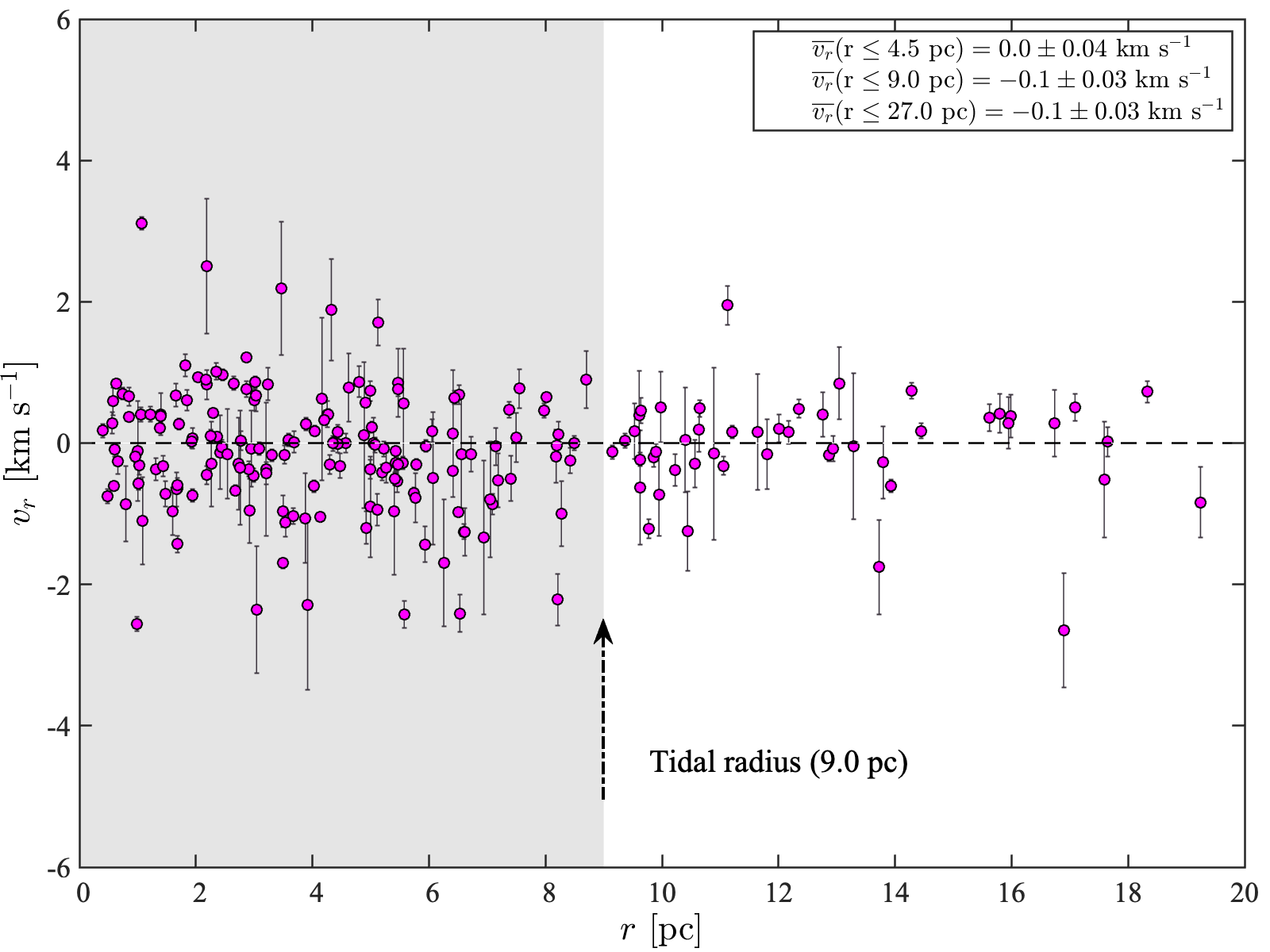}
\caption{Same as Figure~\ref{fig:radial_pleiades}, but for the Hyades cluster.}
\label{fig:radial_hyades}
\end{figure}
%%%%%%%%%%%%%%%%%%%%%%%%%%%%%%%%%%%%%%%%%%%%%

%
In the sample, 163~member stars lie within the tidal radius of the 
Hyades cluster. The median absolute value of their rotational 
velocities~$v_{\varphi}$~is~$\sim$0.5~km~s$^{-1}$ with a standard 
deviation of~$\sim$0.5 km s$^{-1}$. 
Similar to the analysis for the Pleiades and the $\alpha$~Per cluster, 
we first eliminate 7 stars with peculiar rotational velocities after inspecting the 
distribution of rotational velocities of the member stars.
Then, using the remaining~156~member stars, we calculate the RMS rotation velocity 
and uncertainty for the members within each distance bin. 
The results are given in Table~\ref{table:rms_hyades} and 
Figure~\ref{fig:rms_hyades}.
Using Equation~(\ref{equ:v-t}), the RMS rotational velocities as a function
of the distances to the cluster center are well fit by a theoretical circular 
motion velocity curve. Therefore, it may be feasible to describe the rotation of 
the member stars within the tidal radius of the Hyades using Newton's theorems. 
Furthermore, with the same method as for the above two clusters,
after using Equation~(\ref{equ:v-t})~to fit the RMS velocities listed 
in Table~\ref{table:rms_hyades}, we infer that the core radius and mass within
the tidal radius of the Hyades cluster are 2.1 $\pm$ 0.3~pc and 569 $\pm$ 
108~${\rm M}_{\odot}$, 
respectively.

We also examine whether the Hyades cluster, an OC with an age of 
about 650 Myr (see Section \ref{intro}), is expanding or contracting
by statistically analyzing the radial components~$v_{r}$~of the member stars. 
As indicated in Figure~\ref{fig:radial_hyades},
the mean radial components~$\overline{v_{r}}$~of the member stars 
within~0.5 $r_{\rm td}$, 1 $r_{\rm td}$, and 3 $r_{\rm td}$~are~0.0 $\pm$ 0.04, 
--0.1 $\pm$ 0.03, and~--0.1 $\pm$ 0.03 km s$^{-1}$, respectively.
Similar to the Pleiades and the $\alpha$~Per cluster, the mean radial components
are also close to zero. These results suggest that there is no significant expansion or 
contraction in the 
Hyades, and this implies that the rotation of the cluster members within the tidal radius 
may exhibit closed-loop motions. 
Figure~\ref{fig:radial_hyades} also shows that the member stars of the Hyades 
cluster show non-negligible acceleration and 
dispersion in their radial components, indicating that there should be additional mass 
in the system to account for the force supporting the radial component acceleration.
Therefore, the estimated mass in this work is a lower limit for the Hyades.

\section{Discussion}
\label{discussion}

%%%%%%%%%%%%%%%%%%%%%%%%%%%%%%%%%%%%%%%%%%%% Table. 3
\setlength{\tabcolsep}{3.0mm}
\renewcommand\arraystretch{1.2}
\begin{table}[ht]
\centering
\caption{Parameters of the Praesepe, Pleiades, $\alpha$~Per, and the Hyades.}
\begin{tabular}{c|cc|ccccc} 
\hline  \hline 
  &  Rotational velocity &  $\vec{l}$ vs. GP & $Dist.$ & Mass & Age & $r_{\rm td}$ ($r_{\rm co}$) &  $|Z|$ \\  
  &  [km~s$^{-1}$]     &  [degree]  &    [pc]   & ${\rm M}_{\odot}$ & [Myr] & [pc] &  [pc]         \\    \hline   
Praesepe       &    0.12 $\pm$ 0.05 &  $34$ $\pm$ $15$  
& 178 $\pm$ 0.8$^{[1]}$ & $\sim$510$^{[3]}$  & 590--700$^{[1, 7]}$  & 10.7(2.6)$^{[1]}$  
& 101 $\pm$ 2.2$^{[1]}$   \\  \hline  
Pleiades        &    0.24 $\pm$ 0.04 &  $62$ $\pm$ $13$  
&  135 $\pm$ 0.4$^{[1]}$ & $\sim$820$^{[4]}$  & 110--160$^{[1, 7]}$  & 11.6(2.0)$^{[1]}$  
& 55 $\pm$ 0.2$^{[1]}$  \\  \hline
$\alpha$~Per &    0.43 $\pm$ 0.08 &  $48$ $\pm$ $14$  
& 187 $\pm$ 3.9$^{[1]}$ & $\sim$920$^{[5]}$  &  50--70$^{[8, 9]}$   & 9.5(2.3)$^{[1]}$    
& 20 $\pm$ 0.5$^{[1]}$  \\  \hline  
Hyades         &    0.09 $\pm$ 0.03 &     $7$ $\pm$ $7$  
& 47 $\pm$ 0.2$^{[2]}$ & $\sim$280$^{[6]}$  &  650--780$^{[2, 10]}$  & 9.0(3.1)$^{[2]}$     
& 17 $\pm$ 0.1$^{[2]}$  \\  \hline  
 \end{tabular}
 \tablecomments{$^{1}$~\citet{lodieu2019}, $^{2}$~\citet{lodieu2019b}, $^{3}$~\citet{loktin2020}, 
 $^{4}$~\citet{almeida2023}, $^{5}$~\citet{nikiforova2020}, 
 $^{6}$~\citet{roser2011}, $^{7}$~\citet{gossage2018}, 
 $^{8}$~\citet{prosser1996}, $^{9}$~\citet{makarov2006},
 $^{10}$~\citet{brandner2023}. $\vec{l}$: rotation axis. GP: Galactic plane. 
 $\vec{l}$ vs. GP: the included angle between the rotation axis and the Galactic plane.
 $Dist.$: distance to the Sun. $r_{\rm td}$: tidal radius. $r_{\rm co}$: core radius. 
 $|Z|$: $z$-scale heights from the Galactic plane.
 }
 \label{table:rotation_oc}
\end{table}
%%%%%%%%%%%%%%%%%%%%%%%%%%%%%%%%%%%%%%%%%%%%%%%

%%%%%%%%%%%%%%%%%%%%%%%%%%%%%%%%%%%%%%%%%%%%% Figure. 17
\begin{figure}[ht]
\centering
\includegraphics[scale=0.40]{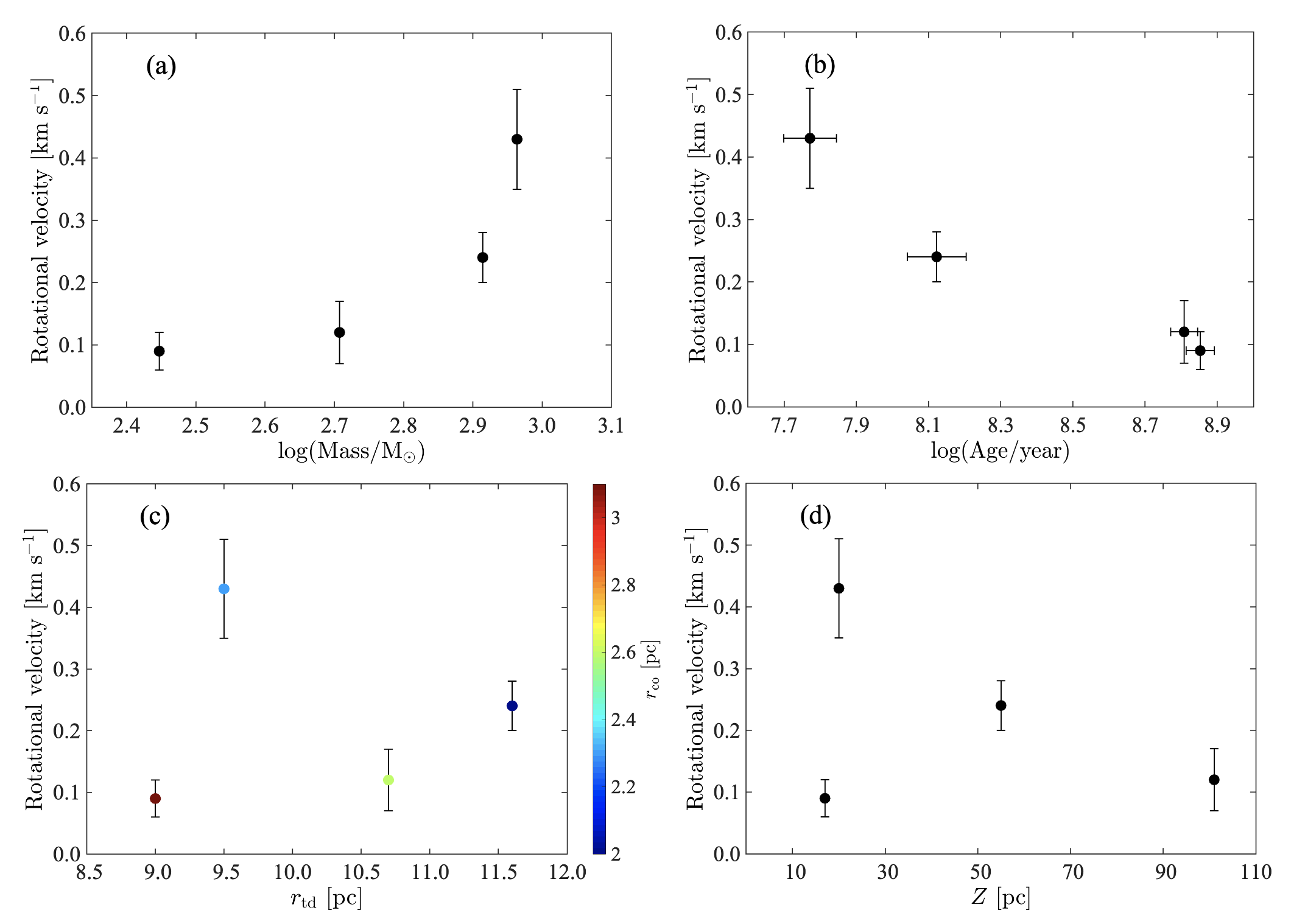}
\includegraphics[scale=0.402]{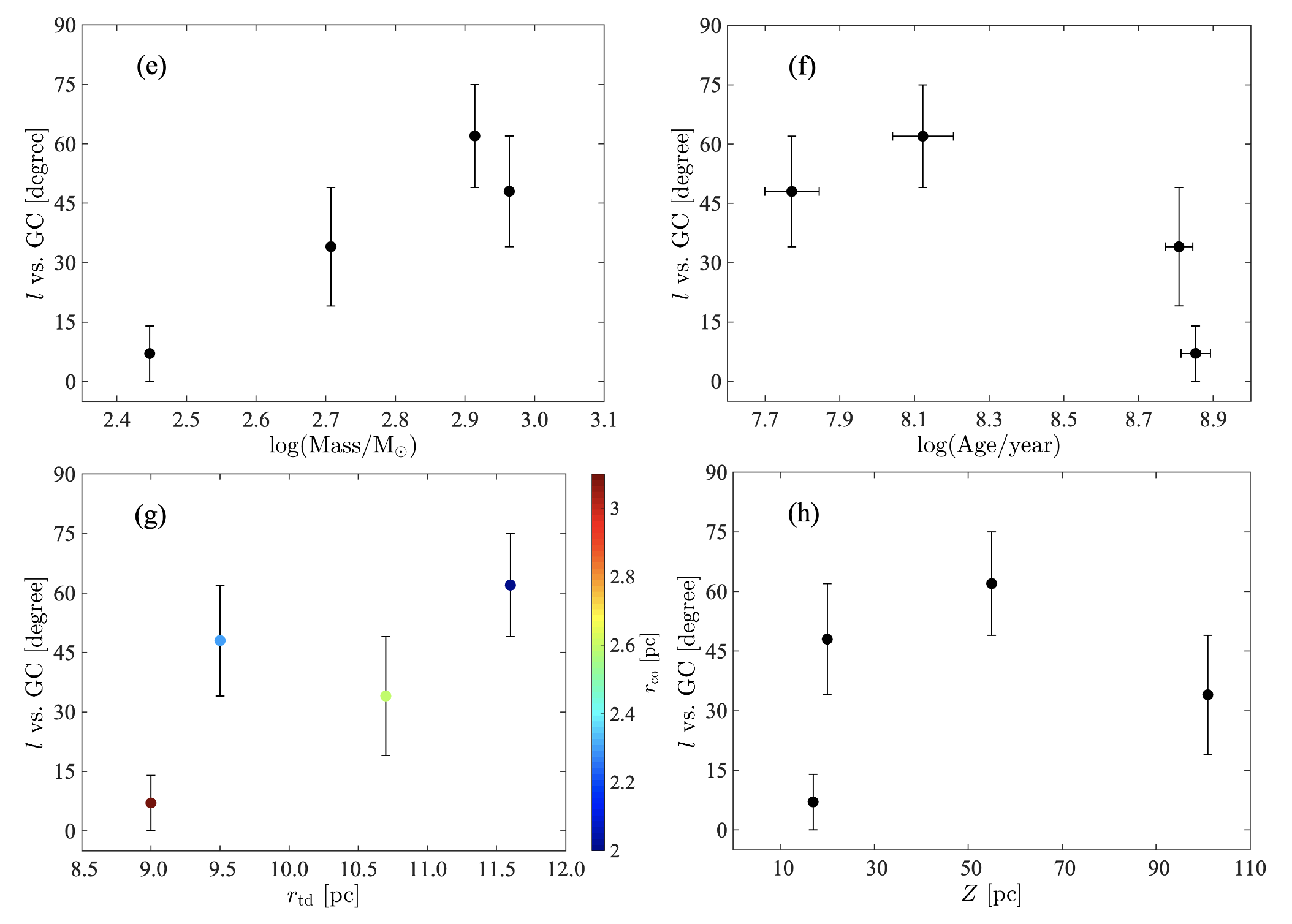}
\caption{Panels (a), (b), (c), and (d): rotational velocities of the Praesepe, 
Pleiades, $\alpha$~Per, and the Hyades clusters as functions of their masses, 
ages, radii ($r_{\rm td}$, $r_{\rm co}$), and $z$-scale heights ($Z$) from the Galactic plane. 
Panels (e), (f), (g), and (h): the same as Panels (a-d), but for the angles of the cluster
rotation axes ($\vec{l}$) with respect to the Galactic plane.}
\label{fig:oc_rotation}
\end{figure}
%%%%%%%%%%%%%%%%%%%%%%%%%%%%%%%%%%%%%%%%%%%%%

Based on \emph{Gaia} DR3, we investigate the 3D internal rotation 
of four nearby member-rich OCs and estimate their rotation axes and mean 
rotational velocities.
Interestingly, Newton's theorem can well interpret the rotation of 
member stars within the tidal radii of these OCs.
In addition, no indication of expansion or contraction has been detected in 
the four OCs, in agreement with the result of~\citet{dellacroce2023}, who
reported that the clusters older than 30 Myr do not experience expansion 
and are mostly compatible with equilibrium configurations.
Although the sample size is small, Praesepe, the Pleiades, $\alpha$~Per, 
and the Hyades are currently the only four OCs with a determined 3D rotation,
which makes it feasible to explore the possible relations between the
rotation of OCs and other cluster parameters.
In Table~\ref{table:rotation_oc}, we compile the distances, masses, 
ages, radii, and $z$-scale heights of these four OCs from the literature.
The cluster rotational velocities and rotation axes given by this work are
also listed.
Clearly, the rotational parameters of the OCs have no obvious relation 
with the cluster distances from the Sun.
The results for the other cluster parameters are given 
in~Figure~\ref{fig:oc_rotation}.

Figure~\ref{fig:oc_rotation}(a-d) present the variation in the rotational velocities 
with other cluster parameters.
It is shown in Figure~\ref{fig:oc_rotation}(a) that OCs with higher masses 
tend to have greater rotational velocities, implying that the rotational velocity 
of OCs may be positively correlated with their masses.
In fact, the Galactic globular clusters, as the stellar populations that are much more 
massive than OCs, are observed to have internal rotation velocities of several 
km s$^{-1}$~\citep[e.g.,][]{lanzoni2018,leanza2022}, which is one order of 
magnitude higher than these four OCs.
In Figure~\ref{fig:oc_rotation}(b), it is interesting to note that young OCs have 
higher rotation velocities than old OCs, which suggests that the rotational
velocity of OCs may gradually decrease as they age.
If this is the case, the present-day rotation of OCs is likely to
originate from their predecessor clusters, whose rotation is found
to be common in hydrodynamical simulations~\citep[e.g.,][]{mapelli2017}.
With the available data, we do not find any clear relation between the 
rotational velocities and the tidal radii or the $z$-scale heights of OCs.

Figure~\ref{fig:oc_rotation}(e-h) present the 
included angles between the cluster rotation axes and the Galactic plane
as a function of cluster parameters.
In Figure~\ref{fig:oc_rotation}(e), we find that the included angles of the
massive OCs appear to be larger than those of the low-mass OCs.
In comparison with old OCs, the younger OCs seem to have larger included
angles, as shown in Figure~\ref{fig:oc_rotation}(f).
The above two relations are not significant when the errors of the included 
angles are taken into account.
Based on these four clusters, no clear correlations can be identified 
between the directions of the rotation axes and the cluster radii or the cluster 
$z$-scale heights, as shown in Figure~\ref{fig:oc_rotation}(g) and (h).
So far, we have only studied the 3D rotation of four star-rich OCs 
within 200~pc from the Sun.
The sample size is very limited. 
The above results need to be further confirmed by the rotational 
characteristics of more OCs.
An accurate detection of 3D rotation in more distant star-rich OCs (such as 
those clusters located within 500~pc) requires high-precision astrometry 
and improved methods.
More efforts will be undertaken to reveal the internal 3D rotation of more
OCs, which will help us to improve our understanding of the formation and
evolution of OCs in the Milky Way.

\section{Summary}
\label{summary}

We report significant detections of rotation in the Pleiades, $\alpha$~Per, 
and Hyades clusters based on the high-quality data of \emph{Gaia} DR3.
In order to unveil the internal kinematics of star clusters, we have slightly
amended the method developed in our previous work \citepalias{hao2022b}.
The analyses on the motions of the member stars not only revealed 3D 
rotation of the Pleiades, $\alpha$~Per, and the Hyades, but also determined 
their rotation axes in the Milky Way, with angles of 
$62^\circ\pm13^\circ$, $48^\circ\pm14^\circ$, and $7^\circ\pm7^\circ$ 
relative to the Galactic plane, respectively.
We also unveil the rotational properties of the member stars in the three OCs 
and suggest that the theorems of Newtonian mechanics can well characterize the 
rotational velocities of the member stars within each cluster tidal radius.
In addition, the statistics of current observations suggest that the three 
clusters do not expand or contract.

Combining the Praesepe cluster analyzed in~\citetalias{hao2022b} with 
the three OCs investigated in this work,
we find a possible relation between the rotational velocity 
and the age and mass of the cluster.
Massive OCs tend to have significant internal rotation, and the rotation 
velocity gradually decreases from young to old OCs. 
However, these results need to be further confirmed when the rotational 
features of more clusters are revealed.
To date, the internal kinematics of almost all known OCs remain elusive, 
leaving a major gap in the study of star clusters in the Milky Way.
The method reported in this work takes a step in the direction of unveiling
the detailed kinematic properties of OCs. 
The 3D rotations of OCs has been measured based on the high-precision 
parallaxes, proper motions, radial velocities from the \emph{Gaia} 
DR3 dataset.
With the improvement of astrometric accuracy and precision in the upcoming 
\emph{Gaia} Data Releases, it is expected that there will be a wealth 
of yet-to-be-revealed interesting kinematic properties of OCs, deserving 
our attention and exploration.

\acknowledgments 
We thank the anonymous referee for the instructive 
comments and suggestions that greatly helped us to improve the paper. 
This work was funded by the NSFC grant 11933011, 11988101, National SKA Program 
of China (grant No. 2022SKA0120103), and the Key Laboratory for 
Radio Astronomy.
L.G.H. acknowledges the support from the Youth Innovation Promotion Association 
CAS. 
Y.J.L. acknowledges support from the NSFC grant 12203104 and the Natural Science 
Foundation of Jiangsu Province (grant No. BK20210999). 
We used data from the European Space Agency mission \textit{Gaia}
(\url{http://www.cosmos.esa.int/gaia}), processed by the \textit{Gaia}
Data Processing and Analysis Consortium (DPAC; see
\url{http://www.cosmos.esa.int/web/gaia/dpac/consortium}). Funding for
DPAC has been provided by national institutions, in particular, the
institutions participating in the \textit{Gaia} Multilateral
Agreement.

\appendix

\setcounter{table}{0}
\setcounter{figure}{0}
\renewcommand{\thetable}{A\arabic{table}}
\renewcommand{\thefigure}{A\arabic{figure}}

\begin{figure}[ht]
\begin{center}
\includegraphics[width=0.50\textwidth]{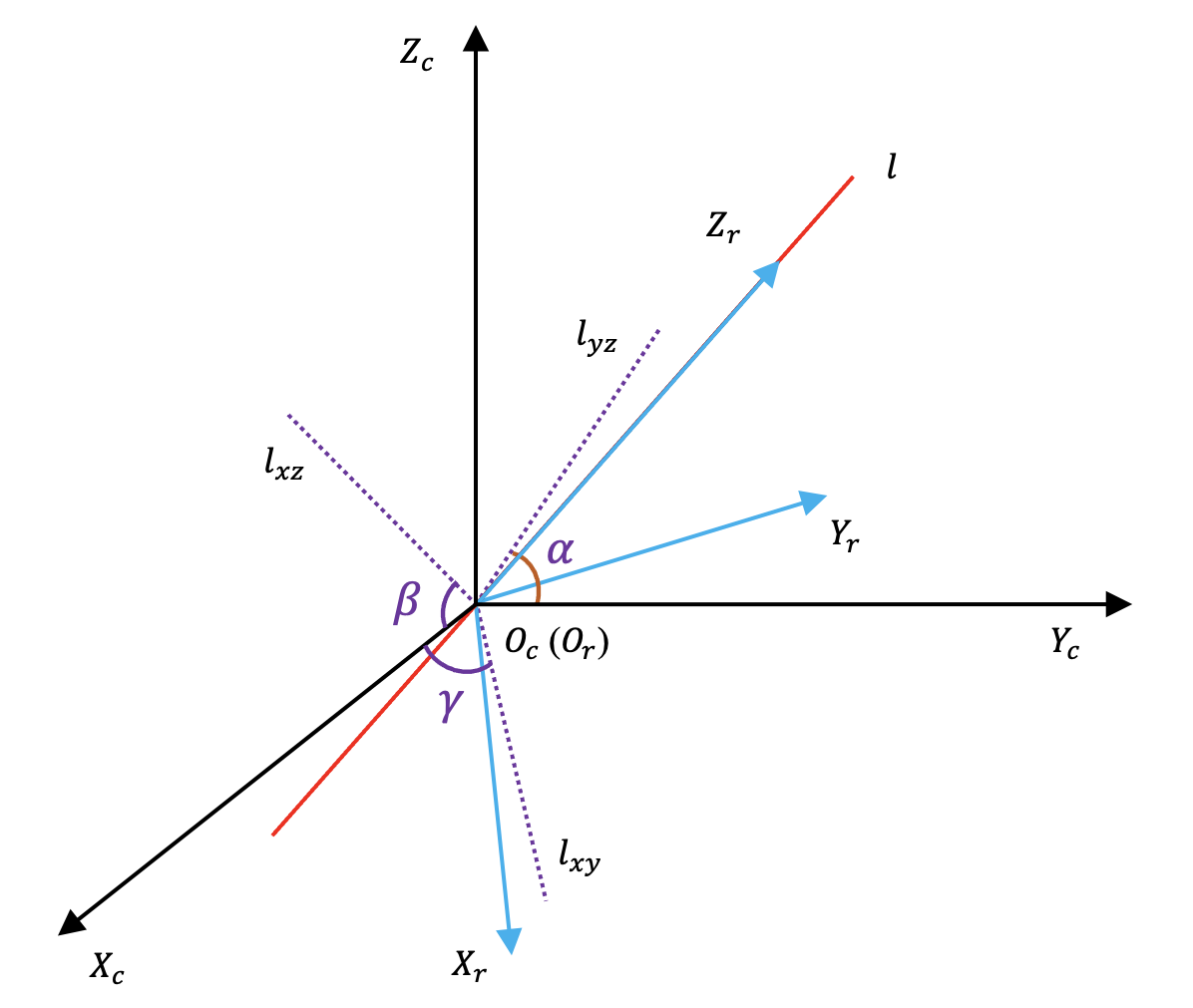}
\caption{Definitions of the 
$O_{\rm c}$--$X_{\rm c}$$Y_{\rm c}$$Z_{\rm c}$ (black) system and 
the $O_{\rm r}$--$X_{\rm r}$$Y_{\rm r}$$Z_{\rm r}$ (blue) 
system (from~\citetalias{hao2022b}).
The cluster rotation axis $\vec{l}$ (red) is in accordance with the $Z_{\rm r}$ axis.
The projections of $\vec{l}$ in the $Y_{\rm c}$--$Z_{\rm c}$, 
$X_{\rm c}$--$Z_{\rm c}$, and $X_{\rm c}$--$Y_{\rm c}$ planes are
$l_{yz}$, $l_{xz}$, and $l_{xy}$ (purple), respectively.
The projection of the $X_{\rm r}$--axis in the $X_{\rm c}$--$Y_{\rm c}$
plane is consistent with $l_{xy}$.
$\alpha$, $\beta$ and $\gamma$ indicate the included angles between
$l_{yz}$ and $Y_{\rm c}$, $l_{xz}$ and $X_{\rm c}$, 
and $l_{xy}$ and $X_{\rm c}$, respectively. 
}
\label{fig:ccs2axis}
\end{center}
\end{figure}

Figure~\ref{fig:ccs2axis} presents the definitions of the 
$O_{\rm c}$--$X_{\rm c}$$Y_{\rm c}$$Z_{\rm c}$ and 
the $O_{\rm r}$--$X_{\rm r}$$Y_{\rm r}$$Z_{\rm r}$ systems.
%

%%%%%%%%%%%%%%%%%%%%%%%%%%%%%%%%%%%%%%% Figure. 1
%%%%%%%%%%%%%%%%%%%%%%%%%%%%%%%%%%%%%%%

%\clearpage
%%%%%%%%%%%%%%%%%%%%%%%%%%%%%%%%%%%%%%
%%%%%%%%%%%%%%%%%%%%%%%%%%%%%%%%%%%%%%%%%%%%

%--------------------------------------------------------------------

\bibliography{sample63}{}
\bibliographystyle{aasjournal}

\end{document}